\documentclass[12pt]{article}

\usepackage{jheppub,hyperref,float,array,adjustbox,mathtools,physics, xcolor}
\usepackage{graphicx,latexsym} 
\usepackage{amsthm,lscape}
\usepackage{comment}

\usepackage{amsmath,amssymb,amsfonts,amsxtra, mathrsfs, makeidx,graphicx,amsthm,epsfig, bm,longtable,float, 
color,tikz,mathtools,xfrac,footnote,rotating, lscape, makecell, environ,mathtools, empheq, physics,cleveref,tensor,
slashed,subfiles,natbib,youngtab,multirow}

\usepackage[font=small]{caption}
\newcommand{\Pf}[1]{\,\mathrm{Pf} #1}
\DeclareMathOperator{\USp}{USp}

\DeclareMathOperator{\SU}{SU}
\DeclareMathOperator{\UU}{U}
\newcommand{\jkint}{\oint\displaylimits_{\mathrm{JK}}}
\newcommand{\ee}{\mathrm{e}}
\newcommand{\mi}{\mathrm{i}}

\title{
\begin{center}
 Dualities from dualities in 2d $\mathcal{N}=(0,2)$
\end{center}
}

\author[a]{Antonio Amariti,}	
\author[a,b]{Pietro Glorioso,}	
\author[c]{Fabio Mantegazza,}
\author[a,b]{Davide Morgante }
\author[a,b]{and Andrea Zanetti}

\affiliation[a]{INFN, Sezione di Milano, Via Celoria 16, I-20133 Milano, Italy}
\affiliation[b]{Dipartimento di Fisica, Università degli studi di Milano, Via Celoria 16, I-20133}
\affiliation[c]{Deutsches Elektronen-Synchrotron DESY, Notkestr. 85, 22607 Hamburg, Germany}

\emailAdd{antonio.amariti@mi.infn.it}
\emailAdd{pietro.glorioso@mi.infn.it}
\emailAdd{fabio.mantegazza@desy.de}
\emailAdd{davide.morgante@mi.infn.it}
\emailAdd{andrea.zanetti@mi.infn.it}

\abstract{
We propose  2d $\mathcal{N}=(0,2)$ dualities between $\SU(N)$ gauge theories 
with fundamental and antisymmetric chiral matter and Landau-Ginzburg theories with chiral and Fermi multiplets.
Many of these dualities can be derived by topologically twisting 4d s-confining gauge theories
on a two-sphere, with integer non-negative $R$ charges. We provide various checks of the dualities, showing that they descend from more “basic" dualities, similarly to analogous derivations in higher dimensions.
The proof is based on the fact that the antisymmetric tensors can be traded with $\USp(2n)$ gauge theories 
with fundamental chirals, mimicking the higher dimensional mechanism known as tensor deconfinement.
The quivers obtained in this way can be shown to be dual to LG models after applying other elementary “basic" dualities.
}

\begin{document}
\maketitle
\flushbottom
\allowdisplaybreaks

\section{Introduction}

In the last years we are experiencing deep developments towards 
the definition and the understanding of a possible 
principle that allows to recast the known supersymmetric infrared dualities 
in terms of fundamental ones.

Restricting to 3d and 4d cases with four supercharges,
the existence of such an organizing principle has allowed the authors of \cite{Pasquetti:2019uop,Bottini:2022vpy,Benvenuti:2020gvy,Bajeot:2022lah,Amariti:2022wae,
Amariti:2023wts,Bajeot:2023gyl,Amariti:2024sde,Amariti:2024gco,Benvenuti:2024glr,Hwang:2024hhy} to derive many of the dualities proposed in the literature in terms of the fundamental bricks worked out in \cite{Seiberg:1994pq,Intriligator:1995ne} in 4d and in \cite{Aharony:1997gp} in 3d.
Furthermore, the fact that the 3d dualities of \cite{Aharony:1997gp} descend from the dimensional reduction of the 4d dualities of \cite{Seiberg:1994pq,Intriligator:1995ne}, through the procedure spelled out in \cite{Aharony:2013dha}, allows to reduce such fundamental bricks
to the 4d ones.

A less explored territory is the one of 2d models with 2d 
$\mathcal{N}=(0,2)$ supersymmetry. Even if this setup has less supercharges, it can be thought as the analog of the 4d $\mathcal{N}=1$ and  3d $\mathcal{N}=2$ cases discussed above, because it is the minimal case where supersymmetry is equipped with holomorphy and because 
it has an abelian $\UU(1)_R$ symmetry that has to be obtained through an extremization procedure \cite{Benini:2012cz}, in order to provide a well defined SCFT.

A well defined prescription to obtain  2d dualities from 4d was worked out in \cite{Gadde:2015wta}.
The prescription consists of compactifying the 4d dualities on a two-sphere while turning on a background flux for the R-symmetry, in order to preserve generically $\mathcal{N}=(0,2)$ supersymmetry.
Such flux is fixed by selecting a 4d non anomalous R-symmetry assigning non-negative integer charges to the 4d 
superfields. In this way it is possible to obtain 2d dualities, inherited from 4d, avoiding finite size effects and sums over theories, that are otherwise quite ubiquitous\footnote{See for example a recent discussion on this topic in \cite{Nardoni:2024sos}, where negative charges have been considered as well.}.
Even if such prescription allows to obtain large families of 2d dualities for special unitary gauge groups (and the triality of \cite{Gadde:2013lxa} in the unitary case), it has been observed that for $\USp(2N)$  only the reduction of the confining case of \cite{Intriligator:1995ne} is consistent, giving rise to a duality between a gauge theory and a Landau-Ginzburg (LG) model. The absence of a fundamental duality for $\USp(2N)$ is an essential reason why 2d $\mathcal{N}=(0,2)$ cases have been less studied so far in order to search for the existence of an organizing principle.

However, restricting to the $\SU(N)$ and the $\USp(2N)$ limiting cases,
i.e. the cases where the dual gauge group vanishes and where the dual description corresponds to a LG model, one can see that there are many similarities with the cases studied in higher dimensions. As stated above,
the 2d dualities for these theories are derived from the 4d ones by a topologically twisted compactification on a two-sphere \cite{Gadde:2015wta}, and by inspection the 2d dualities share common features with their 4d parents. Such similarity inspires the attempt of finding an organizing principle also in 2d. 

Such line of thought has motivated the analysis of \cite{Sacchi:2020pet}, where it was shown that 2d $\mathcal{N}=(0,2)$ $\USp(2N)$ with an antisymmetric and four fundamental chirals is dual to a LG model with a tower of Fermi multiplets interacting with a set of chirals corresponding to the dressed electric mesons of the gauge theory.
This duality can be derived from an analogous 4d $\mathcal{N}=1$ confining duality with an  $\USp(2N)$ gauge group, an antisymmetric and six fundamentals \cite{Cho:1996bi,Csaki:1996eu}. The relevant fact is that, through the same techniques used in 3d and in 4d, it can be proven
that the duality can be derived directly in 2d in terms of the fundamental brick, corresponding to $\USp(2N)$ with $2N+2$ fundamentals, originally worked out in \cite{Gadde:2015wta}.

Motivated by this result, here we aim to enlarge the web of 2d 
$\mathcal{N}=(0,2)$ gauge/LG duals, considering $\SU(N)$ gauge groups with antisymmetric chirals.
This is a rather natural way to proceed, indeed the 2d dualities we look for descend from 4d, where (at least in absence of superpotential and for a single gauge group) a full classification of s-confining gauge theories was provided in \cite{Csaki:1996zb}. Ignoring  possible sporadic cases, there are two main 4d dualities with $\SU(N)$ gauge group that have to be considered:
in the first case with an antisymmetric, four fundamentals and $N$ anti-fundamentals, while in the second case with an antisymmetric flavor and 3 fundamental flavors\footnote{Where an $\SU(N)$ flavor corresponds to a pair of conjugated representations.}.

\begin{figure}[H]
            \centering
              \includegraphics[width=0.45\linewidth]{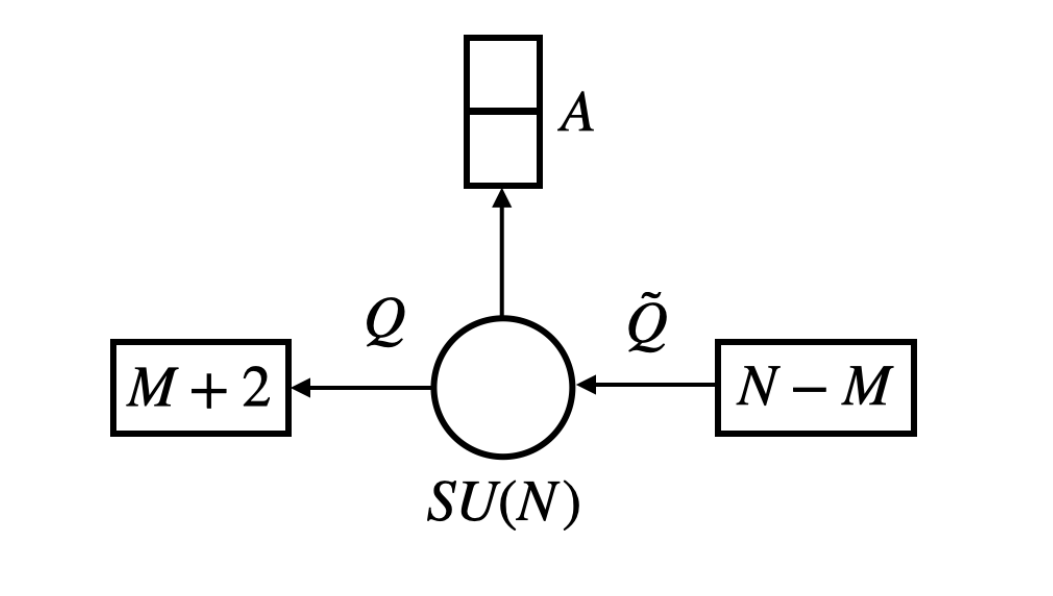}
              \caption{General quiver for the models studied in this section. $N=2n$ in \ref{Case1even}, \ref{Case2even} and \ref{Case3even}; $N=2n+1$ in \ref{Case1odd}, \ref{Case2odd} and \ref{Case3odd}. $M=0,\,1,\,2$.}
              \label{summary1}
          \end{figure}

We start in 4d by considering $\SU(N)$ with 4 fundamentals and $N$ anti-fundamentals and one antisymmetric, distinguishing the even and the odd case for the rank of the gauge group.
We focus on various consistent $R$ charge assignments, basically fixing $R=0$ for most of the (anti)-fundamental chiral multiplets, except for two fields that have $R=1$, while the antisymmetric is always fixed at $R=0$.
There are three possibilities, corresponding to $N-M$ anti-fundamentals   and $M+2$  fundamentals with $M=0,1,2$. The six gauge theories found in this way are summarized in Figure \ref{summary1}. In each case we have only 2d chirals in the matter content and we then expect that the 4d s-confined descriptions reduce to $\mathcal{N}=(0,2)$ LG models with both chiral and Fermi multiplets.
The dualities obtained in this way calls for a series of checks that we perform for each case. The most convincing analysis regards the ``derivation" of these dualities in terms of other simpler ones already conjectured and studied in the literature. Such dualities regards $\SU(N)$ and $\USp(2N)$ gauge theories with (anti)-fundamental matter. Then, the validity of our dualities follows from the validity of such fundamental ``bricks", in analogy with the analysis of \cite{Sacchi:2020pet} for the case of $\USp(2N)$ with four fundamentals and an antisymmetric.
A similar exploration is then carried out for the second class of families, corresponding to $\SU(N)$ with 3 fundamental  and one antisymmetric flavor.
In this case, with the same  $R$ charge assignment as above, we obtain (up to conjugation) two possibilities, either we have 3 fundamentals and one anti-fundamental or 2 fundamental flavors, in addition to the antisymmetric flavor. Again these dualities are shown to follow from the basic $\SU(N)$ and $\USp(2N)$ ones. 

Remarkably, the power of the approach adopted here, allowed us to derive new 2d $\mathcal{N}=(0,2)$ dualities that could not have been guessed by the topological twist of any 4d s-confining gauge theory in the classification of \cite{Csaki:1996zb}. A first model corresponds to $\SU(2n)$ with an antisymmetric flavor and four fundamentals.  A second model, corresponds to $\USp(4)$ with two antisymmetric tensors and two fundamentals.

Actually a comment here is in order. The dualities proposed for these two models without an s-confining known 4d parent are justified by the tensor deconfinement technique, in analogy with the other cases that possess a 4d parent 
s-confining duality. In principle the validity of the deconfinement technique at the level of the elliptic genus is subtle because 
it relies on the JK-prescription, and there is not a general proof that it can be applied partially in a product gauge group model, i.e. a model that is obtained after deconfining a tensor or by dualizing the original node in the deconfined theory.
While for the models with a 4d parent a justification of such mechanism follows from the relation between the elliptic genus and the $S^2 \times T^2$ partition function with the corresponding topological twist, the two models without a 4d parent do not 
have a similar justification and further studies are required. The model with an $\SU(2n)$ gauge group, an antisymmetric flavor and four fundamentals has been further investigated in \cite{Amariti:2025jvi}, where  a 4d parent duality, that leads to the 2d confining duality studied here, was found.
On the other hand we did not find yet a 4d parent of the gauge/LG duality involving $\USp(4)$ with two antisymmetric tensors and two fundamentals. For this reason we will further study this model below, by explicitly evaluating the elliptic genus and matching such explicit calculation with the one obtained from the tensor deconfinement technique.

The paper is organized as follows. In Section \ref{sec:Review} we give a brief review of some basic tools that we have used in the rest of the paper for the study of 2d $\mathcal{N}=(0,2)$ theories.
First we survey the superspace, the representations of the matter fields, the action, the gauge and the 't Hooft anomalies. Then we discuss the main aspects of the derivation of the 2d dualities from the topological twist 
and the relation to localization computations. Focusing on the latter we review the main aspects of the 2d elliptic genus and fix the notations that we adopt.
Then in Section \ref{sec:SU1AS} we study the first class of examples, corresponding to 2d $\SU(N)$ gauge theories with fundamental and anti-fundamental chiral flavors and one antisymmetric chiral field.
We have found six models that give rise to a 2d duality with an LG model, and we have provided for each case 
the derivation of the duality from other basic dualities already proposed in the literature. In each case we have corroborated the claim by deriving them from 4d, by computing the 't Hooft anomalies, and by providing the matching of the elliptic genera using the more fundamental identities associated to the basic dualities.
A similar analysis is performed in Section \ref{sec:SU2AS} for 2d $\SU(N)$ gauge theories with fundamental and anti-fundamental chirals and one antisymmetric flavor. While, similarly to the case with a single antisymmetric, we have found (four) cases with a 4d origin, here we have also found a case that cannot be obtained from the topological twist of a 4d $\mathcal{N}=1$ s-confining gauge theory. This case corresponds to $\SU(2n)$ with four fundamentals and one antisymmetric flavor. Even if this example does not have a 4d origin we have seen that all the other checks are satisfied by the duality.
In Section \ref{sec:usp4} we extend the discussion to the symplectic case, discussing another 2d duality without 4d origin
consisting of $\USp(4)$ with two fundamentals and two antisymmetrics.

In Section \ref{Andrea1} we provide some general comments on the relation between the tensor deconfinement technique and the (partial) evaluation of the integrals by using the JK prescription.
In Section \ref{Andrea2} we evaluate the index for $\USp(4)$ with two antisymmetric tensors and two fundamentals.

In Section \ref{commentsR} we comment on the c-extremization procedure in the various models discussed in the paper and to the fact that the presence of a non-compact target space forces us to fix the $R$ charges to be vanishing for the chirals.
In Section \ref{sec:remarks} we provide some general remarks of our analysis.
Speculations and possible further directions are discussed in Section \ref{sec:conclusions}.
We have also added an appendix \ref{sec:basic} reviewing the basic dualities used in the body of the paper in order to prove the 2d dualities in Section \ref{sec:SU1AS}, \ref{sec:SU2AS} and \ref{sec:usp4}.

\section{Review}
\label{sec:Review}

In this Section we briefly review some aspects of 2d $\mathcal{N}=(0,2)$ theories, focusing on the derivation of such models from 4d and on
the relation with the elliptic genus.

The superspace is parameterized by the coordinates $(x^0,x^1,\theta^+,\overline \theta^+)$ and the field content  consists of vector, chiral and Fermi multiplets.
The vector multiplet contains a gauge boson, two adjoint chiral
fermions and an auxiliary field:
\begin{equation}
    V_- = v_- - 2 \mi \theta^+\chi_- - 2 \mi \bar{\theta}^+\bar{\chi}_- + 2 \theta^+ \bar{\theta}^+ D.
\end{equation}
The chiral multiplet is defined as
\begin{equation}
\Phi = \phi + \sqrt{2}\theta^+\psi_+ - \mi \theta^+ \overline \theta^+ \partial_+ \phi , \quad \overline{\mathcal{D}}_+ \Phi = 0, 
\end{equation}
where $\phi$ is a complex scalar, $\psi_+$ is a chiral fermion and $ \overline{\mathcal{D}}_+ $ is the super-covariant derivative, and they are the on shell degrees of freedom.
The last type of multiplet is the Fermi multiplet
\begin{equation}
\label{FermiLambda}
\Psi = \psi_- - \sqrt{2}\theta^+ G - \mi \theta^+ \bar{\theta}^+ \partial_+ \psi_- - \sqrt{2}\bar{\theta}^+ E,\quad
\overline{\mathcal{D}}_+ \Psi = E(\Phi),
\end{equation}
where $E(\Phi)$ is a holomorphic function of the chiral multiplets.
In this case $G$ is an auxiliary field and $\psi_-$ is a chiral left-moving fermion, that is the only on-shell degree of freedom.

The $E$-term in (\ref{FermiLambda}) introduces an interaction between the Fermi and the chiral multiplets through the kinetic term for the Fermi multiplet in the lagrangian.
The other way to introduce an interaction corresponds to the introduction in the lagrangian of a J-term
\begin{equation}
\mathcal{L}_J =  \int \dd \theta^+\, \,\Psi\,J(\Phi)\Big|_{\overline \theta^+=0} + \text{h.c}.\,,
\end{equation}
where $J(\Phi)$ is an holomorphic function of the chiral multiplet. For a model with $n_F$ Fermi fields the $E$-terms and the $J$-terms must satisfy the relation 
\begin{equation}
\sum_{a=1}^{n_F} \Tr [E_a(\Phi) J_a(\Phi)] =0.
\end{equation}

In this paper we have considered only models with vanishing $E$-terms, such that the constraint is automatically satisfied.

A relevant role in the analysis is played by the anomalies.
Anomalies are quadratic in 2d and the contribution of the multiplets does not depend on the conjugation of the representation.
Anomalies depend on the chirality matrix $\gamma_3$ in 2d and, given two abelian symmetries $\UU(1)_a$ and $\UU(1)_b$, the mixed anomaly is given by 
\begin{equation}
\label{anodm}
\kappa_{ab} \equiv \Tr \gamma_3 \UU(1)_a \UU(1)_b\,.
\end{equation}
Furthermore, there are anomalies involving non-abelian symmetries. The right-moving central charge is given by the quadratic anomaly for the $R$ symmetry, $c_R = 3 \Tr \gamma_3 R^2$ while the left central charge is obtained from the gravitational anomaly, from the relation $c_R-c_L = \Tr \gamma_3$.
The gauge anomaly is given by
\begin{equation}
\Tr \gamma_3 G^2 = \sum_{i \in\text{Chirals}} T(\mathfrak{R}_i)-\sum_{i \in\text{Fermi}} T(\mathfrak{R}_{i}) - T(\mathrm{Adj})
\end{equation}
where $T(\mathfrak{R})$ is the quadratic Casimir of the representation $\mathfrak{R}$ of each charged chiral and Fermi multiplet under the gauge group $G$.
Here we will consider only $\SU(N)$ and $\USp(2N)$ gauge groups, such that there are no mixed anomalies involving the gauge symmetry and the abelian flavor symmetries. 
In all the examples below we will study the matching of the 't Hooft anomalies for the global symmetries.

There is a general procedure to construct 2d $\mathcal{N}=(0,2)$ gauge theories starting from 4d $\mathcal{N}=1$
gauge theories, compactifying them on a two-sphere.  
At the level of the 2d theory half of the supersymmetry is preserved if one  turns on a background $R$ symmetry gauge field with unit magnetic flux through the two-sphere \cite{Closset:2013sxa,Benini:2015noa,Honda:2015yha}.  
The non-anomalous $R$ charge need to be quantized and, depending on its value, we are left in 2d with  \cite{Closset:2013sxa,Kutasov:2013ffl}
\begin{itemize}
\item $r-1$ Fermi multiplets if we consider a 4d superfield with  $R$ charge $r>1$
\item $1-r$ Chiral multiplets  if we consider a 4d superfield with  $R$ charge $r<1$
\item no multiplets if we consider a 4d superfield with  $R$ charge $r=1$
\end{itemize}
On the other hand a vector multiplet reduces to a vector multiplet. 
Furthermore, the interactions can be read from the 4d ones.

The field theoretical reduction can be also studied by reducing the corresponding 4d identity for the 
topologically twisted index \cite{Benini:2015noa} to 2d. Such reduction gives rise in general to a sum over the flux sectors, which is understood as the fact that one theory in 4d reduces to a direct sum of theories in 2d.
On the other hand the subclass of reduction with integer non-negative $R$ charge for  all the chiral multiplets allows to reduce to the zero-flux sector \cite{Gadde:2015wta}, implying that one reduces the topologically twisted index to the elliptic genus of a single 2d model.

Here we conclude by reviewing some basic aspects  of the elliptic genus.
The elliptic genus was computed in the RR sector in \cite{Benini:2013nda,Benini:2013xpa}
and in the NSNS sector in \cite{Gadde:2013dda,Gadde:2013wq}. Here we adopt the conventions in the NSNS sector.
The index is defined as
\begin{equation}
I(\vec u;q) \equiv I(\vec u) \equiv  \Tr_{\text{NSNS}}(-1)^F q^{L_0}\prod_{a} u_a^{c_a},
\end{equation}
where $q=\ee^{2 \pi \mi \tau}$ and $\tau$ is the complex structure of the torus. The elliptic genus corresponds to the Witten index refined by the flavor fugacities $u_a$.
If we consider a gauge theory with gauge group $G$ the elliptic genus can be equivalently associated to the following matrix integral over the maximal abelian torus of $G$, parameterized by the fugacity $z$
\begin{equation}
I(u) = \frac{1}{|W|} \jkint \prod_{i=1}^{\text{rk} \, G} \frac{\dd z_i}{2 \pi \mi z_i} I_{V}(\vec z) I_{\chi} (\vec z,\vec u) I_{\psi} (\vec z,\vec u),
\end{equation}
where $|W|$ is the dimension of the Weyl group and JK defines the Jeffrey-Kirwan contour, which specifies a sum of JK residues \cite{Benini:2013nda, Benini:2013xpa} for the evaluation of the integral.
The contribution of the vector, chiral and Fermi multiplets are
\begin{eqnarray}
I_{V}(\vec z) &=& (q;q)_{\infty}^{2 \text{rk}\,G}\prod_{\alpha_G} \theta \left(z^{\alpha_G}\right),\nonumber \\
I_\chi(\vec z,\vec u) &=& \prod_{\rho_G,\rho_F} \frac{1}{\theta \left(q^{\frac{R_\chi}{2}} z^{\rho_G} u^{\rho_F}\right)},  \\
I_{\psi}(\vec z,\vec u) &=&\prod_{\rho_G,\rho_F} \theta\left( q^{\frac{R_{\psi}+1}{2}} z^{\rho_G} u^{\rho_F}  \right),\nonumber
\end{eqnarray}
where $\theta(x) = (x;q)_{\infty}  (q x^{-1} ;q)_{\infty}$ and $(x;q)_\infty = \prod_{j=0}^{\infty}(1-x q^j)$.

In the rest of the paper we mainly refer to $\SU(N)$ and $\USp(2N)$ 
gauge theories with chiral and Fermi multiplets in the (anti-)fundamental 
and in addition chirals in the antisymmetric representation.
For this reason here  we summarize the various 
conventions that we have adopted below.

The index of an $\SU(N)$ gauge theory with $F$ fundamentals $Q$,
$\tilde F$ anti-fundamentals $\tilde Q$, $H$ fundamental Fermi $\Lambda$\footnote{Actually either fundamentals or anti-fundamentals, because the
two representations are equivalent for a Fermi multiplet.},
$K$ antisymmetrics $A$ and $\tilde K$ conjugate antisymmetrics $\tilde A$ is denoted as
\begin{align}
  I_{\SU(N)}^{(F,\tilde F;H;K;\tilde K)}(\vec m;\vec n; \vec h;\vec r,\vec s)&=\frac{(q;q)_{\infty}^{2(N-1)} }{N!}\nonumber\\
  &\times \jkint \prod_{i=1}^{N} \frac{\dd z_i}{2 \pi \mi z_i} 
  \frac{\prod_{i < j } \theta \left((z_i/ z_j)^{\pm 1}\right)
  \prod_{i=1}^{N} \prod_{a=1}^{H}  \theta \left(q^{\frac{R_\Lambda+1}{2}} z_i h_a \right)}{\prod_{i=1}^{N} \left(\prod_{a=1}^{F}
  \theta \left(q^{\frac{R_Q}{2}} z_i m_a \right) \cdot \prod_{a=1}^{\tilde F} \theta \left(q^{\frac{R_{\tilde Q}}{2}} z_i^{-1} n_a \right)\right)}\nonumber\\[5pt]
  &\times \frac{\delta(1-\prod_{i=1}^{N} z_i)}{\prod_{i < j } \left(\prod_{a=1}^{K} \theta \left( q^{{\frac{R_A}{2}}} r_a z_i z_j\right) \cdot \prod_{a=1}^{\tilde K} \theta \left( q^{\frac{R_{\tilde A}}{2}} s_a z_i^{-1} z_j^{-1}\right)\right)} \,.
\end{align}
If one of more of the fields are absent we omit the relative fugacity using a $\cdot$ symbol.
For example the index of $\SU(N)$ with $F$ fundamentals and $F$ anti-fundamentals is denoted as
$I_{\SU(N)}^{(F;F;\cdot;\cdot;\cdot)}(\vec m;\vec n;\cdot,\cdot;\cdot)$.

On the other hand, the index of an $\USp(2N)$ gauge theory with $F$ fundamentals $Q$,
$H$ fundamental Fermi multiplets  $\Lambda$ and one (totally)
antisymmetric chiral $A$ is denoted as
\begin{eqnarray}
I_{\USp(2N)}^{(F;H;1)} (\vec m; \vec h;r)
 &=& 
 \frac{(q;q)_\infty^{2N}}{2^N N!  \, \theta \left(q^{r_A}r\right)^{N-1}}
 \jkint \prod_{i=1}^{N} \frac{\dd z_i}{2 \pi \mi z_i} 
\frac{\prod_{i < j } \theta \left(z_i^{\pm 1} z_j^{\pm 1}\right)
 \prod_{i=1}^{N}  \theta (z_i^{\pm 2 })}
 {\prod_{i < j } \theta \left(q^{\frac{R_A}{2}} z_i^{\pm 1} z_j^{\pm 1} r\right)}
\nonumber \\
 &\times& 
 \prod_{i=1}^{N}  \frac{\prod_{a=1}^{H}  \theta \left(q^{\frac{R_\Lambda+1}{2}} z_i ^{\pm 1} h_a \right)}{
\prod_{a=1}^{F} \theta \left(q^{\frac{R_Q}{2}} z_i^{\pm 1} m_a \right)}\,.
\end{eqnarray}
In absence of the antisymmetric we refer to the index as $I_{\USp(2N)}^{(F;H;\cdot)} (\vec m; \vec h;\cdot)$.

Observe also that the $R$ symmetry that appears in these indices does not necessarily represent the exact $R$ symmetry, and  we adopted the convention (compatible with all the examples studied here) that the $R$ charges of the chiral multiplets are vanishing and that the $R$ charges of the Fermi are $R_\Lambda=1$.

We conclude this section by commenting on a very useful relation that follows from the definition of the theta function. The relation is 
\begin{equation}
\label{inversion}
\theta(x) = \theta (q/x)\,.
\end{equation}
While this relation is mathematically trivial, it is physically meaningful, as it is commonly interpreted as the statement that one can conjugate a Fermi multiplet by exchanging a $\mathrm{J}$-term with an $\mathrm{E}$-term. 
Here it will play also a crucial in implementing the flipping of an operator in a duality.
The flipping procedure corresponds to moving a singlet from one side of the equality to the other, by introducing a singlet with opposite statistics.
By using the relation (\ref{inversion}) we can reproduce the field theory argument, where such flip is implemented through a J-term, at the level of the elliptic genus.
For example, let us consider on the electric side a gauge invariant operator $\mathcal{O}$ corresponding to a combination of chiral fields $\phi_i$ and possibly one Fermi field $\psi$ (entering at linear level),  i.e. $\mathcal{O}= f(\phi_i,\psi)$.
If this operator is a singlet of a dual phase, either corresponding to a gauge theory or to a LG model, we can flip such singlet
by adding a $J$ term involving the operator $\mathcal{O}$. Such flipper corresponds to adding in the 2d superpotential either a term $\Delta W_{2\mathrm{d}}^{(\text{dual})} = \Psi \mathcal{O}_\chi$, if the singlet is a chiral field, or $\Delta W_{2\mathrm{d}}^{(\text{dual})} = \Phi \mathcal{O}_F$, if the singlet is a Fermi. In both cases this corresponds to a mass term, and we can integrate out the fields $(\mathcal{O}_\chi,\Psi)$ or 
$(\mathcal{O}_F,\Phi)$ from the dual model.
On the other hand, in the electric theory we add the 2d superpotential deformation $\Delta W_{2\mathrm{d}}^{(\text{ele})} = \Psi f(\phi_i,\psi)$ or $\Delta W_{2\mathrm{d}}^{(\text{ele})} = \Phi f(\phi_i,\psi)$.
At the level of the elliptic genus the singlet dual to the operator $\mathcal{O}$ can either be represented as $\mathcal{I}_\chi=\theta(x)^{-1}$ or $\mathcal{I}_F=\theta(q/ x)$, depending on its bosonic or fermionic nature, using the conventions of the R-charges described above, i.e. all the Fermi have charge $1$ and all the chirals have $R$-charge 0 (actually this discussion can be relaxed here, by turning on a $q$-behavior for generic fugacity $x$, but we prefer to keep the notations simpler in the discussion at hand).
The contribution to the elliptic genus of the flipper is dictated by the J-term $\Delta W_{2\mathrm{d}}^{(\text{dual})} $ and it either corresponds to $\mathcal{I}_\Psi=\theta (q/ x)$ or $\mathcal{I}_\Phi= \theta(x)^{-1}$.
On the dual side this simplifies the contribution of $\mathcal{I}_\chi$ or $\mathcal{I}_F$  thanks to the formula (\ref{inversion}). On the other hand, on the electric side we multiply the elliptic genus by $\mathcal{I}_{\Psi}$ or $\mathcal{I}_{\Phi}$.

\section{$\SU(N)$ with one antisymmetric}
\label{sec:SU1AS}

In this section we consider the first class of examples, corresponding to $\SU(N)$ gauge theories with $N-M$ anti-fundamental, and $M+2$ fundamental chirals (with $M=0,1,2$) and one antisymmetric tensor.
These are anomaly free gauge theories and we are going to support the claim that they are dual to LG theories.
The details of the LG descriptions require to separate the discussion in each case for $N=2n$ and  $N=2n+1$.

\subsection{$\SU(2n)$ with $2n$ $\overline \square$, $2$ $\square$}
\label{Case1even}
In this case the LG is given by five chiral fields $\Phi_I$ corresponding to the gauge invariant combinations
\begin{eqnarray}
\label{combo1}
\Phi_1 = A^{n-1}  Q^2,\quad
\Phi_2 = A \tilde Q^2,\quad
\Phi_3 = Q \tilde Q     ,\quad
\Phi_4 = \tilde Q^{2n},\quad
\Phi_5 = \text{Pf}\, A,    
\end{eqnarray}
and two Fermi multiplets $\Psi_{1,2}$ interacting with the chirals through a J-term (a.k.a. a 2d superpotential)
\begin{equation}
\label{guess}
W=\Psi_1 (\Phi_2 ^{n-1} \Phi_3^2 + \Phi_1 \Phi_4 )
+
\Psi_2 (\Phi_2^n + \Phi_5 \Phi_4).
\end{equation}
We want to show that this duality  descends from the two basic gauge/LG dualities reviewed in the Appendix \ref{sec:basic}, i.e. $\USp(2n)$ with $2n+2$ fundamentals and 
$\SU(n)$ with $n$ fundamental flavors.

\begin{figure}
\begin{center}
  \hspace*{-1cm}\includegraphics[width=12cm]{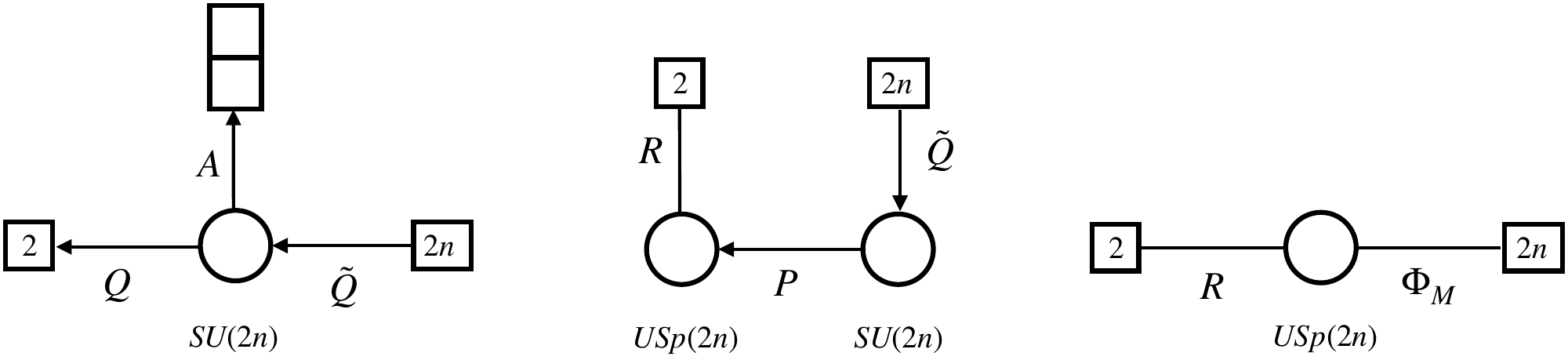}
  \end{center}
  \caption{ The first quiver  represents  the $\SU(2n)$ gauge theory with an antisymmetric $A$, $2n$ anti-fundamentals $\tilde Q$ and two fundamentals $Q$. The second quiver is obtained by trading the antisymmetric $A$ and the two fundamentals $Q$ with an auxiliary $\USp(2n)$ gauge node with the bifundamental $P$ and the fundamentals $R$. The third quiver is obtained by dualizing the $\SU(2n)$ node into a LG theory. Observe that in the quivers we did not represent the singlets that arise in the various steps, as they are discussed in detail in the discussion appearing in the paper.}
    \label{firstcase}
\end{figure}

In order to simplify our analysis we add a J-term to the electric theory\footnote{With an abuse of terminology, from now on we refer to the gauge theories we start with as ``electric", borrowing the 4d nomenclature.}
corresponding to 
\begin{equation}
\label{flipped}
W = \psi_A \epsilon_{\alpha_1 \alpha_2 }\epsilon^{i_1\ldots i_{2n}}A_{i_1\,i_2}\cdots A_{i_{2n-3}\,i_{2n-2}} Q_{i_{2n-1}}^{\alpha_1} Q_{i_{2n}}^{\alpha_2}.
\end{equation}
In the rest of the paper, gauge and flavor contractions are going to be understood when not specified otherwise. 
Then we trade the antisymmetric with an $\USp(2n)$ gauge theory
as in Figure \ref{firstcase}
with superpotential
\begin{equation}
W = \Psi_R R^2.
\end{equation}
This process is the 2d counterpart of the Berkooz deconfinement technique for two-index tensor matter fields in 4d \cite{Berkooz:1995km,Luty:1996cg}. Such procedure has been already used in 2d $\mathcal{N}=(0,2)$ theories in \cite{Sacchi:2020pet}. In the rest of the paper we will often refer to the mechanism as ``deconfinement'' of a tensor. 

The next step consists of dualizing the $\SU(2n)$ gauge node that has
$2n$ fundamentals $P$ and $2n$ anti-fundamentals $\tilde Q$, using the duality in Appendix \ref{sec:SUNNN}.
Defining the following $\SU(2n)$ gauge invariant chiral fields $\Phi_M = P \tilde Q$, $\Phi_B = P^{2n}$ and $\Phi_{\tilde B} = \tilde Q^{2n}$
the superpotential of the dual theory is
\begin{equation}
W = \Psi_R R^2 + \Psi_{\SU(2n)} (\det \Phi_M + \Phi_{\tilde B} \Phi_{\phantom{\tilde{B}}\!\!\!\!\! B}  )
\end{equation}
The last step corresponds to dualizing the $\USp(2n)$ node with $2$ fundamentals $R$ and $2n$ fundamentals $\Phi_M$, using the results reviewed in Appendix \ref{sec:USP2Np2}.
The gauge invariant combinations in this case are $\Phi_{s} = R^2$,
$\Phi_{A} = \Phi_M^2$ and $\Phi_q = R \Phi_M$.
The superpotential for this theory is
 \begin{equation}
 \label{almostfinal}
W = \Psi_R \Phi_s + \Psi_{\SU(2n)} (\Phi_{A}^{n} + \Phi_{\tilde B} \Phi_{\phantom{\tilde{B}}\!\!\!\!\! B}  )
+\Psi_{\USp(2n)} \text{Pf} \left(
\begin{array}{cc}
\Phi_A&\Phi_q \\
-\Phi_q^T&\Phi_s
\end{array}
\right).
\end{equation}
We can integrate out the massive field $\Phi_s$ using the first J-term in (\ref{almostfinal}) and in this way we are left with 

\begin{equation}
 \label{tfinal}
W = \Psi_{\SU(2n)} (\Phi_{A}^{n} + \Phi_{\tilde B} \Phi_{\phantom{\tilde{B}}\!\!\!\!\! B}  )
+\Psi_{\USp(2n)} \Phi_{A}^{n-1} \Phi_q^2.
\end{equation}
This superpotential corresponds to (\ref{guess}) with flipped $\Phi_1$ 
with the dictionary
\begin{equation}
    \Psi_1 \leftrightarrow  \Psi_{\USp(2n)},\quad \Psi_2 \leftrightarrow  \Psi_{\SU(2n)},\quad\Phi_2 \leftrightarrow  \Phi_A,\quad\Phi_3 \leftrightarrow  \Phi_q,\quad \Phi_4 \leftrightarrow   \Phi_B,\quad \Phi_5 \leftrightarrow  \Phi_{\tilde B}.
\end{equation}
Even if the presence of the flipper $\psi_A$ in (\ref{flipped}) reproduces only partially
the superpotential (\ref{guess}), setting $\Phi_1=0$, one can engineer a different derivation, using another flipped
superpotential instead of  (\ref{flipped}). For example, by flipping the operator $\Pf A$, we have obtained, through a similar analysis, the  superpotential (\ref{guess}), but this time with $\Phi_5=0$.
Similar comments holds in many of the examples below and we will not discuss them further.

One may wonder if it is also possible to reconstruct the full superpotential (\ref{guess}) by using our procedure 
\footnote{We are extremely grateful to the referee for suggesting us to complete our analysis in this direction.}.
This is possible by observing that the Fermi $\psi_A$ on the electric side corresponds to the combinations $\Psi_1 \Phi_4$ 
in the dual LG. At this level we can add a chiral flipper,  that we denote as $\varphi$ in the dual LG description, corresponding to the deformation $\Delta W = \varphi \Phi_4 \Psi_1$. This reflects on the gauge theory side of the duality in the addition of the 2d superpotential $\Delta W = \psi_A \varphi$.
Integrating out the massive fields the electric superpotential is vanishing, but the equations of motion also fix the relation $\varphi = A^{n-1} Q^2$, i.e. they identify the singlet $\varphi$ with the singlet $\Phi_1$ in the dual WZ model.
In this way we have reconstructed the expected 2d superpotential term $ \Psi_1 \Phi_1 \Phi_4$ that was missing in the analysis above.
Observe that similar comments apply to all the examples discussed below. However, we cannot start our analysis choosing $W=0$ in the electric phase, because in the dual LG phases it implies the presence of Fermi fields that do not have a clear 
interpretation in terms of gauge invariant operators of the gauge theory. On the other hand, if we flip the electric theory
such Fermi in the dual can be related through the duality map to the electric Fermi flipper, by mixing them with other gauge invariant combinations of chirals. 
Once such correspondence is established we can ``flip the flipper" and finding the expected duality for electric theories with vanishing $W_{2\mathrm{d}}$.

The 2d duality can be derived by topologically twisting the 4d s-confining duality involving an $\SU(2n)$ gauge theory with $2n$ fundamentals, $4$ fundamentals and one antisymmetric derived in \cite{Csaki:1996zb}.
The twist is done along the 4d non-anomalous $R$ symmetry that assigns $R$ charge 0 to the antisymmetric, the anti-fundamentals and two fundamentals and $R$ charge 1 to the remaining two fundamentals.
The confined degrees of freedom are
\begin{equation}
\label{CSAKIFIELDS}
\Sigma_1= A^{n-1}  Q^2,\quad
\Sigma_2= A \tilde Q^2,\quad
\Sigma_3 = Q \tilde Q,\quad
\Sigma_4= \tilde Q^{2n},\quad
\Sigma_5 = \text{Pf}\, A,\quad    
\Sigma_6 = A^{n-2}  Q^4,
\end{equation}
interacting through the superpotential
\begin{equation}
\label{WCSAKI}
W = \Sigma_5 \Sigma_3^4 \Sigma_2^{n-2}
+
\Sigma_1 \Sigma_3^2 \Sigma_2^{n-1}
+
\Sigma_6 \Sigma_2^n
+
\Sigma_4 \Sigma_5 \Sigma_6
+ 
\Sigma_4 \Sigma_1^2.
\end{equation}
When we twist this WZ model with the $R$ symmetry assignment 
discussed above we see that the 4d superfield $\Sigma_1$, in the antisymmetric representation of $\SU(4)$ splits into a chiral and a Fermi,
denoted respectively as $\Phi_1$ and $\Psi_1$ above, that are singlets under the surviving $\SU(2)$ flavor symmetry.
The components that survive in the 4d superfields $\Sigma_{2,\dots,5}$
are the 2d chirals $\Phi_{2,\dots,5}$ while the 4d superfield $\Sigma_6$ becomes the 2d Fermi $\Psi_2$.
One can also check that the first term in (\ref{WCSAKI}) does not survive in 2d while the other four terms  in (\ref{WCSAKI})  reconstruct the 2d superpotential (\ref{guess}).

We proceed by checking the anomaly matching of the global symmetries.
The charges of the fields in the electric and in the dual LG theory, including the flipper $\psi_A$ in (\ref{flipped}), are
\begin{equation}
\label{TableFields1}
\begin{array}{c|ccccc|c}
&\UU(1)_Q&\UU(1)_{\tilde Q} & \SU(2)& \SU(2n) & \UU(1)_A&\UU(1)_{R_0} \\
Q&1&0&\square&\cdot&0&0\\
\tilde Q&0&1&\cdot&\square&0&0\\
A&0&0&\cdot&\cdot&1&0\\
\psi_A&-2&0&\cdot&\cdot&1-n&1\\
\hline
\Phi_2&0&2&\cdot&\begin{array}{c} \square\vspace{-3mm} \\ \square\end{array}&1&0\\
\Phi_3&1&1&\square&\square&0&0\\
\Phi_4 &0&2n&\cdot&\cdot&0&0\\
\Phi_5&0&0&\cdot&\cdot&n&0\\
\Psi_1&-2&-2n&\cdot&\cdot&1-n&1\\
\Psi_2&\cdot&-2n&\cdot&\cdot&-n&1
\end{array}
\end{equation}
The anomalies of the global symmetries are given by
\begin{equation}
\begin{array}{lll}
&
\kappa_{QQ} = 4n-4,& \quad
\kappa_{\tilde Q \tilde Q} = 4n^2 , \\
&
\kappa_{AA}= n^2+n-1,& \quad
\kappa_{Q A}= 2(1-n),  \\
&
\kappa_{\tilde Q Q }= 0, & \quad
\kappa_{\tilde Q A }=  0,   \\
&
\kappa_{ \SU(2)^2} = n, & \quad
\kappa_{\SU(2n)^2  } = n,\\
&\kappa_{R_0R_0}= n(3+2n), &\quad
\kappa_{R_0A}=-2n^2+2n-1,\\
&\kappa_{R_0Q}= 2-4n, &\quad
\kappa_{R_0 \tilde Q }=-4n^2,\\
\end{array}
\end{equation}
and we checked that they match across the dual phases.

We conclude our analysis by providing a derivation of the duality 
from the elliptic genus.
The identity that we want to prove in this case is 
\begin{eqnarray}\label{toprove1}
&&
\theta\left(q/(x^2 t^{2n-2})\right)
I_{\SU(2n)}^{(2;2n;\cdot;1;\cdot)}(x \vec u;y \vec v;\cdot;t^2;\cdot)=
\prod_{1\leq a<b \leq 2n} \frac{1}{\theta(v_a v_b y^2 t^2)}
 \nonumber \\
&\times&
\frac{\theta(q/ (y^{2n} x^2 t^{2n-2})) \theta(q/(y t)^{2n})}{\theta( y^{2n})\theta(t^{2n}) }
\prod_{a=1}^{2} \prod_{b=1}^{2n} \frac{1}{\theta(u_a v_b x y )},
\end{eqnarray}
where $\prod_{a=1}^{2} u_a = \prod_{a=1}^{2n} v_a=1$.

Following the discussion above we have flipped the operator $ A^{n-1}  Q^2$ using the Fermi field $\psi_A$ in the superpotential (\ref{flipped}).
This amounts to considering the theta function $\theta\left(q/(x^2 t^{2n-2})\right)$ on the LHS of (\ref{toprove1}).

The next step corresponds to deconfine the antisymmetric tensor using the identity (\ref{ellipticdualUSP}).
This boils down to the following substitution in the integrand in the LHS of 
(\ref{toprove1}) 
\begin{eqnarray}
&&
\frac{\theta\left(q/(x^2 t^{2n-2})\right)}{\prod_{i=1}^{2n} 
\prod_{b=1}^{2}
\theta (z_i u_b x)
\prod_{i<j} \theta (z_i z_j t^2) }
\rightarrow \theta(q t^2/x^2) I_{\USp(2n)}^{(2,2n;\cdot;\cdot)}(x/t \vec u,t \vec z).
\end{eqnarray}
After this substitution we apply formula (\ref{ellipticdualSU}), corresponding to the $\SU(2n)$ duality.
We are left with an $\USp(2n)$ theory with $2n+2$ fundamentals and various chiral and Fermi singlets. The elliptic genus for this theory is
\begin{equation}
\frac{\theta(q/(y t)^{2n})}{\theta( y^{2n})\theta(t^{2n})}
 I_{\USp(2n)}^{(2,2n;\cdot;\cdot)}(x/t \vec u,t y \vec v).
\end{equation}
To conclude the proof we use (\ref{ellipticdualUSP}) in this integral, obtaining the RHS of (\ref{toprove1}).

\subsection{$\SU(2n+1)$ with $2n+1$ $\overline \square$, $2$ $\square$}
\label{Case1odd}
In this case the LG is given by four chiral fields $\Phi_I$ corresponding to the gauge invariant combinations
\begin{eqnarray}
\Phi_1 = A^{n}  Q,\quad
\Phi_2 = A \tilde Q^2,\quad
\Phi_3 = Q \tilde Q     ,\quad
\Phi_4 = \tilde Q^{2n+1},\quad 
\end{eqnarray}
and one Fermi multiplet $\Psi$ interacting with the chirals through
a superpotential
\begin{equation}
\label{expectedsecond}
W=\Psi (\Phi_2 ^{n} \Phi_3+ \Phi_1 \Phi_4 )\,.
\end{equation}

\begin{figure}
\begin{center}
  \includegraphics[width=12cm]{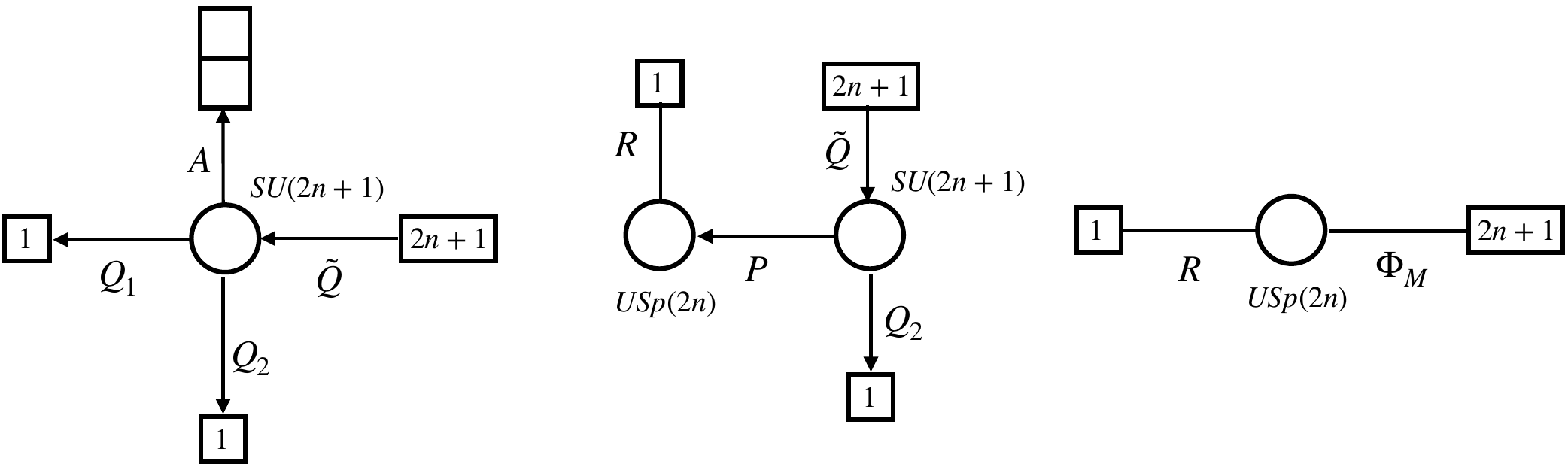}
  \end{center}
  \caption{The first quiver  represents  the $\SU(2n+1)$ gauge theory with an antisymmetric $A$, $2n+1$ anti-fundamentals $\tilde Q$ and two fundamentals $Q_{1,2}$. Observe that we split these two fundamentals in the figure because in the second quiver we traded the antisymmetric $A$ and just one of these two fundamentals (here  $Q_1$)  with an auxiliary $\USp(2n)$ gauge node with the bifundamental $P$ and the fundamental $R$. The third quiver is obtained by dualizing the $\SU(2n+1)$ node into a LG theory. Again we did not represent the various singlets in these figures.}
    \label{secondcase}
\end{figure}
In order to simplify our analysis we add a J-term to the electric theory
corresponding to 
\begin{equation}
\label{flipped2}
W = \psi_A A^{n}  Q_{2}\,.
\end{equation}
Then we trade the antisymmetric with an $\USp(2n)$ gauge theory
as in Figure \ref{secondcase}
with superpotential
\begin{equation}
W = 0\,.
\end{equation}
The next step consists of dualizing the $\SU(2n+1)$ gauge node that has
$2n$ fundamentals $P$, one fundamental $Q_2$ and $2n+1$ anti-fundamentals $\tilde Q$.
Defining the following gauge invariant chiral fields $\Phi_M = \tilde Q P $, $\Phi_s= Q \tilde Q_2$, $\Phi_B = P^{2n} Q_2$ and
$\Phi_{\tilde B} = \tilde Q^{2n+1}$ the superpotential of the dual theory is
\begin{equation}
W = \Psi_{\SU(2n+1)} \left(
\det 
 \left(
 \begin{array}{c}
\Phi_M \\
\Phi_s
\end{array}
\right)
+  \Phi_{\tilde B}\Phi_{\phantom{\tilde{B}}\!\!\!\!\! B} \right)\,.
\end{equation}
The last step corresponds to dualizing the $\USp(2n)$ node with $1$ fundamental $R$ and $2n+1$ fundamentals $\Phi_M$.
The gauge invariant combinations in this case are  $\Phi_A = \Phi_M^2$ and $\Phi_a = \Phi_M \tilde Q$
and the superpotential for this theory is
\begin{equation}
\label{finalWfromdec}
W = \Psi_{\SU(2n+1)} \left(\Phi_A^n \Phi_s + \Phi_{\tilde B} \Phi_{\phantom{\tilde{B}}\!\!\!\!\! B}\right)
+\Psi_{\USp(2n)} \Phi_A^n \Phi_a \,.
\end{equation}

We can compare this superpotential with the one guessed above in formula (\ref{expectedsecond}).
Notice that the addition of the flipper \eqref{flipped2} explicitly breaks the $\SU(2)$-flavor doublets.
In the dual LG  they are decomposed as 
\begin{equation}
    \Phi_1=(\Phi_1^{(1)},\Phi_1^{(2)})\,, \quad
    \Phi_3=(\Phi_3^{(1)},\Phi_3^{(2)})\,, \quad
    \Psi=(\Psi^{(1)},\Psi^{(2)})\,,
\end{equation}
and $\Phi_1^{(2)}$ is flipped because of \eqref{flipped2}.

Observe that in the table of charges (\ref{table325}) we decomposed the $\SU(2) \times \UU(1)$ flavor symmetry associated to the $\SU(2)$ doublet $Q$ as $U(1)_{Q_{1,2}}$ by mixing the abelian generator $J_3$ of the $\SU(2)$  symmetry and the extra abelian generator $I$ of the $\UU(1)$ flavor symmetry (assigning charge 1), by identifying $I/2+J_3$ as $\UU(1)_{Q_1}$ and $I/2-J_3$ as $\UU(1)_{Q_2}$.

The effect of this last is to modify (\ref{expectedsecond}) as
\begin{equation}
\label{expectedsecondmodified}
W=\Psi^{(1)} (\Phi_2 ^{n} \Phi_3^{(1)}+ \Phi_1^{(1)} \Phi_4 )
+\Psi^{(2)} \Phi_2 ^{n} \Phi_3^{(2)}\,.
\end{equation}
This superpotential is identical to (\ref{finalWfromdec}),  found by the deconfinement procedure, provided the dictionary
\begin{align}
    \Psi^{(1)} \leftrightarrow \Psi_{\SU(2n+1)},\quad &\Psi^{(2)} \leftrightarrow \Psi_{\USp(2n)},\quad \Phi_1^{(1)}  \leftrightarrow \Phi_B,\quad \Phi_2 \leftrightarrow \Phi_A,\nonumber\\[5pt]
    & \hspace{-30pt}\Phi_3^{(1)} \leftrightarrow \Phi_s,\quad\Phi_3^{(2)} \leftrightarrow \Phi_a,\quad \Phi_4 \leftrightarrow \Phi_{\tilde B}.
\end{align}

We proceed by checking the anomaly matching of the global symmetries. The charges of the fields in the electric and in the dual LG theory, including the flipper $\psi_A$ in (\ref{flipped2}), are
\begin{equation}
\label{table325}
\begin{array}{c|ccccccc}
            &    \UU(1)_{Q_1}  & \UU(1)_{Q_2} &  \UU(1)_{\tilde Q} & \SU(2n+1) & \UU(1)_A & \UU(1)_{R_0} \\
            \hline
Q _1         & 1 &     0    &   0     &  \cdot      &  0  &  0  \\
Q _2         & 0&    1     &   0     &  \cdot      &  0  &  0  \\
\tilde Q     & 0 &    0      &   1    &  \square &  0  &   0  \\
A              & 0  &    0     &  0      & \cdot       &  1  &   0   \\
\psi_A      &-1  &    0    &   0      &  \cdot      &  -n &  1   \\
\hline
\Phi_1^{(1)}    & 0         &     1     &    0        &      \cdot        &          n     &         0 \\
\Phi_2             &         0         &      0    &              2               &      \begin{array}{c} \square\vspace{-3mm} \\ \square\end{array}        &       1     &          0 \\
\Phi_3^{(2)}    &   1      &      0    &     1  &       \square       &       0        &           0\\
\Phi_3^{(1)}    &    0     &     1     &     1  &       \square       &     0          &           0\\
\Phi_4            &  0               &     0    &              2n+1      &   \cdot   &    0          &            0\\
\Psi^{(2)}        &    -1              &       0   &           -2n-1                  &         \cdot     &       -n        &           1\\
\Psi^{(1)}        &      0            &       -1   &           -2n-1                  &      \cdot        &         -n      &           1\\          
\end{array}
\end{equation}
The anomalies of the global symmetries are given by
\begin{equation}
    \begin{array}{lll}
    &\kappa_{11} =2n,& \quad \kappa_{22} = 2n+1,  \\
    &\kappa_{12} = 0,& \quad\kappa_{1\tilde Q} = 0, \\ 
    &\kappa_{\tilde Q \tilde Q } = (2n+1)^2,& \quad \kappa_{2\tilde Q} = 0,  \\
    &\kappa_{1A} = -n,& \quad\kappa_{R_0 R_0} = 2n^2+5n+2, \\
    &\kappa_{R_0 1} = -2n,& \quad \kappa_{R_0 2} = -2n-1,  \\
    &\kappa_{R_0 \tilde Q} = -2n-1,& \quad\kappa_{R_0 A} = -2n^2, \\
    &\kappa_{A2}=0,& \quad \kappa_{A\tilde Q}=0, \\
    &\kappa_{AA} = n(n+1),&  \\
    \end{array}
\end{equation}
and we checked that they match across the dual phases.

The duality can be derived by topologically twisting the 4d s-confining duality involving an $\SU(2n+1)$ gauge theory with $2n+1$ fundamentals, $4$ fundamentals and one antisymmetric derived in \cite{Csaki:1996zb}.
The twist is done along the 4d non-anomalous $R$ symmetry that assigns $R$ charge 0 to the antisymmetric, the anti-fundamentals and two fundamentals and $R$ charge 1 to the remaining two fundamentals.
The confined degrees of freedom are
\begin{equation}
\label{Csaki2fields}
\Sigma_1 = A^{n}  Q,\quad
\Sigma_2 = A \tilde Q^2,\quad
\Sigma_3 = Q \tilde Q     ,\quad
\Sigma_4 = \tilde Q^{2n+1},\quad 
\Sigma_5 =A^{n-1} Q^3,
\end{equation}
interacting through the 4d superpotential
\begin{equation}
\label{WCSAKI2}
W = \Sigma_1 \Sigma_3^3 \Sigma_2^{n-1}
+\Sigma_5 \Sigma_3 \Sigma_2 ^n+
\Sigma_4 \Sigma_1 \Sigma_5 \,.
\end{equation}
When we twist this WZ model with the $R$ symmetry assignation 
discussed above we see that the 4d superfields $\Sigma_{1,3}$, in the fundamental representation of $\SU(4)$ survive as 2d chirals 
denoted above as $\Phi_{1,3}$ in the fundamental of $\SU(2)$. 
The 4d superfields $\Sigma_{2,4}$ become the 2d chirals  $\Phi_{2,4}$
as well. 
On the other hand the 4d superfield $\Sigma_5 $ has $R$ charge $R=2$ and it becomes the 2d Fermi $\Psi$, in the fundamental representation of $\SU(2)$.
One can also check that the first term in (\ref{WCSAKI2}) does not survive in 2d while the other two terms  in (\ref{WCSAKI2})  reconstruct the 2d superpotential (\ref{expectedsecond}).

We conclude our analysis by providing a derivation of the duality 
from the elliptic genus. The identity that we want to prove in this case is 
\begin{eqnarray}\label{toprove2}
&&I_{\SU(2n+1)}^{(2;2n+1;\cdot;1;\cdot)}(x \vec u;y \vec v;\cdot;t^2;\cdot)
=
\frac{1}{ \theta( y^{2n+1})}\cdot
\prod_{a<b}\frac{1}{\theta(v_a v_b y^2 t^2)}
\nonumber \\
\times &&\prod_{a=1}^{2}
\frac{\theta(q/(x u_a y^{2n+1}t^{2n}) )}{\theta(t^{2n} x u_a ) }
\prod_{a=1}^{2}\prod_{b=1}^{2n+1}
\frac{1}{\theta(u_a  v_b  x y)}\,,
\end{eqnarray}
where $\prod_{a=1}^{2} u_a = \prod_{a=1}^{2n} v_a=1$.

Following the discussion above we are actually proving the relation 
(\ref{toprove2}) by flipping the operator $ A^{n} Q_{2}$ using the Fermi
 field $\psi_A$ in the superpotential (\ref{flipped2}).
Such a flip corresponds to \emph{moving} the theta function 
$\theta(t^{2n} x u_1)$ to the LHS of (\ref{toprove2}). The corresponding Fermi $\psi_A$ is correctly identified by using the  relation (\ref{inversion}).

The next step corresponds to deconfine the antisymmetric tensor using the identity (\ref{ellipticdualUSP}).
This boils down to the following substitution in the integrand in the LHS of 
(\ref{toprove2}) 
\begin{eqnarray}
\frac{
\theta(x u_1 t^{2n})
}{\prod_{i=1}^{2n+1}
\theta (z_i u_1 x)
\prod_{i<j} \theta (z_i z_j t^2) }
\rightarrow 
I_{\USp(2n)}^{(2n+1,1;\cdot;\cdot)}( t \vec z ,u_1x/t;\cdot;\cdot)\,.
\end{eqnarray}
Once this substitution is done we must apply formula (\ref{ellipticdualSU}), corresponding to the $\SU(2n+1)$ duality.
We are left with an $\USp(2n)$ theory with $2n+2$ fundamentals and various chiral and Fermi singlets. The elliptic genus for this theory is
\begin{equation}
\frac{\theta(q /(u_2 y^{2n+1} t^{2n} x ))}{\theta( y^{2n+1})\theta(t^{2n} x u_2 )\prod_{a=1}^{2n+1} \theta(u_2 v_a x y)}
I_{\USp(2n)}^{(2n+1,1;\cdot;\cdot)}( y t \vec v ,u_1x/t;\cdot;\cdot)\,.
\end{equation}
To conclude the proof we apply (\ref{ellipticdualUSP}) to this integral, obtaining the RHS of (\ref{toprove2}), except the contribution of the flipped singlet corresponding to the $\theta(x u_1 y^{2n+1}t^{2n})$ as discussed above.

\subsection{$\SU(2n)$ with $2n-1$ $\overline \square$, $3$ $\square$}
\label{Case2even}

In this case the LG is given by five chiral fields $\Phi_I$ corresponding to the gauge invariant combinations
\begin{eqnarray}
\label{ginv2ev}
\Phi_1 = Q \tilde Q,\quad
\Phi_2 =  \text{Pf}\, A   ,\quad
\Phi_3 = A^{n-1} Q^2     ,\quad
\Phi_4 = A \tilde Q^{2},
\end{eqnarray}
and one Fermi multiplet $\Psi_{1}$ interacting with the chirals through
a superpotential
\begin{equation}
\label{Wguess32nm1}
W=\Psi_1 (\Phi_1 \Phi_3 \Phi_4^{n-1} + \Phi_2 \Phi_1^3 \Phi_4^{n-2})\,.
\end{equation}

In order to simplify our analysis we add a J-term to the electric theory
corresponding to 
\begin{equation}
\label{flippedev3}
W = \psi_A \Pf A\,.
\end{equation}
Then we trade the antisymmetric with an $\USp(2n-2)$ gauge theory
with superpotential
\begin{equation}
W = 0\,.
\end{equation}
In this case we did not represent the various steps with the help of a quiver description because we are just exchanging the 
antisymmetric $A$ with an $\USp(2n-2)$ gauge node connected to $\SU(2n)$ through a bifundamental that we denote as $P$.

Then we dualize the $\SU(2n)$ gauge node that has $2n+1$ fundamentals and $2n-1$ anti-fundamentals using the results of Appendix \ref{confGWR}.
Defining the $\SU(2n)$ gauge invariant combinations
$\varphi_1 = Q \tilde Q$,
$\varphi_2 =P^{2n-2} Q^2$,
$\varphi_3 =P \tilde Q$, and
$\varphi_4 =P^{2n-3} Q^3$
the dual superpotential becomes
\begin{equation}
W = \Psi_{\SU(2n)} (\varphi_1 \varphi_2+\varphi_3 \varphi_4)\,,
\end{equation}
where the field $\Psi_{\SU(2n)} $ is a 2d Fermi.
The $\USp(2n-2)$ gauge group has now $2n-1$ fundamentals denoted as $\varphi_3$ and one fundamental denoted as $\varphi_4$.
It can be dualized in terms of a LG model using the results of Appendix \ref{sec:USP2Np2}. There are two gauge invariant combinations that arise in this case that we denote as
$\rho = \varphi_3 \varphi_4$ and $\chi = \varphi_3^2$, in the fundamental and in the antisymmetric of the $\SU(2n-1)$ flavor symmetry respectively.
The superpotential of the LG model becomes
\begin{equation}
\label{finalW2ev}
W = \Psi_{\SU(2n)} (\varphi_1 \varphi_2+\rho) + \Psi_{\USp(2n-2)} \rho \chi^{n-1} \rightarrow  \Psi_{\USp(2n-2)} \varphi_1 \varphi_2 \chi^{n-1} \,,
\end{equation}
where in the second part of the formula we have integrated out the massive fields.

Reading the dictionary arising from the various duality step we 
can associate the singlets in (\ref{ginv2ev}) to the ones in (\ref{finalW2ev}) as
$\Phi_1 =\varphi_1$, $\Phi_3 =\varphi_2$ and
$\Phi_4 =\chi$ 
 while the Fermi $\Psi_{\USp(2n-2)}$ is mapped to the Fermi $\Psi_1$ in the superpotential (\ref{Wguess32nm1}).

The 2d duality can be derived by topologically twisting the 4d confining duality involving an $\SU(2n)$ gauge theory with $2n$ fundamentals, $4$ fundamentals and one antisymmetric derived in \cite{Csaki:1996zb}.
The twist is done along the 4d non anomalous $R$ symmetry that assigns $R$ charge 0 to the antisymmetric, $(2n-1)$ anti-fundamentals and three fundamentals and $R$ charge 1 to the remaining fundamental and anti-fundamental.
The confined degrees of freedom are
given in (\ref{CSAKIFIELDS})
interacting through the superpotential (\ref{WCSAKI}).
When we twist this WZ model with the $R$ symmetry assignation 
discussed above we see that the 4d 
superfield $\Sigma_1$, in the antisymmetric representation of $\SU(4)$ survives as the 2d chiral field $\Phi_3$ in the anti-fundamental representation of $\SU(3)$, the field 4d superfield $\Sigma_2$ becomes the 2d chiral $\Phi_4$, while the 4d superfield $\Sigma_3$ splits into 
the 2d chiral $\Phi_1$ and the 2d Fermi $\Psi_1$.
The other field that survives upon the twist is the 4d superfield $\Sigma_5$ that becomes the 2d chiral $\Phi_2$.
The other two 4d superfield $\Sigma_4$ and $\Sigma_6$ have $R$ charge 1 and they do not survive in 2d.
One can also check that the 4d superpotential (\ref{WCSAKI}) becomes the 2d superpotential (\ref{Wguess32nm1}).

We proceed by checking the anomaly matching of the global symmetries.
The charges of the fields in the electric and in the dual LG theory, including the flipper $\psi_A$ in (\ref{flippedev3}), are
\begin{equation}
\begin{array}{c|ccccc|c}
&\UU(1)_Q&\UU(1)_{\tilde Q} & \SU(3)& \SU(2n-1) & \UU(1)_A&\UU(1)_{R_0} \\
Q&1&0&\square&\cdot&0&0\\
\tilde Q&0&1&\cdot&\square&0&0\\
A&0&0&\cdot&\cdot&1&0\\
\psi_A&0&0&\cdot&\cdot&-n&1\\
\hline
\Phi_4&0&2&\cdot&\begin{array}{c} \square\vspace{-3mm} \\ \square\end{array}&1&0\\
\Phi_1&1&1&\square&\square&0&0\\
\Phi_3 &2&0&\overline \square&\cdot&n-1&0\\
\Psi_1&-3&1-2n&\cdot&\cdot&2-2n&1\\
\end{array}
\end{equation}
The anomalies of the global symmetries are given by
\begin{equation}
\begin{array}{lll}
&
\kappa_{QQ} =6n ,& \quad
\kappa_{\tilde Q \tilde Q} = 2n(2n-1), \\
&
\kappa_{AA}=n(n-1),& \quad
\kappa_{Q A}= 0, \\
&
\kappa_{\tilde Q Q } = 0, & \quad
\kappa_{\tilde Q A } =   0,  \\
&
\kappa_{ \SU(3)^2} = n, & \quad
\kappa_{\SU(2n-1)^2  }  =n, \\
&\kappa_{R_0R_0}= n(3+2n), &\quad  \kappa_{R_0A}  =-2n(n-1),\\
&\kappa_{R_0Q}= -6n, &\quad  \kappa_{R_0 \tilde Q }=2n(1-2n),\\
\end{array}
\end{equation}
and we checked that they match across the dual phases.

At the level of the elliptic genus the identity that we need to prove is
\begin{equation}
\label{TOPROVE3}
\theta(q t^{-2n})
I_{\SU(2n)}^{(3,2n-1;\cdot;1;\cdot)}(x \vec u;y \vec v;\cdot;t^2;\cdot)
=
\frac{
\theta(q/y^{2n-1}x^3t^{4n-4})}{
\prod_a \theta(x^2u_a^{-1}t^{2n-2})
\prod_{a,b}\theta(x y u_a v_b)
\prod_{a<b}\theta(v_a v_b y^2 t^2)
}\,,
\end{equation}
with $\prod_{a=1}^{3} u_a =\prod_{a=1}^{2n-1} v_a = 1$. 
Observe that the $\theta$ function in the LHS of (\ref{TOPROVE3})  refers to the flipper $\psi_A$ in (\ref{flippedev3}).
The next step corresponds to deconfine the antisymmetric tensor using the identity (\ref{ellipticdualUSP}).
This boils down to the following substitution in the integrand in the LHS of 
(\ref{TOPROVE3}) 
\begin{eqnarray}
\frac{\theta(q t^{-2n})}{\prod_{i<j} \theta(z_i z_j t^2)}
\rightarrow
I_{\USp(2n-2)}^{(2n+2;\cdot;\cdot)}
(t \vec z;\cdot;\cdot)\,.
\end{eqnarray}
Then we dualize the $\SU(2n)$ gauge group using the identity (\ref{chirallSU}) obtaining the index of an $\USp(2n-2)$ gauge theory with elliptic genus
\begin{equation}
\frac{\prod_{b=1}^{2n-1} \theta(q/(v_b x^3 y t^{2n-2}))}{\prod_{a=1}^{3} \prod_{b=1}^{2n-1} \theta(u_a v_b x y)
\prod_{1\leq a<b\leq 2n-1} \theta(u_a u_b x^2 t^{2n-2})}
I_{\USp(2n-2)}^{(1,2n-1;\cdot;\cdot)}( x^3 t^{2n-3},y t \vec v;\cdot;\cdot)\,.
\end{equation}
We conclude  by applying (\ref{ellipticdualUSP}), dualizing the $\USp(2n-2)$ gauge group and arriving, after applying the  formula  (\ref{inversion}), to the RHS of 
(\ref{TOPROVE3}).

\subsection{$\SU(2n+1)$ with $2n$ $\overline \square$ and $3$ $\square$}
\label{Case2odd}

\begin{figure}
\begin{center}
  \includegraphics[width=12cm]{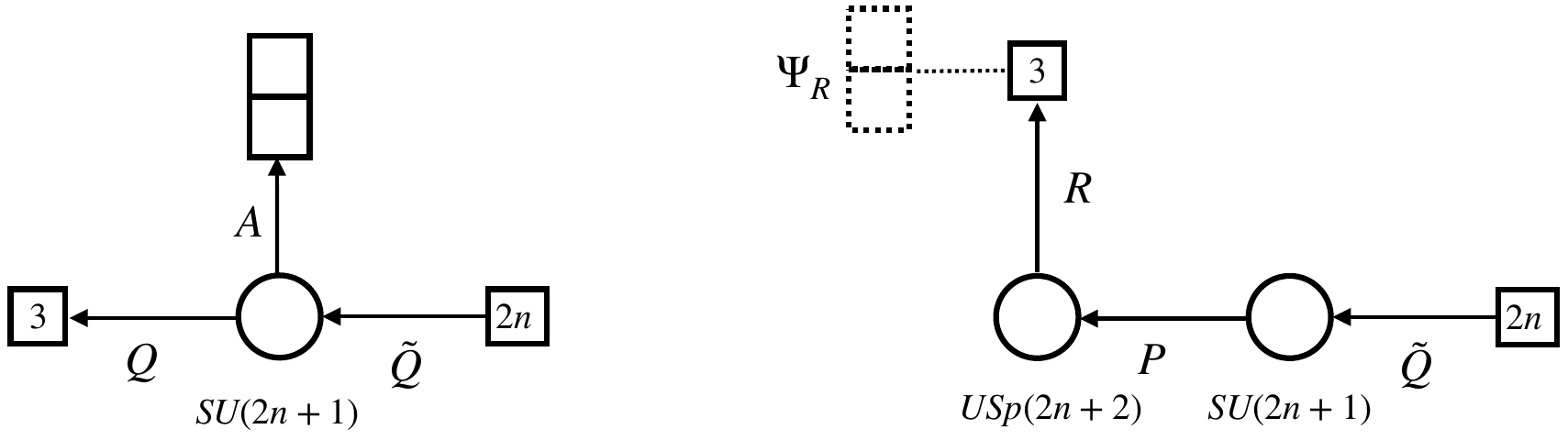}
  \end{center}
  \caption{The first quiver  represents  the $\SU(2n+1)$ gauge theory with an antisymmetric $A$, $2n$ anti-fundamentals $\tilde Q$ and three fundamentals $Q$. In the second quiver we  exchanged the antisymmetric $A$ and the three fundamentals with an auxiliary $\USp(2n)$ gauge node with the bifundamental $P$ and the fundamentals $R$. In this case we also represent the Fermi field $\Psi_R$ in the figure, in the anti-fundamental (if we want to write a J-term in the action) representation of the $\SU(3)$ flavor symmetry. In this case we did not represent a third quiver obtained by dualizing the $\SU(2n+1)$ node.}
    \label{fourhcase}
\end{figure}

In this case the LG is given by five chiral fields $\Phi_I$ corresponding to the gauge invariant combinations
\begin{eqnarray}
\label{ginv3odd}
\Phi_1 = Q \tilde Q,\quad
\Phi_2 =  A^n Q   ,\quad
\Phi_3 = A^{n-1} Q^3     ,\quad
\Phi_4 = A \tilde Q^{2},
\end{eqnarray}
and one Fermi multiplet $\Psi_{1}$ interacting with the chirals through
a superpotential
\begin{equation}
\label{expectedfourth}
W=\Psi_1 (\Phi_4^{n} \Phi_3 + \Phi_2 \Phi_1^2 \Phi_4^{n-1}).
\end{equation}

In order to simplify our analysis we add a J-term to the electric theory
corresponding to 
\begin{equation}
\label{flippedev4}
W = \psi_A A^{n-1} Q^3\,.
\end{equation}
Then we trade the antisymmetric with an $\USp(2n+2)$ gauge theory with three new $\USp(2n+2)$ fundamentals $R$ as in Figure \ref{fourhcase}
with superpotential
\begin{equation}
W = \Psi_R R^2\,.
\end{equation}
The next step consists of using the  duality discussed in appendix \ref{confGWR}
for $\SU(2n+1)$ with $2n$ anti-fundamentals $\tilde Q$ and $2n+2$ anti-fundamentals $Q$.
Defining the $\SU(2n+1)$ gauge invariant combinations $\Phi_M = P \tilde Q$ and $\Phi_B=P^{2n+1}$ the theory has superpotential 
\begin{equation}
\label{case4Wterz}
W = \Psi_R R^2+\Psi_{\SU(2n+1)} \Phi_M \Phi_B\,,
\end{equation}
where $\Psi_{\SU(2n+1)}$ is a Fermi field in the anti-fundamental of the $\SU(2n)$ flavor symmetry.

The last step consists in dualizing $\USp(2n+2)$ that has $2n$ fundamentals $\Phi_M$, one fundamental $\Phi_B$ and three fundamentals $R$.
Some of the components of the antisymmetric meson of this duality, i.e. 
the $\USp(2n+2)$ gauge invariant combinations $ \Phi_M \Phi_B$ 
and $ \Phi_R^2$ become massive because of the superpotential
(\ref{case4Wterz}) and the other components $\Phi_A = \Phi_M^2$,
$\Phi_{MR} = \Phi_M \Phi_R$ and 
$\Phi_{RB} =  \Phi_R   \Phi_B$ interact through the superpotential
\begin{equation}
\label{case4Wfinal}
W=\Psi_{\USp(2n+2)}\Phi_{MR}^2 \Phi_{RB} \Phi_A^{n-1}\,,
\end{equation}
consistently with the expectation above, by the map of the singlets that can be read from the various steps, i.e.
$\Phi_1 = \Phi_{MR}$,
$\Phi_2 =  \Phi_{RB} $,
$\Phi_4 = \Phi_A$ and 
$\Psi_1 = \Psi_{\USp(2n)}$.

The duality can be derived by topologically twisting the 4d s-confining duality involving an $\SU(2n+1)$ gauge theory with $2n+1$ fundamentals, $4$ fundamentals and one antisymmetric derived in \cite{Csaki:1996zb}.
The twist is done along the 4d non anomalous $R$ symmetry that assigns $R$ charge 0 to the antisymmetric, $2n$ anti-fundamentals and  three fundamentals and $R$ charge 1 to the remaining  fundamental and anti-fundamental.
The confined degrees of freedom are given in (\ref{Csaki2fields}),
interacting through the superpotential
(\ref{WCSAKI2}).
It follows that the field $\Sigma_4$ does not survive as a massless 
field in 2d and that the surviving components of $\Sigma_3$ are 
the 2d chiral $\Phi_1$ and the Fermi $\Psi_1$. The other singlets $\Sigma_{1,2,5}$ survive as the 2d chiral fields $\Phi_{2,4,3}$ respectively. 
One can also check that the last term in (\ref{WCSAKI2}) does not survive in 2d while the other two terms  in (\ref{WCSAKI2})  reconstruct the 2d superpotential (\ref{expectedfourth}).

We proceed by checking the anomaly matching of the global symmetries.
The charges of the fields in the electric and in the dual LG theory, including the flipper $\psi_A$ in (\ref{flippedev4}), are
\begin{equation}
\begin{array}{c|ccccc|c}
&\UU(1)_Q&\UU(1)_{\tilde Q} & \SU(3)& \SU(2n) & \UU(1)_A&\UU(1)_{R_0} \\
Q&1&0&\square&\cdot&0&0\\
\tilde Q&0&1&\cdot&\square&0&0\\
A&0&0&\cdot&\cdot&1&0\\
\psi_A&-3&0&\cdot&\cdot&1-n&1\\
\hline
\Phi_1&1&1&\square&\square&0&0\\
\Phi_2&1&0&\square&\cdot&n&0\\
\Phi_4 &0&2&\cdot&\begin{array}{c} \square\vspace{-3mm} \\ \square\end{array}&1&0\\
\Psi_1&-3&-2n&\cdot&1&1-2n&1\\
\end{array}
\end{equation}
The anomalies of the global symmetries are given by
\begin{equation}
\begin{array}{lll}
&
\kappa_{QQ}  = 6(n-1),& \quad
\kappa_{\tilde Q \tilde Q} = 2n(2n+1), \\
&
\kappa_{AA}=n^2+3n-1,& \quad
\kappa_{Q A}=3(1-n),  \\
&
\kappa_{\tilde Q Q } = 0, & \quad
\kappa_{\tilde Q A } = 0,   \\
&
\kappa_{ \SU(3)^2} = n+\frac{1}{2}, & \quad
\kappa_{\SU(2n)^2  }=  n+\frac{1}{2},\\
&\kappa_{R_0R_0}= 2n^2+5n+2, &\quad  \kappa_{R_0A}=-2n^2-1,\\
&\kappa_{R_0Q}= -6n, &\quad  \kappa_{R_0 \tilde Q } =-2n(1+2n),\\
\end{array}
\end{equation}
and we checked that they match across the dual phases.

At the level of the elliptic genus the identity that we need to prove is
\begin{equation}
\label{TOPROVE4}
\theta(q t^{2-2n} x^{-3})
I_{\SU(2n+2)}^{(3,2n;\cdot;1;\cdot)}(x \vec u;y \vec v;\cdot;t^2;\cdot)
=
\frac{
\theta(qy^{-2n}x^{-3}t^{2-4n})}{
\prod_a \theta(x u_a t^{2n})
\prod_{a,b}\theta(x y u_a v_b)
\prod_{a<b}\theta(v_a v_b y^2 t^2)
}\,,
\end{equation}
with $\prod_{a=1}^{3} u_a =\prod_{a=1}^{2n-1} v_a = 1$. 
Observe that the $\theta$ function in the LHS of (\ref{TOPROVE4})  refers to the flipper $\psi_A$ in (\ref{flippedev4}).
The next step corresponds to deconfine the antisymmetric tensor using the identity (\ref{ellipticdualUSP}).
This boils down to the following substitution in the integrand in the LHS of 
(\ref{TOPROVE4}) 
\begin{equation}
\frac{
\theta(q t^{2-2n} x^{-3})
}{\prod_{a=1}^{3} \prod_{i=1}^{2n+1} \theta (u_a x z_i) \prod_{i<j} \theta(z_i z_j t^2)}
\rightarrow
\prod_{a=1}^{3} \theta(q t u_a x^{-2}) \cdot  
I_{\USp(2n+2)}^{(2n+1,3;\cdot;\cdot)}
(t \vec z,xt^{-1} \vec u;\cdot;\cdot)\,.
\end{equation}
Then we dualize the $\SU(2n+1)$ gauge group using the identity (\ref{chirallSU}) obtaining the index of an $\USp(2n+2)$ gauge theory with elliptic genus
\begin{equation}
\prod_{a=1}^{3} \theta(q t u_a x^{-2}) 
\cdot
\prod_{a=1}^{2n} \theta(q t^{-2n-2}v_a^{-1} y^{-1})
\cdot
I_{\USp(2n+2)}^{(1,2n,3;\cdot;\cdot)}( t^{2n+1},y t \vec v,x/t \vec u;\cdot;\cdot)\,.
\end{equation}
We conclude  by applying (\ref{ellipticdualUSP}), dualizing the $\USp(2n+2)$ gauge group and arriving, after applying the  formula (\ref{inversion}), to the RHS of 
(\ref{TOPROVE4}).

\subsection{$\SU(2n)$ with $2n-2$ $\overline \square$, $4$ $\square$}
\label{Case3even}

\begin{figure}
\begin{center}
  \includegraphics[width=12cm]{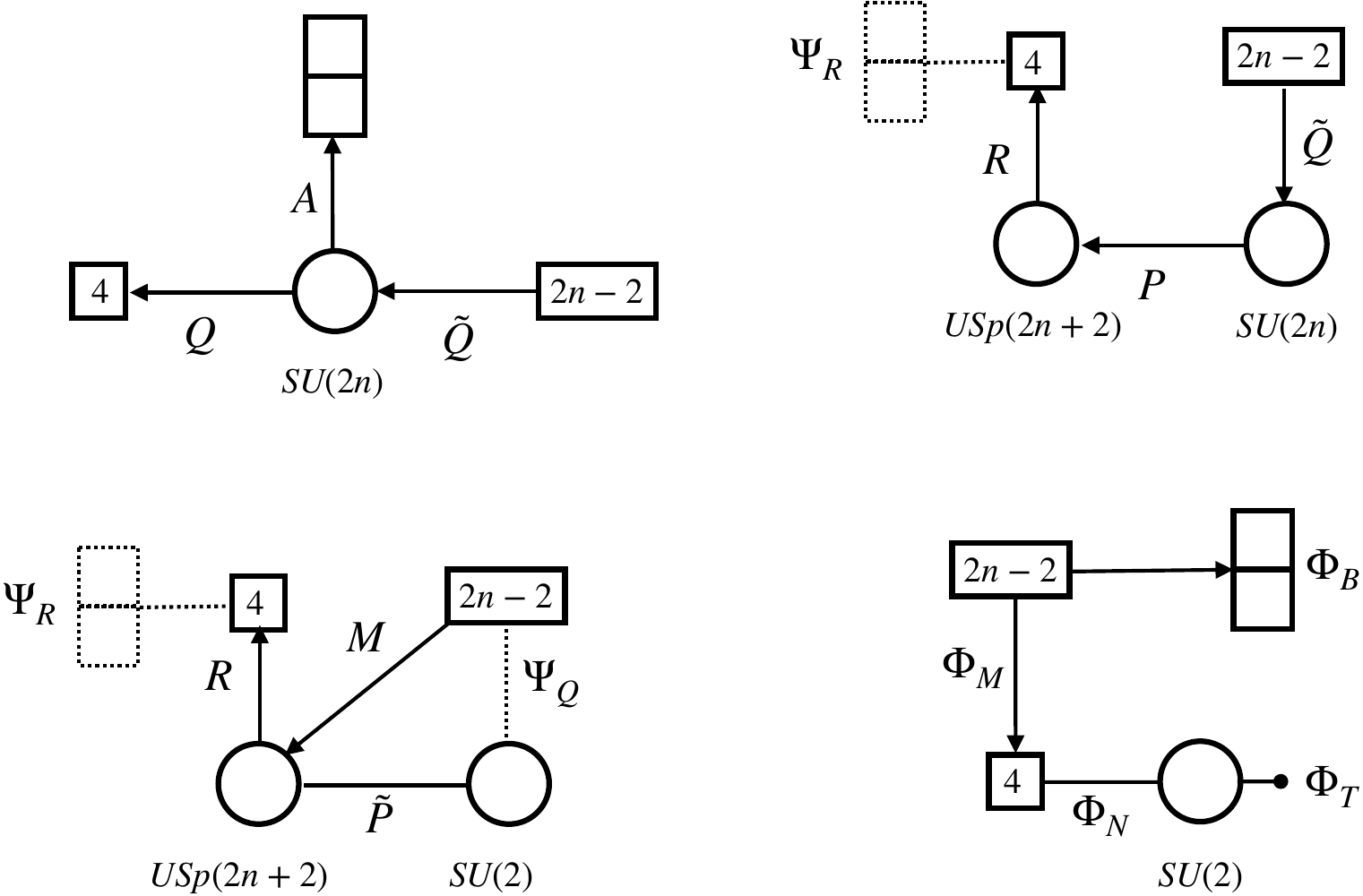}
  \end{center}
  \caption{The first quiver  represents  the $\SU(2n)$ gauge theory with an antisymmetric $A$, $2n-2$ anti-fundamentals $\tilde Q$ and four fundamentals $Q$. In the second quiver we  exchanged the antisymmetric $A$ and the four fundamentals with an auxiliary $\USp(2n+2)$ gauge node with the bifundamental $P$ and the fundamentals $R$. In this case we also represent the Fermi field $\Psi_R$ in the figure, in the antisymmetric representation of the $\SU(4)$ flavor symmetry. In the third quiver we represent the theory obtained after the duality on $\SU(2n)$, that gives an $\SU(2)$ gauge theory. Then the fourth quiver is obtained by dualizing $\USp(2n+2)$, leaving just an $\SU(2)$ gauge theory. Observe that in this case we represented in the various quivers the gauge singlets in non-trivial representations of the flavor symmetry group, while the others are omitted and can be found in the discussion in the body of the paper.}
    \label{fifthcase}
\end{figure}

In this case the LG is given by five chiral fields $\Phi_I$ corresponding to the gauge invariant combinations
\begin{eqnarray}
\label{fieldssu2n4}
\Phi_1 = Q \tilde Q,\quad
\Phi_2 = A^n,\quad
\Phi_3 = A^{n-1} Q^2     ,\quad
\Phi_4 =A^{n-2} Q^4,\quad
\Phi_5 =A \tilde Q^2,
\end{eqnarray}
and two Fermi multiplets $\Psi_{1,2}$ interacting with the chirals through
a superpotential
\begin{equation}
\label{guess5}
W=\Psi_1 ( \Phi_5^{n-2}\Phi_1^2 \Phi_3 + \Phi_5^{n-3} \Phi_2 \Phi_1^4)
+
\Psi_2 (\Phi_3^2+\Phi_2\Phi_4).
\end{equation}

In order to simplify our analysis we add a J-term to the electric theory
corresponding to 
\begin{equation}
\label{Case3even1}
W = \psi_A A^{n-2} Q^4\,.
\end{equation}
Then, we trade the antisymmetric with an $\USp(2n+2)$ gauge theory and four auxiliary $\USp(2n+2)$ fundamentals,  as in Figure \ref{fifthcase}, with superpotential
\begin{equation}
\label{Case3even2}
W = \Psi_R R^2\,.
\end{equation}
where $\Psi_R$ is a Fermi in the antisymmetric of the $\SU(4)$
flavor symmetry.

Next, we proceed by dualizing the $\SU(2n)$ node into $\SU(2)$ as explained in Appendix \ref{appdualGRW}. Following the rules of such duality we are left with the third quiver in Figure \ref{fifthcase} with superpotential 
\begin{equation}
\label{Case3even3}
W = \Psi_R R^2+\Psi_{Q} M \tilde P\,,
\end{equation}
where the meson  $M=\tilde Q P$ and the dual $\SU(2)$ fundamental $\tilde P$ are two chiral fields and $\Psi_Q$ is a Fermi, in the fundamental of the dual $\SU(2)$ gauge group .
At this point of the discussion we can dualize the $\USp(2n+2)$ gauge group, because it has $2n+4$ fundamentals, denoted as $R,M$ and $\tilde P$ in the quiver.
Some components of the antisymmetric meson of this duality are massive because of the superpotential (\ref{Case3even3}), and they are $R^2$ and $ M \tilde P$. The other singlets are 
$\Phi_T = P^2$, $\Phi_N = \tilde P R$,
$\Phi_{ M} = MR$ and $\Phi_{ B} = M^2$, where the last two are in the fundamental and in the  antisymmetric of the $\SU(2n-2)$ flavor symmetry respectively.
The superpotential for this $\SU(2)$ gauge theory is
\begin{equation}
\label{Case3even4}
W = \Psi_{\USp(2n+2)} (\Phi_{ B}^{n-2} \Phi_{ M}^2\Phi_{N}^2+\Phi_T \Phi_{ M}^4\Phi_{ B}^{n-3})\,,
\end{equation}
where the Fermi $\Psi_{\USp(2n+2)}$ is generated by the duality.

The last step of the derivation consists of studying the $\SU(2)$ gauge theory with four fundamentals $\Phi_N$. This theory is dual to a LG where the gauge singlets are $\Phi_{C} = \Phi_N^2$
and a Fermi $\Psi_{\SU(2)}$ with superpotential 
\begin{equation}
\label{Case3even5}
W = \Psi_{\USp(2n+2)} (\Phi_{ B}^{n-2} \Phi_{ M}^2\Phi_{C}+\Phi_T \Phi_{ M}^4\Phi_{ B}^{n-3})
+\Psi_{\SU(2)} \Phi_C^{2}\,,
\end{equation}
which is equivalent to (\ref{guess5}) after removing the flipped field $\Phi_4$ because of the dictionary 
\begin{equation}
   \Phi_T \leftrightarrow \Phi_2,\quad \Phi_B \leftrightarrow \Phi_5,\quad \Phi_M \leftrightarrow \Phi_1,\quad \Phi_C \leftrightarrow \Phi_3,\quad \Psi_{\USp(2n+2)}\leftrightarrow \Psi_1,\quad  \Psi_{\SU(2)} \leftrightarrow \Psi_2. 
\end{equation}

We proceed by checking the anomaly matching of the global symmetries.
The charges of the fields in the electric and in the dual LG theory, including the flipper $\psi_A$ in (\ref{Case3even1}), are
\begin{equation}
\begin{array}{c|ccccc|c}
&\UU(1)_Q&\UU(1)_{\tilde Q} & \SU(4)& \SU(2n-2) & \UU(1)_A&\UU(1)_{R_0} \\
Q&1&0&\square&\cdot&0&0\\
\tilde Q&0&1&\cdot&\square&0&0\\
A&0&0&\cdot&\cdot&1&0\\
\psi_A&-4&0&\cdot&\cdot&2-n&1\\
\hline
\Phi_1 &1&1&\square&\square&0&0\\
\Phi_2&0&0&\cdot&\cdot&n&0\\
\Phi_3&2&0&\begin{array}{c} \square\vspace{-3mm} \\ \square\end{array}&\cdot&n-1&0\\
\Phi_5&0&2&\cdot&\begin{array}{c} \square\vspace{-3mm} \\ \square\end{array}&1&0\\
\Psi_1&-4&2-2n&\cdot&\cdot&3-2n&1\\
\Psi_2&-4&0&\cdot&\cdot&2-2n&1\\
\end{array}
\end{equation}
The anomalies of the global symmetries are given by
\begin{equation}
\begin{array}{lll}
&
\kappa_{QQ}  =8(n-2) ,& \quad
\kappa_{\tilde Q \tilde Q} =4n(n-1),  \\
&
\kappa_{AA} =(n+4)(n-1),& \quad
\kappa_{Q A}= 4(2-n), \\
&
\kappa_{\tilde Q Q } =0 , & \quad
\kappa_{\tilde Q A } =  0,  \\
&
\kappa_{ \SU(3)^2} =n , & \quad
\kappa_{\SU(2n-2)^2  } =n,  \\
&\kappa_{R_0R_0}= n(3+2n), &\quad \kappa_{R_0A} =-2(n^2-n+1),\\
&\kappa_{R_0Q}= 4-8n, &\quad \kappa_{R_0 \tilde Q }=-4n(n-1),\\
\end{array}
\end{equation}
and we checked that they match across the dual phases.

The last check consists of showing that the identity between the elliptic genera of the gauge theory 
and of the  LG model follows from the other basic identities that do not involve the antisymmetric matter. The expected identity in this case is given by
\begin{eqnarray}
\label{ToProve5}
&&
\theta(q/(t^{2n-4}x^4 ))
I_{\SU(2n)}^{(4;2n-2;\cdot;1;\cdot)}(x \vec u;y \vec v;\cdot; t^2;\cdot)
=
\frac{\theta(q/(x^4 t^{4n-4})) \theta(q/(x^4 t^{4n-6} y^{2n-2}))}{\theta(t^{2n})} 
\nonumber \\
\times &&
\prod_{1\leq a <b \leq 4} \frac{1}{\theta( u_a u_b x^2 t^{2n-2})} \prod_{1\leq a <b \leq 2n-2} \frac{1}{\theta( v_a v_b y^2 t^{2})}\cdot \prod_{a=1}^{4} \prod_{b=1}^{2n-2} \frac{1}{\theta(u_a v_b x y )}\,.
\end{eqnarray}
In order to prove this relation we apply the following substitution involving the antisymmetric and the four fundamentals in the integrand on the LHS of (\ref{ToProve5}) 
\begin{equation}
\frac{\theta(q/(t^{2n-4}x^4 ))}{\prod_{i=1}^{2n} \theta(u_a x z_i) \prod_{1\leq i<j\leq 2n}\theta(z_i z_j t^2)}\rightarrow
\prod_{1\leq a<b \leq 4}\theta(q t^2/(u_a u_b x^2))\,\cdot\,
I_{\USp(2n+2)}^{(4,2n-2;\cdot;\cdot)}\left(\frac{x}{t} \vec u,t \vec z;\cdot;\cdot\right)\,.
\end{equation}
Then we use the relation (\ref{idbrutta}) (or equivalently (\ref{idbrutta2})) transforming the $\SU(2n)$ integral into  $\SU(2)$ and then we apply the relation (\ref{ellipticdualUSP}) to the $\USp(2n+2)$ integral.
We are then left with the integral associated to the model with an $\SU(2)$ gauge group, corresponding to the last quiver in Figure \ref{fifthcase}.
The elliptic genus of this theory can be again computed using (\ref{ellipticdualUSP}) for $N=1$ and in this way we arrive to the RHS of (\ref{ToProve5}).

\subsection{$\SU(2n+1)$ with $2n-1$ $\overline \square$, $4$ $\square$}
\label{Case3odd}

\begin{figure}
\begin{center}
  \includegraphics[width=15cm]{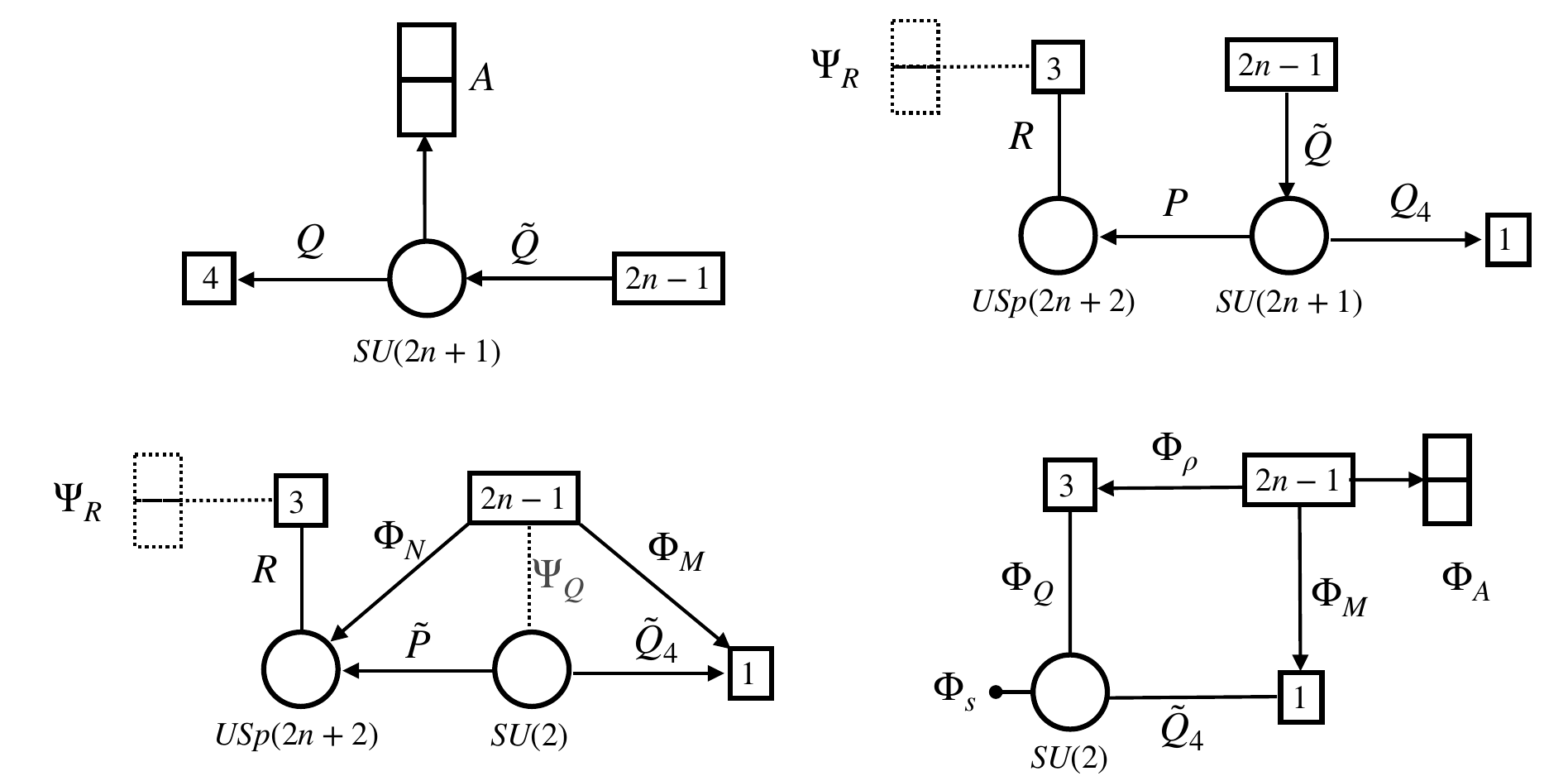}
  \end{center}
  \caption{The first quiver  represents  the $\SU(2n+1)$ gauge theory with an antisymmetric $A$, $2n-1$ anti-fundamentals $\tilde Q$ and four fundamentals $Q$. In the second quiver we  traded the antisymmetric $A$ and the three out of the four fundamentals, here $Q_{1,2,3}$,  with an auxiliary $\USp(2n+2)$ gauge node with the bifundamental $P$ and the fundamentals $R$. In this case we also represent the Fermi field $\Psi_R$ in the figure, in the antisymmetric representation of the $\SU(3)$ flavor symmetry. In the third quiver we represent the theory obtained after the duality on $\SU(2n+1)$, that gives an $\SU(2)$ gauge theory. Then the fourth quiver is obtained by dualizing $\USp(2n+2)$, leaving just an $\SU(2)$ gauge theory. Observe that in this case we represented in the various quivers the gauge singlets in non-trivial representations of the flavor symmetry group, while the others are omitted and can be found in the discussion in the body of the paper.}
    \label{sixthcase}
\end{figure}

In this case the LG is given by five chiral fields $\Phi_I$ corresponding to the gauge invariant combinations
\begin{equation}
\label{singlets6}
\Phi_1= Q \tilde Q,\quad
\Phi_2= A^n Q,\quad
\Phi_3= A^{n-1} Q^3,\quad
\Phi_4= A \tilde Q^2\,,
\end{equation}
and two Fermi multiplets $\Psi_{1,2}$ interacting with the chirals through
a superpotential
\begin{equation}
\label{guess6}
W=\Psi_1( \Phi_{4}^{n-1} \Phi_1 \Phi_2 + \Phi_{4}^{n-2}\Phi_1^3 \Phi_3 )
+
\Psi_2 (\Phi_2\Phi_3)\,.
\end{equation}

In order to simplify our analysis we add a J-term to the electric theory
corresponding to 
\begin{equation}
\label{Case3odd1}
W = \psi_A A^{n-1} Q_1 Q_2 Q_3\,,
\end{equation}
that breaks the $\SU(4)$ flavor symmetry to $\SU(3) \times \UU(1)_4 $
Then we trade the antisymmetric with an $\USp(2n+2)$ gauge theory and three auxiliary $\USp(2n+2)$ fundamentals $R$,  as in Figure \ref{sixthcase}, with superpotential
\begin{equation}
\label{Case3odd2}
W = \Psi_R R^2\,,
\end{equation}
where $\Psi_R$ is a Fermi in the antisymmetric of the leftover $\SU(3)$
flavor symmetry.

Then we proceed by dualizing the $\SU(2n)$ node into $\SU(2)$ as explained in Appendix \ref{appdualGRW}. Following the rules of such duality we are left with the third quiver in Figure \ref{sixthcase} with superpotential 
\begin{equation}
\label{Case3odd3}
W = \Psi_R R^2+ \Psi_{Q} (\Phi_N \tilde P +\Phi_M \tilde Q_4)\,,
\end{equation}
where we defined the $\SU(2n+1)$ mesonic combinations $\Phi_M = \tilde Q Q_4$ and  $\Phi_N = \tilde Q P$.
The $\SU(2)$ charged fields in this case are the anti-fundamental chirals $\tilde P$ and $\tilde Q_4$ and the Fermi $\Psi_Q$.
The $\USp(2n+2)$ gauge group has then three fundamental $R$, $(2n-1)$ fundamentals $\Phi_N$  and two fundamentals $\tilde P$.
It can be then dualized in terms of a LG model and the superpotential in this case becomes
\begin{equation}
\label{Case3odd4}
 W=\Psi_{\USp(2n+2)} ( \Phi_{A}^{n-1} \Phi_\rho  \Phi_Q^2 + \Phi_{A}^{n-2} \Phi_\rho^3 \Phi_s+\Phi_A^{n-2} \Phi_\rho^2 \tilde Q_4 \Phi_M \Phi_Q)\,,
\end{equation}
where the massless $\USp(2n+2)$ gauge invariant combinations are
$\Phi_{A} = \Phi_N^2$, $\Phi_{\rho}= \Phi_N R$, $\Phi_{Q}= R \tilde P$ and $\Phi_{s}= \tilde P^2$.
The last step consist of dualizing the $\SU(2)$ gauge node, with 1 fundamentel chiral $\tilde Q_4$ and three fundamentals $\Phi_Q$.

The two gauge invariant combinations in this case are $\Phi_B =\tilde Q_4 \Phi_Q$ and $\Phi_C = \Phi_Q^2$.
The superpotential of the LG model is 
\begin{equation}
\label{Case3odd5}
 W=\Psi_{\USp(2n+2)} ( \Phi_{A}^{n-1} \Phi_\rho  \Phi_C + \Phi_{A}^{n-2} \Phi_\rho^3 \Phi_s+\Phi_A^{n-2} \Phi_\rho^2 \Phi_M \Phi_B)
 + \Psi_{\SU(2)} \Phi_B \Phi_C.
\end{equation}
We can compare the superpotential (\ref{Case3odd5}) with the one guessed above  in (\ref{guess6}) after the addition of the flipper $\psi_A$ in  
(\ref{Case3odd1}).
In this case, it is necessary to split the gauge invariant combinations $\Phi_{1,2,3}$ in (\ref{singlets6}) into two components $\Phi_{1,2,3}^{(1,2)}$
as in (\ref{table6}).
Then we consider the superpotential (\ref{guess6}) splitting the fields $\Phi_{1,2,3}$  and setting to zero the component $\Phi_3^{(1)}$ because of the flipper $\psi_A$.
The superpotential obtained after this procedure coincides with (\ref{Case3odd5}) with the dictionary, that we can read from the various
duality steps, given by
\begin{align}
    &\hspace{25pt}\Phi_A = \Phi_4,\quad \Phi_B = \Phi_2^{(1)},\quad \Phi_C = \Phi_3^{(2)},\quad\Phi_\rho = \Phi_1^{(1)},\nonumber\\[5pt]
    &\Phi_s = \Phi_2^{(2)},\quad \Phi_M = \Phi_1^{(2)},\quad \Psi_1 = \Psi_{\USp(2n+2)},\quad \Psi_2 =  \Psi_{\SU(2)}\nonumber.\\
\end{align}

The duality can be derived by topologically twisting the 4d s-confining duality involving an $\SU(2n+1)$ gauge theory with $2n+1$ fundamentals, $4$ fundamentals and one antisymmetric derived in \cite{Csaki:1996zb}.
The twist is done along the 4d non anomalous $R$ symmetry that assigns $R$ charge 0 to the antisymmetric, $2n-1$ anti-fundamentals and  the four fundamentals and $R$ charge 1 to the remaining two  anti-fundamentals.
The confined degrees of freedom are given in (\ref{Csaki2fields}),
interacting through the superpotential
(\ref{WCSAKI2}).
It follows that the field $\Sigma_4$  survives as the massless Fermi
field $\Psi_1$ in 2d.
All the other fields $\Sigma_i$ give rise to massless chiral in 2d, with the dictionary $\Sigma_{1,2,3,5} \rightarrow \Phi_{2,4,1,3}$ and in addition the field $\Sigma_2$ give rise to the 2d Fermi fields $\Psi_2$.  
One can also check that the 4d superpotential (\ref{WCSAKI2}) becomes  the 2d superpotential (\ref{guess6}) after the twisted compactification accordingly to the rules explained above.
\newpage
We proceed by checking the anomaly matching of the global symmetries.
The charges of the fields in the electric and in the dual LG theory, including the flipper $\psi_A$ in (\ref{Case3odd1}), are
\begin{equation}
\label{table6}
\begin{array}{c|ccccccc} 
&\UU(1)_Q&\UU(1)_4&\SU(3)&\SU(2n-1)&\UU(1)_{\tilde Q}& \UU(1)_A&\UU(1)_{R_0} \\
\hline 
Q_{1,2,3} &1&  1& \square & 0&   0& 0&0\\ 
Q_4       &   1 & -3  & \cdot &0&   0 & 0 &0\\
\tilde Q&0&0&\cdot &\square & 1  &0  &0 \\ 
A&0&0& \cdot  &\cdot   &0 &1&0  \\ 
\psi_A&-3&-3&\cdot & \cdot& 0  &1-n  &1  \\ 
\hline
\Phi_1^{(1)}&1&1 &\square & \square  &1 &0&0   \\ 
\Phi_1^{(2)}&1&-3&\cdot &\square &1   &0&0   \\ 
\Phi_2^{(1)}&1&1&\square &\cdot &0   &n&0   \\ 
\Phi_2^{(2)}&1&-3&\cdot & \cdot&0   &n&0   \\ 
\Phi_3^{(2)}&3&-1&\overline \square & \cdot& 0 &n-1&0   \\ 
\Phi_4&0&0&\cdot &
\begin{array}{c} \square\vspace{-3mm} \\ \square\end{array}
 & 2  &1&0   \\ 
\Psi_1&-4&0&\cdot &\cdot & 1-2n  &2-2n&1  \\ 
\Psi_2&-4&0& \cdot& \cdot&  0 &1-2n&1   \\ 
\end{array}
\end{equation}
The anomalies of the global symmetries are given by
\begin{equation}
\begin{array}{lll}
&
\kappa_{44} =24 n+3 ,& \quad
\kappa_{4Q} =-9,  \\
&
\kappa_{4 \tilde Q} = 0,& \quad
\kappa_{4A}=3-3n,  \\
&
\kappa_{QQ}= 8n-5,& \quad
\kappa_{\tilde Q\tilde Q} =4n^2-1,  \\
&
\kappa_{AA} = n^2+3n-1,& \quad
\kappa_{QA}=3-3n,  \\
&
\kappa_{Q \tilde Q} =0 , & \quad
\kappa_{A \tilde Q}=  0,  \\
&
\kappa_{\SU(3)^2} =n+\frac{1}{2} , & \quad
\kappa_{\SU(2n-1)^2} =n+\frac{1}{2}, \\
&\kappa_{R_0 R_0}= 2n^2+5n+2, &\quad \kappa_{A R_0}=-2n^2-1,\\
&\kappa_{Q R_0}= -1-8n, &\quad \kappa_{\tilde Q R_0}=1-4n^2,\\
&\kappa_{4 R_0}= 3, &
\end{array}
\end{equation}
and we checked that they match across the dual phases.

We conclude the analysis by studying the identity relating the elliptic genera of the dual phases. In this case the expected identity is
\begin{equation}
\label{TOProve6}
I_{\SU(2n+1)}^{(4;2n-1;\cdot;1;\cdot)} (x \vec u;y \vec v;\cdot;t^2;\cdot)
=
\frac{
\theta(q/(x^4 t^{4n-4} y^{2n-1}))\theta(q/(x^4 t^{4n-2}))
}
{\prod_{a,b}\theta(u_a v_b x y ) \prod_a \theta(t^{2n} u_a)\theta(t^{2n-2} u_a^{-1} x^3) \prod_{a<b}\theta(t^2 v_a v_b)
}\,,
\end{equation}
and we are going to prove that it follows from the basic identities for $\SU(n)$ and $\USp(2n)$ gauge groups with (anti-)fundamental chiral multiplets.
Actually in this case we do not start by considering the LHS of (\ref{TOProve6}) but we consider the addition of the Fermi flipper 
$\psi_A$ in the superpotential (\ref{Case3odd1}). This boils down to multiply both sides of the conjectured identity  (\ref{TOProve6}) 
by the term $\theta(q u_4/(t^{2n-2}  x^3))$.  Using the  relation  (\ref{inversion}) this terms simplify with the term  $\theta(t^{2n-2}  x^3 u_4^{-1})$
in the denominator in the RHS of  (\ref{TOProve6}).

We proceed by deconfining the antisymmetric through the substitution in the integrand on the LHS of (\ref{TOProve6})
\begin{equation}
\frac{\theta(q/(t^{2n-2} x^3 u_4^{-1}))}{\prod_{i=1}^{2n+1} \prod_{a=1}^{3} \theta (u_a x z_i) \prod_{i<j} \theta(z_i z_j t^2)} 
\rightarrow
I_{\USp(2n+2)}^{(3,2+1;\cdot;\cdot)} (u_1 x/t,u_2 x/t,u_3 x/t,t \vec z;\cdot;\cdot)\,.
\end{equation}
Then we use the relation 
(\ref{idbrutta}) (or equivalently (\ref{idbrutta2})) transforming the $\SU(2n+1)$ integral into  $\SU(2)$ and then we apply the relation (\ref{ellipticdualUSP}) to the $\USp(2n+2)$ integral.
We are then left with the integral associated to the model with an $\SU(2)$ gauge group, corresponding to the last quiver in Figure \ref{sixthcase}.
The elliptic genus of this theory can be again computed using (\ref{ellipticdualUSP}) for $N=1$ and in this way we arrive to the RHS of (\ref{TOProve6}), except the missing 
term $\theta(t^{2n-2}  x^3 u_4^{-1})$ corresponding to the presence of the flipper $\psi_A$ on the gauge theory side.

\section{$\SU(N)$ with one antisymmetric flavor}
\label{sec:SU2AS}

\begin{minipage}{\linewidth}
      \centering
      \begin{minipage}{0.3\linewidth}
          \begin{figure}[H]
              \includegraphics[width=\linewidth]{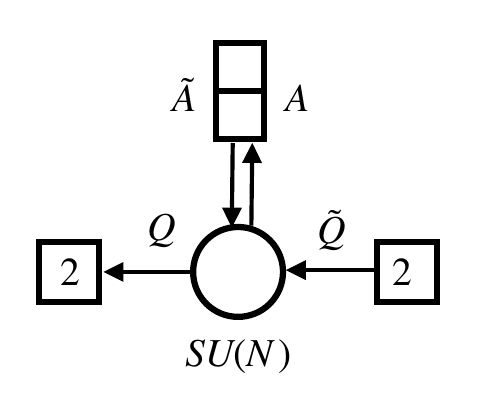}
             \caption{$N=2n$ in \ref{Case4even}; $N=2n+1$ in \ref{Case4odd}.}
          \end{figure}
      \end{minipage}
      \hspace{0.06\linewidth}
      \begin{minipage}{0.26\linewidth}
          \begin{figure}[H]
              \includegraphics[width=\linewidth]{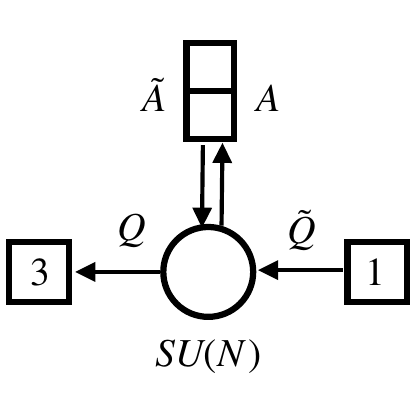}
              \caption{$N=2n$ in \ref{Case5even}; $N=2n+1$ in \ref{Case5odd}.}
          \end{figure}
      \end{minipage}
       \hspace{0.06\linewidth}
       \begin{minipage}{0.26\linewidth}
          \begin{figure}[H]
              \includegraphics[width=\linewidth]{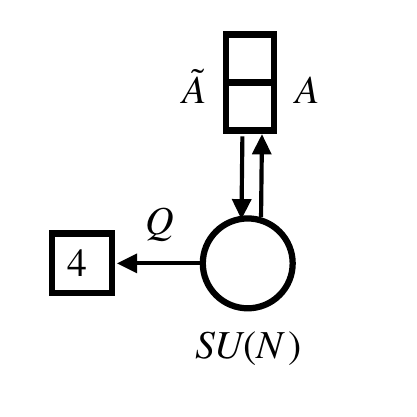}
             \caption{$N=2n$ in \ref{Case6even}.}
          \end{figure}
      \end{minipage}
  \end{minipage}\\

In this section we consider an $\SU(N)$ gauge theory with $2-M$ anti-fundamental chirals $\tilde Q$, $2+M$ fundamental chirals $Q$ (with $M=0,1,2$) one antisymmetric tensor $A$
and one conjugate antisymmetric tensor $\tilde A$. 
These are anomaly free gauge theories and we are going to support the claim that each model is dual to a LG theory.
Again the details of the LG description require to separate the discussion for each $M$ distinguishing the case of $N=2n$ and the case $N=2n+1$.

\subsection{$\SU(2n)$ with $2$ fundamental flavors}
\label{Case4even}

\begin{figure}
\begin{center}
  \includegraphics[width=12cm]{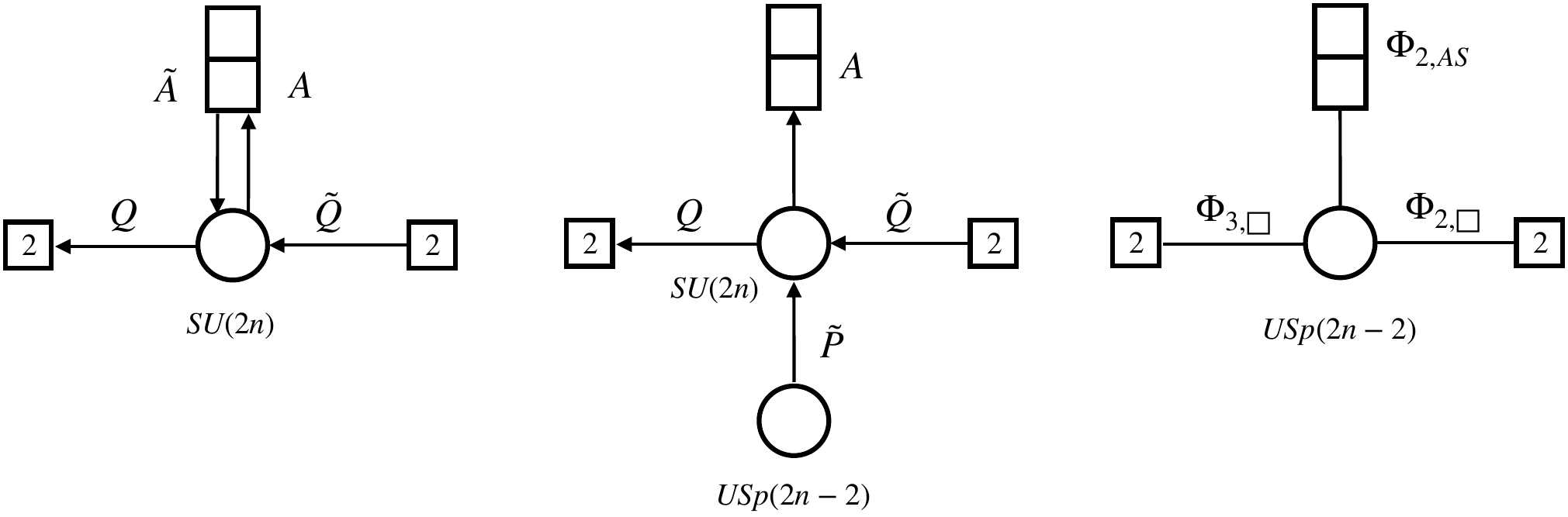}
  \end{center}
  \caption{In this figure we illustrate the process leading to the proof that the duality originates from other basic dualities. The first quiver represents the original $\SU(2n)$ gauge theory with two fundamental flavor
  and one antisymmetric flavor. Then we trade  a conjugate antisymmetric  with an $\USp(2n-2)$ gauge group, with a new bifundamental  $\tilde P$ between this gauge node and the original $\SU(2n)$.
  The we dualize $\SU(2n)$ using the result derived in subsection \ref{Case1even} obtaining the third quiver.
 }
    \label{thirdcase}
\end{figure}

We start by considering the case of $\SU(2n)$ with two fundamentals $Q$, two anti-fundamentals $\tilde Q$, one antisymmetric $A$ and one conjugate antisymmetric $\tilde A$.
This theory is dual to a LG  where the chiral fields $\Phi_I$ correspond to the gauge invariant combinations
\begin{equation}
    \begin{aligned}
        \varphi_1 &= \Pf A, \\
        \varphi_{5,k} &= Q(A \tilde A)^k \tilde Q,
    \end{aligned}
    \qquad
    \begin{aligned}
        \varphi_2 &=  \text{Pf} \tilde A ,\\
        \varphi_{6,m} &= \tilde A(A \tilde A)^m Q^2,
    \end{aligned}
    \qquad
    \begin{aligned}
        \varphi_3 &=  A^{n-1} Q^2 ,\\
        \varphi_{7,m} &=  A(A \tilde A)^m \tilde Q^2 ,
    \end{aligned}
    \qquad
    \begin{aligned}
        \varphi_4 &= \tilde A^{n-1} \tilde Q^2 ,\\
        \varphi_{8,\ell} &=  (A \tilde A)^\ell,
    \end{aligned}
\end{equation}
with $k=0,\dots,n-1$, $m=0,\dots,n-2$ and $\ell=1,\dots,n-1$.
In addition there are  $n$ Fermi $\Psi_{0,\dots,n}$.
The superpotential in this case is a complicated function of the 
chiral fields, where the number of terms increases with the
rank of the gauge group.
However, we claim that by flipping some of the operators in the electric theory, through the superpotential 
\begin{equation}
\label{guessed3}
W = \psi_A \text{Pf} A + \psi_{\tilde A} \text{Pf} {\tilde A}
+\sum_{\ell=1}^{n-1} \tilde \psi_{\ell} \,\mathrm{Tr}\, (A \tilde A)^\ell\,,
\end{equation}
the dual superpotential becomes cubic in the remaining $\Phi_{3,\dots,7}$ chiral bosons.
\begin{equation}
\label{expected3}
W
=
\Psi_{n-1}  \varphi_3  \varphi_4 
+
\sum_{i,j,k=1}^{n-1} \Psi_{i} \varphi_{6,j-1}\varphi_{7,k-1}
\delta_{j_1+j_2+j_3,2n-1}
+
\sum_{i,j,k=0}^{n-1} \Psi_{i} \varphi_{5,j} \varphi_{5,k} \delta_{i+j+k,2n-2}\,.
\end{equation}
Then we trade the tensor  $\tilde A$ with an $\USp(2n-2)$ gauge group
as in Figure \ref{thirdcase} and the superpotential becomes
\begin{equation}
W =  \psi_A \Pf A  
+\sum_{\ell=1}^{n-1} \tilde \psi_{\ell} \,\mathrm{Tr}\, (A \tilde P^2)^\ell\,.
\end{equation}
The next step consists of dualizing the $\SU(2n)$ gauge node using the duality discussed in 
subsection \ref{Case1even}.

Actually here the original  $\SU(2n)$ flavor symmetry is partially gauged and we need to split the representations of $\SU(2n)$ singlets accordingly.
The two fields $\Phi_1$ and $\Phi_4$ in (\ref{combo1}) are not charged under the $\SU(2n)$ flavor symmetry and we keep on referring to them with the same terminology. Such fields correspond to the combinations  
$\Phi_1 = A^{n-1} Q^2$ and $\Phi_{4} = \tilde Q^2 \tilde P^{2n-2}$.
On the other hand the fields $\Phi_2$ decompose into an $\USp(2n-2)$ singlet
$\Phi_{2,\cdot} = A \tilde Q^2$, two fundamentals $\Phi_{2,\square} = A \tilde Q \tilde P$
and an antisymmetric $\Phi_{2,AS} = A  \tilde P^2$.
Analogously $\Phi_3$ decomposes as a singlet $\Phi_{3,\cdot} = Q \tilde Q$ and two fundamentals
$\Phi_{3,\square} = Q \tilde P$.

In this way the superpotential $W = \hat \Psi_1 (\Phi_2^{n-1} \Phi_3^2 + \Phi_1 \Phi_4)+\hat \Psi_2 \Phi_2^n$ becomes
\begin{eqnarray}
\label{split4p1}
W &=& \hat \Psi_1 \Big(
\sum_{\ell=0}^{n-3} \left((\Phi_{3, \square} \Phi_{2,AS}^{n-3-\ell} \Phi_{2 ,\square} )(\Phi_{3, \square} \Phi_{2,AS}^{\ell} \Phi_{2, \square})+( \Phi_{3, \square} \Phi_{2,AS}^{n-3-\ell} \Phi_{3, \square} ) (\Phi_{2, \square} \Phi_{2,AS}^{\ell} \Phi_{2, \square}) \right)\nonumber \\
&+& \Phi_1 \Phi_4 +
\Phi_{3,\square} \Phi_{3,\cdot} \Phi_{2,AS}^{n-2}\Phi_{2,\square}
+
\Phi_{3,\cdot} \Phi_{3,\cdot} \Phi_{2,AS}^{n-1}
+
\Phi_{3,\square} \Phi_{3,\square}\Phi_{2,\cdot} \Phi_{2,AS}^{n-2} 
\Big)
\nonumber \\
&+&
\hat \Psi_2
\Big( \Phi_{2,AS}^{n-1} \Phi_{2,\cdot}
+
\Phi_{2,AS}^{n-2} \Phi_{2,\square} \Phi_{2,\square}
\Big)
+
\sum_{\ell=1}^{n-1} \tilde \psi_{\ell} \,\mathrm{Tr}\, \Phi_{2,AS}^\ell\,.
\end{eqnarray}
We are then left with an  $\USp(2n-2)$ gauge theory with a totally (traceless
\footnote{Such trace is indeed set to zero by the flipper $\tilde{\psi}_1$}
) antisymmetric $\Phi_{2,AS}$, and four fundamentals, 
where two 
of them are denoted as $\Phi_{2,\square}$ and the other two are denoted as $\Phi_{3,\square}$. This theory is dual to a LG model, where the fields are the mesonic combinations
\begin{equation}
\mathcal{L}_{ab}^{j} = \Phi_{a,\square}\Phi_{b,\square}\Phi_{2,AS}^{j-1}\quad j=1,\dots,n-1 \quad\text{and}\quad a,b=2,3\,.
\end{equation}
The superpotential of the LG model is obtained from (\ref{sacchiUSp4ASspot}) in addition to the deformations that can be read from (\ref{split4p1}). We have
\begin{eqnarray}
\label{LG4p1}
W&=& \hat \Psi_1\qty(\Phi_1 \Phi_4+ \mathcal{L}_{33}^{n-1} \Phi_{2,\cdot}+\mathcal{L}_{23}^{n-1}\Phi_{3,\cdot}+
\sum_{\ell=0}^{n-3} (\mathcal{L}_{23}^{(n-2-\ell)} \mathcal{L}_{23}^{(\ell+1)}+ \mathcal{L}_{22}^{(n-2-\ell)} \mathcal{L}_{33}^{(\ell+1)} ))
\nonumber \\
&+&
\sum_{i,j,k=1}^{n-1} \Psi_{\USp(2n-2)}^{(i)} (\mathcal{L}_{23}^{(j)}\mathcal{L}_{23}^{(k)}+\mathcal{L}_{22}^{(j)}\mathcal{L}_{33}^{(k)}) \delta_{i+j+k,2n-1}\Big|_{\mathcal{L}_{22}^{(n-1)}=0}\,,
\end{eqnarray}
where  $\Psi_{\USp(2n-2)}^{(j)}$ are Fermi fields.

We can compare this expression with (\ref{expected3}) by spelling out the explicit dictionary between the composites.
We start observing that the composite are mapped as
\begin{equation}
\mathcal{L}_{22}^{(i)} = \varphi_{7,i},\,\,\,
\mathcal{L}_{33}^{(j)} =\varphi_{6,j-1},\,\,\,
\mathcal{L}_{23}^{(j)} = \varphi_{5,j},\,\,\, 
\Phi_{2,\cdot} = \varphi_{7,0},\,\,\,
\Phi_{3,\cdot} = \varphi_{5,0},\,\,\,
\Phi_1 = \varphi_3,\,\,\,
\Phi_4 = \varphi_4\,,
\end{equation} while the Fermi are mapped as
$\hat \Psi_1 = \Psi_{n-1}$ and 
$\Psi_{\USp(2n-2)}^{(j)} = \Psi_{j-1}$
with $i=1,\dots,n-2$, $j=1,\dots,n-1$. Using this dictionary we have checked that (\ref{expected3}) and (\ref{LG4p1}) become identical.

We proceed by checking the anomaly matching of the global symmetries. The charges of the fields in the electric and in the dual LG theory, including the flippers in (\ref{guessed3}) 
\begin{equation}
\begin{array}{c|ccccccc}
 & \UU(1)_{Q}& \SU(2)_Q& \UU(1)_{\tilde Q}& \SU(2)_{\tilde Q}&\UU(1)_A&\UU(1)_{\tilde A} &\UU(1)_R \\
Q &1&\square&0&\cdot&0&0&0\\
\tilde Q&0&\cdot&1&\square&0&0&0\\
A&0&\cdot&0&\cdot&1&0&0\\
\tilde A&0&\cdot&0&\cdot&0&1&0\\
\psi_A&0&\cdot&0&\cdot&-n&0&1\\
\psi_{\tilde A}&0&\cdot&0&\cdot&0&-n&1\\
\tilde \psi_\ell&0&\cdot&0&\cdot&-\ell&-\ell&1\\
\hline
\varphi_{3}&2&\cdot&0&\cdot&n-1&0&0\\
\varphi_{4}&0&\cdot&2&\cdot&0&n-1&0\\
\varphi_{5,k}&1&\square&1&\square&k&k&0\\
\varphi_{6,m}&2&\cdot&0&\cdot&m&m+1&0\\
\varphi_{7,m}&0&\cdot&2&\cdot&m+1&m&0\\
\Psi_j&-2&\cdot&-2&\cdot&2-2n+j\quad\quad&2-2n+j&1\\
\end{array}
\end{equation}
with $\ell=1,\dots,n-1$, $k=0,\dots,n-1$, $m=0,\dots,n-2$ and $j=0,\dots,n-1$.
The anomalies of the global symmetries are given by
\begin{equation}
\begin{array}{lll}
&
\kappa_{QQ} =4n,& \quad
\kappa_{\tilde Q \tilde Q} =4n,  \\
&
\kappa_{AA}  =-\frac{n(n-1)  (2 n-7)}{6} ,& \quad
\kappa_{QA}= 0, \\
&
\kappa_{\tilde Q Q} = 0, & \quad
\kappa_{\tilde QA} = 0,   \\
&\kappa_{\tilde A \tilde A} =-\frac{n(n-1)  (2 n-7) }{6} ,& \quad
\kappa_{\tilde A  A} =-\frac{n (n-1)  (2 n-1)}{6},  \\
&\kappa_{Q \tilde A} = 0,& \quad
\kappa_{\tilde A \tilde Q} =0, \\
&
\kappa_{\SU(2)_Q^2} =n, & \quad
\kappa_{\SU(2)_{\tilde Q}^2}=n,
\end{array}
\end{equation}
and we checked that they match across the dual phases.

In this case we provide a derivation of the duality from a 4d parent duality.
The electric theory is a $\SU(2n)$ gauge theory with one antisymmetric flavor $(\mathcal{A},\tilde{\mathcal{A}})$ and three fundamental flavors $(Q,\tilde{Q})$,
with the superpotential
\begin{equation}
\label{elebajeot}
W_{4\mathrm{d}}^{(\text{ele})} =
\sum_{i=1}^{n-1} \alpha_i \Tr \big(\mathcal{A} \tilde{\mathcal{A}}\big)^i + \beta \Pf \mathcal{A} + \tilde \beta \Pf \tilde{\mathcal{A}}\,.
\end{equation}
In this way the confining theory corresponds to a WZ model described by the 4d superfields
\begin{equation}
    \begin{aligned}
        B_2&= \mathcal{A}^{n-1} Q^2,\\
        \tilde B_2&= \tilde {\mathcal{A}}^{n-1} \tilde Q^2,
    \end{aligned}
    \qquad
    \begin{aligned}
        h_{m+1} &= \tilde{\mathcal{A}}(\mathcal{A} \tilde{\mathcal{A}})^m  Q^2, \\
        \tilde h_{m+2}&=\mathcal{A}(\mathcal{A} \tilde{\mathcal{A}})^m  \tilde Q^2,
    \end{aligned}
    \qquad
    \begin{aligned}
        M_{j+1} &= Q (\mathcal{A}  \tilde{\mathcal{A}})^{j} \tilde Q, \\
        &
    \end{aligned}
\end{equation}
with $j=0,\dots,n-1$ and $m=0,\dots,n-2$ and with superpotential
\begin{equation}
\label{WBB}
W = B_2 \tilde B_2 M_n + \sum_{i,k,j=1}^{n} (\tilde h_i M_j h_k + M_i M_j M_k) \delta_{i+j+k,2n+1}\Big|_{h_n=\tilde h_1 = 0}\,.
\end{equation}
This flipped duality was derived originally in \cite{Bajeot:2022kwt} from the deconfinement technique in four dimensions.
The  2d duality can be derived by topologically twisting such flipped 4d confining duality.
The twist is done along the 4d non anomalous $R$ symmetry that assigns $R$ charge 0 to the antisymmetric, its conjugate, two  anti-fundamentals and two fundamentals and $R$ charge 1 to the remaining  fundamental and anti-fundamental.
Furthermore the flippers $\alpha_i$, $\beta$ and $\tilde \beta$ have 
R-charge 2.
Such charge assignation provides the same field content of the 2d theory discussed above, where the flippers become the Fermi fields in the superpotentials (\ref{guessed3}).
Some of the 4d singlets survive as 2d chirals, some  as 2d Fermi and other have $R$ charge 1 and they disappear from the 2d dynamics. The precise $4d/2d$ map for the fields that survives is
\begin{eqnarray}
B_2 \rightarrow \varphi_3,  \quad
\tilde B_2 \rightarrow \varphi_4,  \quad
M_{j+1} \rightarrow \{\varphi_{5,j},\Psi_{j}\},  \quad
\tilde h_m\rightarrow  \varphi_{7,m-2},  \quad
h_m\rightarrow \varphi_{6,m-1}\,.
\end{eqnarray}
By applying this map one can also check that the 2d superpotential (\ref{expected3}) is recovered from the 4d one (\ref{WBB}).

We conclude the analysis by studying the identity relating the elliptic genera of
the dual phases. In this case the index of the original theory is given by
\begin{equation}
\label{original41}
I=
 \theta\left(\frac{q}{ t^{2n}}\right) \theta\left(\frac{q}{ w^{2n}}\right) 
\prod_{\ell=1}^{n-1} \theta\left(\frac{q}{ (tw)^{2j}}\right)     
I_{\SU(2n)}^{(2;2;\cdot;1;1)}(x \vec u;y \vec v; \cdot;t^2;w^2)\,.
\end{equation}

We proceed by considering  a substitution involving also the $\theta$ in the integrand  of (\ref{original41}), trading the $\theta$ associated to the antisymmetric $\tilde A$ with an $\USp(2n-2)$ integral. Explicitly the substitution is 
\begin{equation}
\frac{ \theta(q/w^{2n}) }
{\prod_{i<j} \theta (z_i z_j w^2) }
\rightarrow 
  I_{\USp(2n-2)}^{(2n;\cdot;\cdot)}(w/ \vec z;\cdot;\cdot)\,.
\end{equation}

Then we dualize the $\SU(2n)$ gauge group using the relation (\ref{toprove1}), with the aid of formula (\ref{inversion}),
 and we obtain the index for the $\USp(2n-2)$  theory with four fundamentals.
\begin{equation}
    \begin{aligned}
        I = &\frac{\theta(q/((wt)^{2n-2} x^2 y^2)) \theta(q/(w^{2n-2} t^{2n} y^2)) \prod_{\ell=1}^{n} \theta(q/(wt)^{2 \ell})}{\theta(w^{2n-2}y^2) \theta(t^2 y^2) \theta(t^{2n-2} x^2) \prod_{a,b=1}^{2} \theta(u_a v_b x y)}
        \\
        \times &I_{\USp(2n-2)}^{(2,2;\cdot;1)}(x w \vec u,t^2 w y \vec v ;\cdot;t^2 w^2)\,.
    \end{aligned}
\end{equation}
Then we use the identity (\ref{USp(2n)A4}) and apply (\ref{inversion}),  such that the final results becomes 
\begin{equation}
I = 
\frac{
\prod_{\ell=0}^{n-1}\!
\theta(q/(x^2 y^2 (w t)^{2(2n-2-\ell)}))
}{
\theta(t^{2n-2} x^2)
\theta(w^{2n-2} y^2)
\prod_{\ell=0}^{n-2}\!
\theta(  x^2 w^{2\ell+2} t^{2\ell}  )
 \theta(  y^2 w^{2\ell} t^{2\ell+2}  )
\prod_{j=0}^{n-1} \!\prod_{a,b=1}^{2}\!\theta( u_a v_b x y (wt)^{2j}  )
}\,,
\end{equation}
corresponding to the collection of $\theta$ functions for the chirals and the Fermi expected 
from the duality obtained in the field theory analysis above.

\subsection{$\SU(2n+1)$ with $2$ fundamental flavors}
\label{Case4odd}

\begin{figure}
\begin{center}
  \includegraphics[width=12cm]{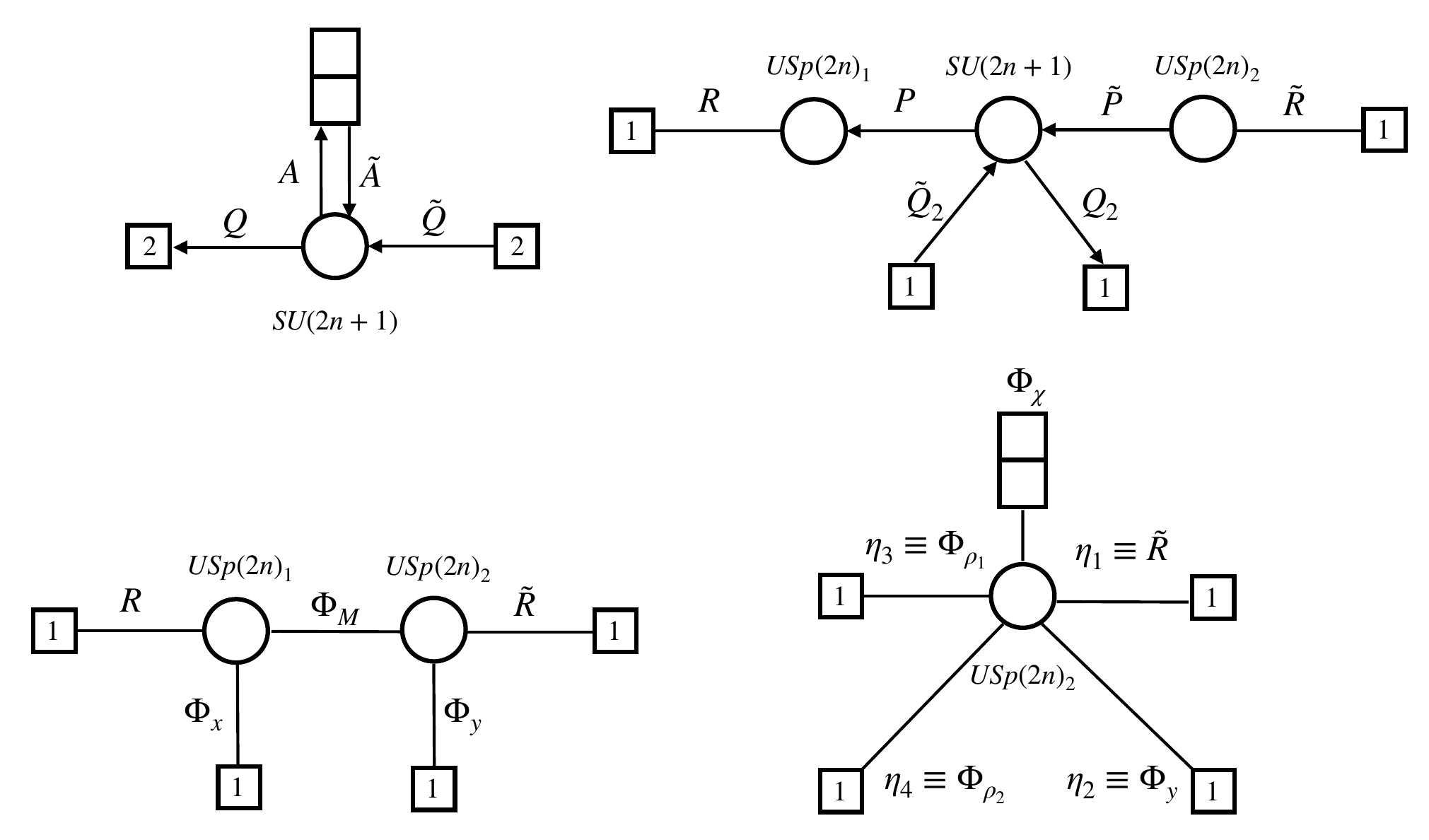}
  \end{center}
  \caption{In this figure we illustrate the process leading to the proof that the duality originates from other basic dualities. The first quiver represents the original $\SU(2n+1)$ gauge theory with two fundamental flavors
  and one antisymmetric flavor. Then we trade  each conjugate pair of antisymmetric and one fundamental with an $\USp(2n)$ gauge group, with a bifundamental  and one fundamental.
  This procedure  breaks (only apparently, due to the structure of the flippers in the superpotential (\ref{flipped4})) the two global $\SU(2)$ symmetries rotating the bifundamentals to the Cartan
  subgroups. 
  Then we dualize $\SU(2n+1)$ obtaining the third $\USp(2n) \times \USp(2n)$ quiver.
  The final quiver is obtained by dualizing one of the two $\USp(2n)$ gauge groups. 
 }
    \label{fourthcase}
\end{figure}

Here we consider the case of $\SU(2n+1)$ with two fundamentals $Q$, two anti-fundamentals $\tilde Q$, one antisymmetric $A$ and one conjugate antisymmetric $\tilde A$.
This theory is dual to a LG  where the chiral fields $\varphi_I$ correspond to the gauge invariant combinations
\begin{equation}
    \label{terms}
    \begin{aligned}
        \varphi_1 &= A^n Q, \\
        \varphi_{4,k}  &= \tilde A(A \tilde A)^k Q^2,
    \end{aligned}
    \qquad
    \begin{aligned}
        \varphi_2&=  \tilde A^n \tilde Q,\\
        \varphi_{4,k}  &= \tilde A(A \tilde A)^k Q^2,
    \end{aligned}
    \qquad
    \begin{aligned}
        \varphi_{3,k} &= Q(A \tilde A)^k \tilde Q,\\
        \varphi_{6,\ell} &=  (A \tilde A)^\ell,
    \end{aligned}
\end{equation}
with $k=0,\dots,n-1$ and $\ell=1,\dots,n$,
in addition to $n$ Fermi $\Psi_{0,\dots,n}$.
Again the superpotential is a complicated function of the 
chiral fields, where the number of terms increases with the
rank of the gauge group.
However, we claim that by flipping some of the operators in the electric theory, through the superpotential 
\begin{equation}
\label{flipped4}
W = \psi_A A^n Q + \psi_{\tilde A} \tilde A^n \tilde Q 
+\sum_{\ell=1}^{n} \tilde \psi_{\ell}\, \mathrm{Tr} \,(A \tilde A)^\ell
+\psi_M Q \tilde Q,
\end{equation}
the dual superpotential becomes cubic in the remaining chiral multiplets.
\begin{eqnarray}
\label{expected4}
W
=\sum_{i,j,k=0}^{n-1}\Psi_{i}\varphi_{4,j}\varphi_{5,k}\delta_{i+j+k,2n-2}+
\sum_{i,j,k=1}^{n-1}\Psi_{i}\varphi_{3,j}\varphi_{3,k}\delta_{i+j+k,2n-1}.
\end{eqnarray}
In order to proceed in this case 
we consider explicitly the flavor structure of the flippers.
The superpotential of the electric theory in this case becomes
\begin{equation}
\label{flipped4}
W = \sum_{i=1}^{2} \left( \hat \psi_A^{(i)}  A^n Q _i+  \check \psi_{\tilde A}^{(i)}  \tilde A^n \tilde Q _i \right)
+\sum_{\ell=1}^{n} \tilde \psi_{\ell} \,\mathrm{Tr}\, (A \tilde A)^\ell+\psi_M Q \tilde Q\,.
\end{equation}
Then we trade the two antisymmetrics with two $\USp(2n)$ gauge theories
as in Figure \ref{fourthcase}. The
 superpotential for this theory is
\begin{eqnarray}
W &=& \hat \psi_A^{(2)}  P^{2n} Q_2 +
 \check \psi_{\tilde A} ^{(2)} \tilde P^{2n} \tilde Q_2
+\sum_{\ell=1}^{n} \tilde \psi_{\ell} \,\mathrm{Tr}\, (P \tilde P)^{2\ell}
\nonumber \\
&+& \psi_{M_{11}} P R \tilde P \tilde R
+ \psi_{M_{12}} P R \tilde Q_2
+ \psi_{M_{21}} Q_2 \tilde P \tilde R
+ \psi_{M_{22}} Q_2 \tilde Q_2.
\end{eqnarray}
The next step consists of dualizing the $\SU(2n+1)$ gauge node that has
$2n$ fundamentals $P$, one fundamental $Q_2$,
$2n$ anti-fundamentals $\tilde P$ and one anti-fundamental $\tilde Q_2$, using the duality reviewed in Appendix \ref{sec:SUNNN}.
Defining the following $\SU(2n+1)$ gauge invariant chiral fields 
$\Phi_M = P \tilde P$, $\Phi_m = Q_2 \tilde Q_2$,
$\Phi_x = P \tilde Q_2$, $\Phi_y = Q_2 \tilde  P$,
$\Phi_B = P^{2n} Q_2$ and $\Phi_{\tilde B} = \tilde P^{2n} \tilde Q_2$
the superpotential of the dual theory is
\begin{eqnarray}
W & =&\hat \psi_A^{(2)}  \Phi_B+
 \check \psi_{\tilde A} ^{(2)} \Phi_{\tilde B} +
\sum_{\ell=1}^{n} \tilde \psi_{\ell} \Tr \Phi_M^{2\ell}
+
\Psi_{\SU(2n)} \left(\det 
\left(\begin{array}{cc}
\Phi_M &\Phi_x   \\
\Phi_y & \Phi_m \\
\end{array}
\right)
+ \Phi_{\tilde B} \Phi_{B}  \right) \nonumber \\
&+& \psi_{M_{11}}  R  \Phi_M \tilde R
+ \psi_{M_{12}}  R  \Phi_x
+ \psi_{M_{21}} \Phi_y \tilde R
+ \psi_{M_{22}} \Phi_m.
\end{eqnarray}
The chirals $\Phi_B$ and $\Phi_{\tilde B}$  and $\Phi_m$ are set to zero by the equation of motion.
We can then dualize one of the two $\USp(2n)$ gauge theories (for example we choose $\USp(2n)_{1}$ in the following), using the duality reviewed in Appendix \ref{sec:USP2Np2}.
There are $(2n+2)$ fundamentals of $\USp(2n)_{1}$, $2n$ identified with $\phi_M$, one with $\phi_x$ and one with $R$.
 The singlets of this gauge theory correspond to 
an antisymmetric of $\USp(2n)_2$ denoted as  $\Phi_\chi = \Phi_M^2$
 two fundamentals of $\USp(2n)_2$ denoted as  $\Phi_{\rho_1} = \Phi_M R$ and  $\Phi_{\rho_2} = \Phi_M \phi_x$ and a singlet $\Phi_s = R  \phi_x$.
 The superpotential is
\begin{eqnarray}
    \label{calenda}
W &=& \sum_{\ell=1}^{n} \tilde \psi_{\ell} \Tr \Phi_\chi^{\ell}
+
\Psi_{\SU(2n+1)}
\Phi_{\rho_2}  \Phi_y \Phi_\chi ^{n-1}
+
\Psi_{\USp(2n)_1}
\text{Pf} \left(
\begin{array}{ccc}
\Phi_\chi &\Phi_{\rho_1}&\Phi_{\rho_2}\\
-\Phi_{\rho_1}&0&\Phi_s\\
-\Phi_{\rho_2}&-\Phi_s&0
\end{array}
\right)
\nonumber \\
&+& \psi_{M_{11}}   \Phi_{\rho_1} \tilde R
+ \psi_{M_{12}}    \Phi_s
+ \psi_{M_{21}} \Phi_y \tilde R,
\end{eqnarray}
where the chiral $\Phi_s$ is set to zero by the equation of motion and the leftover in the Pfaffian is then
$\epsilon \cdot \left( \Phi_\chi ^{n-1 }\Phi_{\rho_1}\Phi_{\rho_2} 
\right)$.
The last step consists of dualizing the $\USp(2n)_2$ gauge theory with an antisymmetric chiral
\footnote{Observe that this field is actually traceless because of the flipper $\tilde{\psi}_1$, \eqref{calenda}, this allows us to dualize the $\USp(2n)_2$ gauge group using the results review in Appendix \ref{appSacchi}.}
$\Phi_{\chi}$ and four fundamental chirals, identified 
with the chirals $\eta\equiv \{\tilde R$, $\Phi_y$, $\Phi_{\rho_1},\Phi_{\rho_2}\}$.
The singlets of the duality are
$\Phi_{\eta_a \eta_b}^{(j)} =\eta_a \Phi_\chi^{j-1}\eta_b$
with $j=1,\dots,n$.
The superpotential becomes
\begin{eqnarray}
\label{spot2asas}
W&=&\sum_{i,j,k} \epsilon_{abcd}  \Psi_{\USp(2n)_2}^{(i)} 
\Phi_{\eta_a \eta_b}^{(j)} \Phi_{\eta_c \eta_d}^{(k)} 
\delta_{i+j+k,2n+1}
+
\Psi_{\SU(2n+1)} \Phi_{\eta_2 \eta_4}^{(n)}
+
\Psi_{\USp(2n)_1} \Phi_{\eta_3 \eta_4}^{(n)}
\nonumber \\
&+& \psi_{M_{11}}   \Phi_{\eta_1 \eta_3}^{(1)}
+ \psi_{M_{21}} \Phi_{\eta_1 \eta_2}^{(1)}
\end{eqnarray}
After integrating out the massive terms we can 
associated the surviving fields with the ones spelled out in formula 
(\ref{terms}).
The precise dictionary is
\begin{equation}
\varphi_{3,(a1)}^{(\ell)}  \leftrightarrow \Phi_{\eta_1 \eta_c}^{(\ell+1)},\quad 
\varphi_{3,(a2)}^{(\ell)}  \leftrightarrow \Phi_{\eta_4 \eta_c}^{(\ell)},\quad 
\varphi_4^{(j)}  \leftrightarrow \Phi_{\eta_2 \eta_3}^{(j+1)},\quad 
\varphi_5^{(j)}  \leftrightarrow \Phi_{\eta_1 \eta_4}^{(j+1)}, \quad
\Psi_{j} \leftrightarrow \Psi_{\USp(2n)_2}^{(j+1)}   
\end{equation}
with $a=1,2$, $c=2,3$, $\ell=1,\dots,n-1$ and $j=0,\dots,n-1$. After using the above dictionary, and integrating out the massive fields in \eqref{spot2asas}, we get the superpotential in \eqref{expected4}.

We can derive the duality from 4d using the results of \cite{Bajeot:2022kwt}.
The 4d $\SU(2n+1)$ electric gauge theory has an antisymmetric $\mathcal{A}$,
a conjugate antisymmetric $\tilde{\mathcal{A}}$, three fundamentals $\mathcal{Q}_{1,2,3}$ and three anti-fundamentals 
$\widetilde{\mathcal{Q}}_{1,2,3}$. Here we further consider the flip of some of the chiral ring operators in the electric superpotential. Our choice of flippers is actually different from the one discussed in \cite{Bajeot:2022kwt}. Namely we have
\begin{equation}
\label{spotdaat}
W = \sum_{a=1}^{3} (s_a \mathcal{A}^{n}  \mathcal{Q}_a
+  \tilde s_a \tilde{\mathcal{A}}^{n} \widetilde{\mathcal{Q}}_a) +\sum_{\ell=1}^{n} \beta_\ell \,\mathrm{Tr}\,(\mathcal{A} \mathcal{\tilde A})^\ell\,.
\end{equation}
The non-vanishing gauge singlets in the chiral ring are
\begin{eqnarray} 
\label{eq:mapbenve}
\Sigma_1^{(k)} = \mathcal{Q}  (\mathcal{A} \tilde{ \mathcal{A}})^k \widetilde{\mathcal{Q}},\quad
\Sigma_2^{(k+1)} =   \mathcal{A} (\mathcal{A} \tilde{ \mathcal{A}})^k \widetilde{\mathcal{Q}}^2,\quad
\Sigma_3^{(k)} =\tilde{\mathcal{A}}  (\mathcal{A} \tilde{ \mathcal{A}})^k \mathcal{Q}^2  ,
\end{eqnarray}
with $k=0,\dots,n-1$ .
The dual superpotential can be read from the analysis of \cite{Bajeot:2022kwt} and it is
\begin{eqnarray}
\label{dualbenve}
W&=&\Sigma_3^{(n-1)} \Sigma_1^{(0)} \Sigma_2^{(n)}
+
\Sigma_2^{(1)} \Sigma_1^{(n-1)} \Sigma_3^{(n-1)}
+
\sum_{\ell=0}^{n-3} \Sigma_2^{(\ell+2)} \Sigma_1^{(n-2-\ell)} \Sigma_3^{(n-1)}
\nonumber \\
&+&
\sum_{i,j,k=1}^{n-1} ( \Sigma_3^{(i-1)} \Sigma_1^{(j)} \Sigma_2^{(k+1)}
+
 \Sigma_1^{(i)} \Sigma_1^{(j)} \Sigma_1^{(k)}) \delta_{i+j+k,2n-1}\,.
\end{eqnarray}
Actually in order to make contact with the 2d model discussed in this section we also add an extra superpotential term to (\ref{spotdaat})
\begin{equation}
\label{defbenve}
\Delta W = \sum_{a,b=1}^{2} \mathcal{M}_{ab} \mathcal{Q}_a \widetilde{\mathcal{Q}}_b\,,
\end{equation}
breaking the $SU(3)^2$ flavor symmetry. The flipper $\mathcal{M}_{ab} $ removes the terms $ \Sigma_{1,ab}^{(0)}$
from the dual superpotential in (\ref{dualbenve}).

Then we assign the $R$ charges to the fields, setting all of  them to zero except for $\mathcal{Q}_3$ and $\widetilde{\mathcal{Q}}_3$, that are set to one. The flippers $s_{1,2}$, $\tilde s_{1,2}$, $\beta_\ell$ and $\mathcal{M}_{ab}$ have $R$ charges $R=2$, while the flippers $s_{3}$ and $\tilde s_{3}$ have $R$ charge 1.
It follows that the 4d dual fields $\Sigma_2^{(k+1)} $ and  $\Sigma_3^{(k)} $ survive as 2d chiral fields (only the ones carrying $\mathcal{Q}_{1,2}$ and  $\widetilde{\mathcal{Q}}_{1,2}$), corresponding to the  2d fields $\varphi_5^{(k)}$ and   $\varphi_4^{(k)}$  while the 
fields $\Sigma_1^{(j)} $ split into chirals corresponding to the fields $\varphi_3^{(j)}$, for $j=1,\dots,n-1$, and Fermi fields,  for $j=0,\dots,n-1$, corresponding to the Fermi $\Psi_{j}$. 
Plugging the fields that survive the twist into the 4d superpotential (\ref{dualbenve}), once the electric deformation (\ref{defbenve}) is added, we recover the superpotential (\ref{spot2asas}).

We proceed by checking the anomaly matching of the global symmetries. The
charges of the fields in the gauge and in the dual LG theory, including the flippers in (\ref{flipped4}),
are
\begin{equation}
\begin{array}{c|ccccccc}
 & \UU(1)_{Q}& \SU(2)_Q& \UU(1)_{\tilde Q}& \SU(2)_{\tilde Q}&\UU(1)_A&\UU(1)_{\tilde A} &\UU(1)_R \\
Q &1&\square&0&\cdot&0&0&0\\
\tilde Q&0&\cdot&1&\square&0&0&0\\
A&0&\cdot&0&\cdot&1&0&0\\
\tilde A&0&\cdot&0&\cdot&0&1&0\\
\psi_A&-1&\square&0&\cdot&-n&0&1\\
\psi_{\tilde A}&0&\cdot&-1&\square&0&-n&1\\
\tilde \psi_\ell&0&\cdot&0&\cdot&-\ell&-\ell&1\\
\psi_M&-1&\square&-1&\square&0&0&1\\
\hline
\varphi_{3}^{(k)}&1&\square&1&\square&k&k&0\\
\varphi_{4}^{(m)}&2&\cdot&0&\cdot&m&m+1&0\\
\varphi_{5}^{(m)}&0&\cdot&2&\cdot&m+1&m&0\\
\Psi_j&-2&\cdot&-2&\cdot&2-2n+j\quad\quad&2-2n+j&1\\
\end{array}
\end{equation}
with $\ell=1,\dots,n$, $k=1,\dots,n-1$, $m=0,\dots,n-1$ and $j=0,\dots,n-1$.
The anomalies of the global symmetries are given by
\begin{equation}
\begin{array}{lll}
&
\kappa_{QQ} =4(n-1),& \quad
\kappa_{\tilde Q\tilde Q} =4(n-1), \nonumber \\
&
\kappa_{AA} =-\frac{n(n-1)  (2 n+5)}{6} ,& \quad
\kappa_{QA}=-2n,\nonumber \\
&
\kappa_{\tilde Q Q} = -4, & \quad
\kappa_{\tilde  Q A}  = 0,  \nonumber \\
&\kappa_{\tilde  A \tilde A}=-\frac{n(n-1)  (2 n+5)}{6},& \quad
\kappa_{\tilde  A A}=-\frac{ n (n+1) (2 n+1)}{6},\nonumber \\
&\kappa_{\tilde  A Q}= 0,& \quad
\kappa_{\tilde  Q \tilde A}=-2n,\nonumber \\
&
\kappa_{\SU(2)_Q^2} =n-1, & \quad
\kappa_{\SU(2)_{\tilde Q}^2}=n-1,
\end{array}
\end{equation}
and we checked that they match across the dual phases.

We conclude by checking the matching of the elliptic genera.
The index of the gauge theory is 
\begin{eqnarray}
\label{original4}
I &=&
\prod_{a,b=1}^{2} \theta(q/(u_a v_b x y ) \cdot
\prod_{\ell=1}^{n} \theta(q/(wt)^{2j})\cdot
\prod_{a=1}^{2} \theta(q/(t^{2n} u_a x))  \theta(q/(w^{2n} v_a y)) 
\nonumber \\
&\times&
I_{SU(2n+1)}^{(2;2;\cdot;1;1)}(x \vec u;y \vec v;\cdot;t^2;w^2)
\end{eqnarray}
with $u_1 u_2=v_1 v_2 = 1$ and 
where the terms in the first line corresponds to the Fermi flippers in (\ref{flipped4}).

We proceed by considering two substitutions involving also the $\theta$ functions  in the integrand in (\ref{original4}). Such substitutions are
\begin{eqnarray}
\frac{
\theta(x u_1 t^{2n})
}{\prod_{i=1}^{2n+1} 
\theta (z_i u_1 x)
\prod_{i<j} \theta (z_i z_j t^2) }
\rightarrow I_{\USp(2n)}^{(2n+1,1;\cdot;\cdot)}(t \vec z,u_1 x/t;\cdot;\cdot)
\end{eqnarray}
and
\begin{eqnarray}
\frac{
\theta(y v_1 w^{2n})
}{\prod_{i=1}^{2n+1} 
\theta (z_i^{-1} v_1 y)
\prod_{i<j} \theta (z_i^{-1} z_j^{-1} w^2) }
 \rightarrow I_{\USp(2n)}^{(2n+1,1;\cdot;\cdot)}(w/\vec z,v_1 y/w;\cdot;\cdot).
\end{eqnarray}

Then we dualize the $\SU(2n+1)$ gauge group using the relation (\ref{ellipticdualSU}), obtaining the integral
\begin{eqnarray}&&
\left( \frac{(q;q)_\infty^{2n}}{2^n n!}\right)^2 
  \theta \left(\frac{q}{t^{2n} w^{2n} u_2 v_2 x y}\right)
  \theta \left(\frac{q}{ x y u_2 v_1}\right)
    \theta \left(\frac{q}{ x y u_1 v_2}\right)
  \theta \left(\frac{q}{ x y u_1 v_1}\right)
 \cdot
\prod_{\ell=1}^{n} \theta \left(\frac{q}{(wt)^{2j}}\right)
\nonumber \\
&&
\jkint 
\prod_{\ell=1}^{n} \frac{d \rho_\ell}{2\pi i \rho_\ell} 
\frac{d \sigma_\ell}{2\pi i \sigma_\ell} 
 \frac{\prod_{\ell < k} 
 \theta\left(\sigma_\ell^{\pm 1} \sigma_k^{\pm 1} \right)
 \theta\left(\rho_\ell^{\pm 1} \rho_k^{\pm 1} \right) 
\prod_{\ell=1}^{n} \theta\left(\rho_\ell^{\pm 2} \right)
 \theta\left(\sigma_\ell^{\pm 2} \right)
}
{\prod_{j,\ell=1}^{n} \theta(t w \sigma_j^{\pm 1} \rho_\ell^{\pm 1})
 \prod_{j=1}^{n}  
\theta\Big( \frac{v_1 y  \sigma_j^{\pm 1} }{w}\Big)
\theta( u_2 x w \sigma_j^{\pm 1} )
\theta\Big(\frac{u_1 x\rho_j^{\pm 1}}{t}\Big) 
\theta(v_2 y t\rho_j^{\pm 1})
},
\nonumber \\
\end{eqnarray}
corresponding to the elliptic genus of the third quiver in Figure \ref{fourthcase}.
We proceed by applying (\ref{ellipticdualUSP}) to the integrals in $\rho_{\ell}$ (or equivalently to the integrals in  $\sigma_{\ell}$).
We choose the first option in order to keep the discussion parallel to the field theory analysis.
Once we apply such formula we are left with

\begin{eqnarray}
\label{Ultimostepfourth}
I&=&
\theta\left(\frac{q}{t^{2n} w^{2n} u_2 v_2 x y}\right)
\theta\left(\frac{q}{t^{2n} w^{2n} u_1 v_2 x y }\right)
\theta\left(\frac{q}{x y u_2 v_1}\right) 
\theta\left(\frac{q}{x y u_1 v_1}\right) 
\nonumber \\
&\times&
\prod_{\ell=2}^{n} \theta\left(\frac{q}{(wt)^{2j}}\right) \,
I_{USp(2n)}^{(4;\cdot;1)}
\Big(\frac{v_1 y}{w},u_2 x w ,v_2 y t^2 w,u_1 x w;\cdot;w^2 t^2\Big).
\end{eqnarray}

The last step consists of applying the identity (\ref{USp(2n)A4}) to the integral (\ref{Ultimostepfourth}).
After massaging the result using formula  (\ref{inversion}) we obtain
\begin{eqnarray}
\frac{\prod_{\ell=1}^{n} \theta(q/(x^2 y^2 (w^2 t^2)^{2n-\ell-1}))}{\prod_{a,b=1}^{2} \left(\prod_{j=1}^{n-1} \theta(u_a v_b x y (wt)^{2j}) \cdot
\prod_{\ell=0}^{n-1}  \theta(x^2 w^{2\ell+2} t^{2\ell}) \theta(y^2 w^{2\ell} t^{2\ell+2})
\right)},
\end{eqnarray}
that corresponds to the expected elliptic genus for the dual theory studied above.

%
%
%
%
%
%
\subsection{$\SU(2n)$ with $3$ fundamentals and $1$ anti-fundamental}
\label{Case5even}
%
%
%
%
%
%
\begin{figure}
\begin{center}
  \includegraphics[width=12cm]{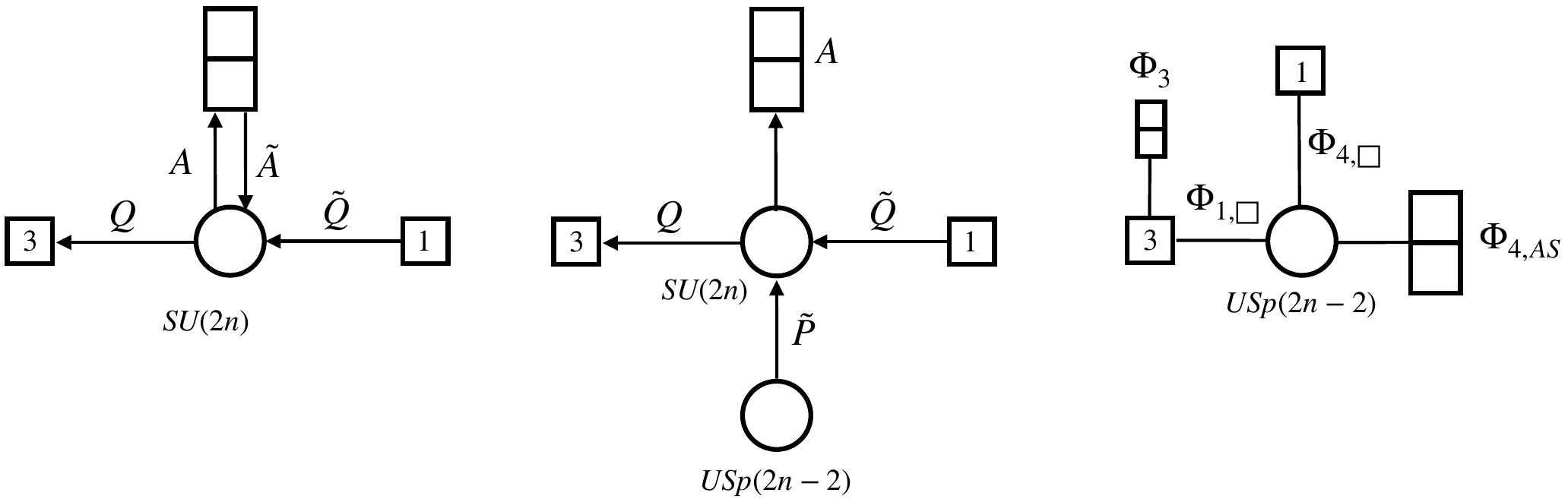}
  \end{center}
 \caption{The first quiver represents the original $\SU(2n)$ gauge theory with three fundamentals, one anti-fundamental  
  and one antisymmetric flavor. Then we trade the conjugate antisymmetric with an $\USp(2n-2)$ gauge group with a bifundamental $\tilde{P}$.
  Then we dualize the $\SU(2n)$ group using the duality derived in subsection \ref{Case2even}, where the $\SU(2n-1)$ global symmetry is partially gauged, obtaining the third $\USp(2n)$ quiver. We represented in this quiver the singlets in non trivial representations of the flavor symmetry group.}
    \label{31AAtev}
\end{figure}

Here we consider the case of $\SU(2n)$ with three fundamentals $Q$, one anti-fundamental $\tilde Q$, one antisymmetric $A$ and one conjugate antisymmetric $\tilde A$.
This theory is dual to a LG  where the chiral fields $\phi_I$ correspond to the gauge invariant combinations
\begin{equation}
    \label{compositesasas}
    \begin{aligned}
        \varphi_{1,k} &= Q(A \tilde A)^k \tilde Q, \\
        \varphi_{4} &= \Pf \tilde A,
    \end{aligned}
    \qquad
    \begin{aligned}
        \varphi_{2,m} &=  \tilde A (A \tilde A)^m  Q^2, \\
        \varphi_{5,\ell} &= (A \tilde A)^\ell,
    \end{aligned}
    \qquad
    \begin{aligned}
        \varphi_{3}&=\Pf A, \\
        \varphi_6 &= A^{n-1} Q^2,
    \end{aligned}
\end{equation}
with $k=0,\dots,n-1$, $m=0,\dots,n-2 $ and $\ell=1,\dots,n-1$
and a set of Fermi multiplets interacting with the chirals through
a superpotential.

Such superpotential in this case is a complicated function of the 
chiral fields, where the number of terms increases with the
rank of the gauge group.
However, we claim that by flipping some of the operators in the gauge theory, through the superpotential 
\begin{equation}
\label{pot1case2AS31}
W =  \psi_A \Pf A + \psi_{\tilde A} \Pf \tilde A +\sum_{\ell=1}^{n-1} \tilde \psi_\ell \Tr  \big(A \tilde A\big)^\ell+\psi_M Q \tilde Q,
\end{equation}
the dual superpotential becomes
\begin{equation}
    \label{eq:initialSS}
    W = \hat{\Psi}\varphi_6 \,\varphi_{1,n-1}+\sum_{j_1,j_2,j_3=1}^{n-1} \Psi_{j_1-1}\, \varphi_{1,j_2}\, \varphi_{2,j_3-1}
    \delta_{j_1+j_2+j_3,2n-1}\,.
\end{equation}
The duality can be proven in presence of the flippers in (\ref{pot1case2AS31}) by trading the conjugate antisymmetric $\tilde A$ with an auxiliary $\USp(2n-2)$ gauge group and a bifundamental $\tilde{P}$ as in the second quiver in Figure \ref{31AAtev}.
The superpotential of this gauge theory is 
\begin{equation}
\label{pot2case2AS31}
W = \psi_A \Pf A+\sum_{\ell=1}^{n-1} \tilde \psi_\ell \Tr \big(A \tilde P^2\big)^\ell+\psi_M Q \tilde Q\,.
\end{equation}

Then we observe that the $\SU(2n)$ gauge group has one antisymmetric, three fundamentals and $2n-1$ anti-fundamentals, split into $2n-2$ fields denoted as $\tilde P$ and one anti-fundamental $\tilde Q$.
It follows that we can use the results of subsection \ref{Case2even} upon taking into account the
$\SU(2n-1)$ symmetry breaking pattern imposed by the partial $ \USp(2n-2)$ gauging.
In this case the $\SU(2n)$ singlets of the duality, defined in formula (\ref{ginv2ev}), become 
\begin{equation}
    \begin{aligned}
        \Phi_1 &= Q \tilde Q \rightarrow \{\Phi_{1,\square}= Q \tilde P, \Phi_{1,\cdot} = Q \tilde Q \}, \\
        \Phi_3 &= A^{n-1} Q^2,
    \end{aligned}
    \qquad
    \begin{aligned}
        \Phi_2 &= \Pf A, \\
        \Phi_4 &= A \tilde Q^2 \rightarrow \{\Phi_{4,\square}= A \tilde Q \tilde P,\Phi_{4,AS} =A\tilde P^2 \}.
    \end{aligned}
\end{equation}
We are left with the theory described by the third quiver in Figure \ref{31AAtev} where the superpotential, obtained after integrating out the massive fields, is
\begin{equation}
\label{pot3case2AS31}
W = \Psi_{\SU(2n)} (\Phi_{1,\square} \,\Phi_3 \,\Phi_{4,\square}\, \Phi_{4,AS}^{n-2})+\sum_{\ell=2}^{n-1} \hat{\psi}_\ell \Tr \Phi_{4,AS}^\ell\,.
\end{equation}

We are then left with an  $\USp(2n-2)$ gauge theory with a totally antisymmetric $\Phi_{4,AS}$, and four fundamentals, 
where three 
of them are denoted as $\Phi_{4,\square}$ and the last one is denoted as $\Phi_{1,\square}$. This theory is dual to a LG model, where the fields are the mesonic combinations
\begin{equation}
\mathcal{L}_j = \Phi_{1,\square} \Phi_{4,\square} \Phi_{4,AS}^{j-1},\quad
\mathcal{M}_j = \Phi_{1,\square} \Phi_{1,\square} \Phi_{4,AS}^{j-1},\quad j=1,\dots,n-1\,,
\end{equation}
where the first combination is in the fundamental representation of the leftover $\SU(3)$ flavor symmetry and the second is in anti-fundamental of $\SU(3)$.
The superpotential of the LG model is
\begin{equation}
\label{pot4case2AS31}
W = \Psi_{\SU(2n)} \Phi_3 \mathcal{L}_{n-1}+\sum_{j_1,j_2,j_3} \Psi_{\USp(2n-2)}^{(j_1-1)} \mathcal{L}_{j_2} \mathcal{M}_{j_3}
\delta_{j_1+j_2+j_3,2n-1}\,,
\end{equation}
where  $\Psi_{\USp(2n-2)}^{(j)}$  are Fermi fields.
The dictionary between the composites in (\ref{pot4case2AS31}) and the  ones in (\ref{compositesasas}) that are not flipped by the superpotential (\ref{pot1case2AS31})
\begin{equation}
\varphi_{1,k} \leftrightarrow \mathcal{L}_k, \quad
\varphi_{2,m} \leftrightarrow \mathcal{M}_{m+1},\quad
\varphi_6 \leftrightarrow \Phi_3,\quad\hat{\Psi}\leftrightarrow\Psi_{\SU(2n)},\quad \Psi_{j}\leftrightarrow \Psi^{(j)}_{\USp(2n-2)},
\end{equation}
where $k=1,\ldots, n-1$, $j,m=0,\ldots, n-2$. By plugging in \eqref{pot4case2AS31} the dictionary above, we get the superpotential \eqref{eq:initialSS}.

We can derive the duality from 4d, by considering the same electric theory as in section \ref{Case4even} with the $R$ charge assignment that sets to one the fields $\widetilde{\mathcal{Q}}_2, \, \widetilde{\mathcal{Q}}_3$ and to zero the others.
The superpotential in this case reads
\begin{equation}
    W = B_2 \tilde B_2 M_n + \sum_{i,k,j=1}^{n} (\tilde h_i M_j h_k) \delta_{i+j+k,2n+1}|_{h_n=\tilde h_1 = 0}.
\end{equation}
The 2d superpotential \eqref{pot4case2AS31} is immediately recovered from the 4d reduction upon employing the dictionary
\begin{equation}
    \varphi_{1,j+1} \leftrightarrow M_{j+2}, \quad \varphi_{2,j}\leftrightarrow h_{j+1}, \quad \Psi^{(j)}_{\mathrm{USp}(2n)}\leftrightarrow \tilde{h}_j, \quad\varphi_6 \leftrightarrow B_2, \quad \Psi_{\mathrm{SU}(2n)} \leftrightarrow \tilde{B}_2,
\end{equation}
with $j=0,\dots,n-2$.\\
We proceed by checking the anomaly matching of the global symmetries. 
The charges of the fields in the electric and in the dual LG theory, including the flippers in (\ref{pot1case2AS31}),
are
\begin{equation}
\begin{array}{c|cccccc}
&\UU(1)_Q &\SU(3) &\UU(1)_{\tilde Q} &\UU(1)_A &\UU(1)_{\tilde A}&\UU(1)_R \\
\hline
Q &1&\square&0&0&0&0\\
\tilde Q&0&\cdot&1&0&0&0\\
A&0&\cdot&0&1&0&0\\
\tilde A&0&\cdot&0&0&1&0\\
\psi_{\tilde A} &0&\cdot&0&0&-n&1\\
\psi_A&0&\cdot&0&-n&0&1\\
\tilde \psi_\ell&0&\cdot&0&\ell&\ell&1\\
\psi_M&-1&\overline \square&-1&0&0&1\\
\hline
 \varphi_{1,k} &1&\square&1&k&k&0\\
 \varphi_{2,m}&2&\overline \square&0&m&m+1&0\\
  \varphi_{6}&2&\overline \square&0&n-1&0&0\\
 \hat{\Psi} &-3&\cdot&-1&2-2n&1-n&1\\
\Psi_{j}&-3&\cdot&-1&3-2n+j \quad\quad&2-2n+j&1\\
\end{array}
\end{equation}
with $\ell=1,\dots,n-1$, $k=1,\dots,n-1$, $m=0,\dots,n-2 $ and $j=0,\dots,n-2$.
The anomalies of the global symmetries are given by
\begin{equation}
\label{anmatching3AAt}
\begin{array}{lll}
&
\kappa_{QQ} =6n-3,& \quad
\kappa_{\tilde Q\tilde Q} =2n-3, \\
&
\kappa_{AA}=\kappa_{\tilde A \tilde A}=-\frac{n (n-1)(2n-7)}{6},& \quad
\kappa_{QA}=\kappa_{\tilde A \tilde Q} =\kappa_{ Q\tilde A}=\kappa_{\tilde Q A}=0, \\
&
\kappa_{\tilde Q  Q}= -3, & \quad
 \kappa_{\SU(3)^2 }=n-\frac{1}{2}, \\
&\kappa_{\tilde A A}=-\frac{n(n-1)   (2 n-1)}{6},  \\
\end{array}
\end{equation}
and we checked that they match across the dual phases.

We conclude by showing that the identity between the elliptic genera of the gauge theory and of the LG dual descends from the basic identities for $\SU(n)$ and $\USp(2n)$ gauge theories with (anti-)fundamental matter.
The identity that we need to prove in this case is
\begin{eqnarray}
\label{ToProveeven2A3}
&&
\theta(q/w^{2n})\theta(q/t^{2n})\prod_{i=1}^{n-1} \theta(q/(wt)^{2i} ) \theta(q/(u_a v x y ))
I_{\SU(2n)}^{(3;1;\cdot;1;1)} (x \vec u;y;\cdot;t^2;w^2)
 \nonumber \\
=&&
\frac{\theta(q/(x^3  y w^{2n-2}t^{4n-4})) \prod_{\ell=0}^{n-2} \theta(q/(x^3 t^{4n-6-2\ell} w^{4n-4-2\ell})) }{
\prod_{a=1}^{3} (\theta(t^{2n-2} u_a^{-1} x^2) \cdot \prod_{\ell=0}^{n-2} \theta (u_a^{-1} x^2 w^{2\ell+2} t^{2\ell} )
\prod_{\ell=1}^{n-1} \theta (u_a x y   (t w)^{2\ell}))
},
\end{eqnarray}
with $u_1 u_2 u_3=1$ and where the $\theta$ functions in the LHS of (\ref{ToProveeven2A3})
refer to the Fermi flippers in the superpotential (\ref{pot1case2AS31}).

We proceed by deconfining the conjugated antisymmetric $\tilde A$ in the integrand on the LHS of   (\ref{ToProveeven2A3}) by using the substitution 
\begin{eqnarray}
\frac{\theta(q/w^{2n})}{\prod_{1\leq i<j \leq 2n} \theta (z_i^{-1} z_j^{-1} w^2)} \rightarrow
I_{\USp(2n-2)}^{(2n;\cdot;\cdot)}(w/\vec{z};\cdot;\cdot).
\end{eqnarray}
Then we proceed by applying the identity (\ref{TOPROVE3}) to the integral associated to the $\SU(2n)$ gauge group.
We are left with the index of the $\USp(2n-2)$ gauge theory that becomes
\begin{eqnarray}
\frac{\prod_{i=2}^{n-1} \theta(q/(wt)^{2i} )}
{\prod_{a=1}^3\theta(t^{2n-2} u_a^{-1} x^2)}I_{\USp(2n-2)}^{(3,1;\cdot;1)}( x w \vec u,yw t^2;\cdot;w^2t^2).
\end{eqnarray}
The last step consists of using the identity (\ref{USp(2n)A4}) and after applying the  formula  (\ref{inversion}) we arrive at the LHS of (\ref{ToProveeven2A3}).
%
%
%
%
%
%
\subsection{$\SU(2n+1)$ with $3$ fundamentals and $1$ anti-fundamental}
\label{Case5odd}
%
%
%
%
%
%
\begin{figure}
\begin{center}
  \includegraphics[width=12cm]{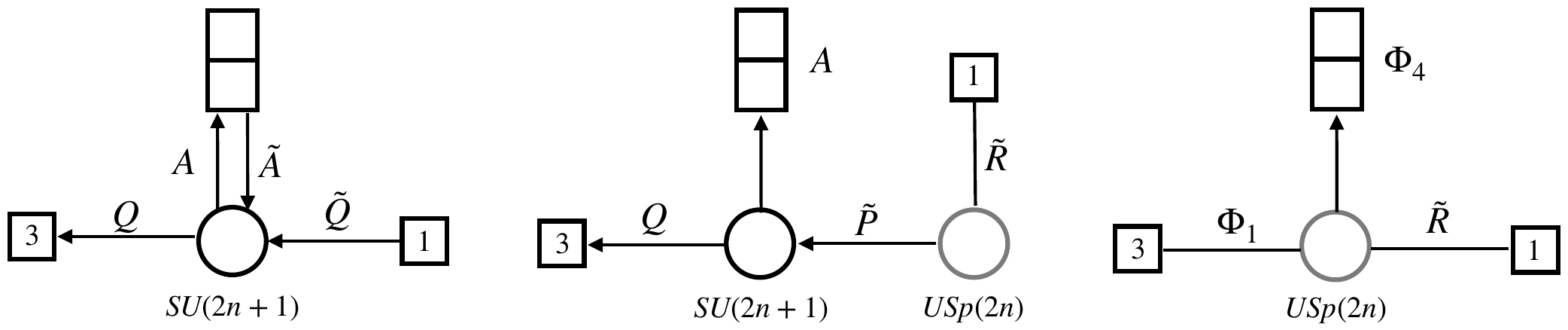}
  \end{center}
 \caption{The first quiver represents the original $\SU(2n+1)$ gauge theory with three fundamentals $Q$, one anti-fundamental $\tilde Q$  
  and one antisymmetric flavor, $(A,\tilde A)$. Then we trade the conjugate antisymmetric $\tilde A$ and the anti-fundamental $\tilde Q$ with an 
  $\USp(2n)$ gauge group with a bifundamental $\tilde P$ and a fundamental $\tilde R$.
  Then we dualize the $\SU(2n+1)$ group using the duality derived in subsection \ref{Case2odd}, where the $\SU(2n)$ global symmetry is  gauged, obtaining the third $\USp(2n)$ quiver, where we did not represent the various singlets that can be read in the analysis in the text.}
    \label{31AAtodd}
\end{figure}

In this case there LG is given by the chiral fields $\varphi_I$ corresponding to the gauge invariant combinations
\begin{equation}
    \begin{aligned}
        \varphi_{1,k} &= Q(A \tilde A)^k \tilde Q, \\
        \varphi_{4} &= \tilde A^n \tilde Q,
    \end{aligned}
    \qquad
    \begin{aligned}
        \varphi_{2,m} &= \tilde A (A \tilde A)^m  Q^2, \\
        \varphi_{5} &= A^{n-1} Q^3,
    \end{aligned}
    \qquad
    \begin{aligned}
        \varphi_{3} &= A^n Q, \\
        \varphi_{6,\ell} &= (A \tilde A)^\ell\,,
    \end{aligned}
\end{equation}

with $k=0,\dots,n-1$, $m=0,\dots,n-1$ and $\ell=1,\dots,n$.
and a set of Fermi multiplets interacting with the chirals through
a superpotential.

Such superpotential in this case is a complicated function of the 
chiral fields, where the number of terms increases with the
rank of the gauge group.
However we claim that by flipping some of the operators in the gauge theory, through the superpotential 
\begin{equation}
\label{pot1case2AS31odd}
W = \sum_{\ell=1}^{n} \tilde\psi_\ell \Tr \big(A \tilde A\big)^\ell+\psi_{\tilde A} \tilde  A^{n} \tilde Q+ \psi_A A^{n-1} Q^3\,,
\end{equation}
the dual superpotential becomes cubic in the remaining  chiral bosons $\varphi_{1,k}$ and $\varphi_{2,m}$ and $\varphi_{3}$
\begin{equation}
    \label{eq:s_init}
    W=\hat{\Psi}\varphi_3\,\varphi_{2,n-1}+\sum_{i,j,k=0}^{n-1}\Psi_{i}\,\varphi_{1,j}\,\varphi_{2,k}\,\delta_{i+j+k,2n-2}\,.
\end{equation}

The duality can be proven in presence of the flippers in (\ref{pot1case2AS31odd}) by trading the conjugate antisymmetric $\tilde A$ using an auxiliary $\USp(2n)$ gauge group as in the second quiver in Figure \ref{31AAtodd}.
The superpotential of this gauge theory is 
\begin{equation}
\label{pot2case2AS31odd}
W = \sum_{\ell=1}^{n} \tilde \psi_\ell \Tr \big(\tilde P^2 A\big)^\ell+ \psi_A A^{n-1} Q^3\,.
\end{equation}

Then we observe that the $\SU(2n+1)$ gauge group has one antisymmetric, three fundamentals and $2n$ anti-fundamentals $\tilde P$, where we can use the results in subsection \ref{Case2odd}.
In this case the $\SU(2n+1)$ singlets of the duality, defined in formula (\ref{ginv3odd}), become 
\begin{equation}
\Phi_1 = \tilde P Q,\quad
\Phi_2 = A^n Q,\quad
\Phi_3 = A^{n-1} Q^3,\quad
\Phi_4 = A \tilde P^2\,.
\end{equation}
We are left with the theory described by the third quiver in Figure \ref{31AAtodd} where the superpotential, obtained after integrating out the massive fields, is
\begin{equation}
\label{pot3case2AS31odd}
W = \Psi_{\SU(2n+1)} \Phi_4^{n-1} \Phi_1^2\Phi_2+\sum_{\ell=2}^{n} \tilde \psi_\ell \Tr \Phi_4^\ell\,.
\end{equation}
We are then left with an  $\USp(2n)$ gauge theory with a totally antisymmetric $\Phi_{4}$, and four fundamentals, 
where three 
of them are denoted as $\Phi_1 $ and the last one is denoted as $\tilde R$. This theory is dual to a LG model, where the fields are the mesonic combinations
\begin{equation}
\mathcal{M}_j = \Phi_1 \Phi_4^{j-1} \tilde R,\quad
\mathcal{L}_\ell = \Phi_1 \Phi_4^{j-1} \Phi_1,\quad j=1,\dots,n\,,
\end{equation}
where the first combination is in the fundamental representation of the leftover $\SU(3)$ flavor symmetry and the second is in anti-fundamental of $\SU(3)$.
The superpotential of the LG model is
\begin{equation}
\label{pot4case2AS31odd}
W_{\text{fin}} = \Psi_{\SU(2n+1)} \mathcal{L}_{n} \Phi_2+ \sum_{j_1,j_2,j_3} \Psi_{\USp(2n)}^{(j_1)}\mathcal{M}_{j_2}\mathcal{L}_{j_3} \delta_{j_1+j_2+j_3,2n+1}\,,
\end{equation}
where $\Psi_{\USp(2n)}^{(j_1)}$ are Fermi fields. The dictionary between the field in \eqref{pot4case2AS31odd} and the ones in \eqref{eq:s_init} can be red through the sequence of dualities discussed above and it is explicitly given by 
\begin{equation}
    \mathcal{M}_i\leftrightarrow\varphi_{1,i-1},\quad \mathcal{L}_i \leftrightarrow \varphi_{2,i-1},\quad \Phi_2\leftrightarrow \varphi_3,\quad \Psi_{\SU(2n+1)}\leftrightarrow\hat{\Psi},\quad \Psi_{\USp(2n)}^{(j)}\leftrightarrow\Psi^{(j-1)}.
\end{equation}

We can derive the duality from 4d starting from the model discussed in subsection \ref{Case4odd}. Starting from the superpotential 
\begin{eqnarray}
    \label{eq:sbenveno}
    W = \sum_{a=1}^{3} (s_a \mathcal{A}^{n-1}  \mathcal{Q}^3
    +  \tilde s_a \tilde{\mathcal{A}}^{n} \widetilde{\mathcal{Q}}_a) +\sum_{\ell=1}^{n} \beta_\ell \,\mathrm{Tr}\, (\mathcal{A} \mathcal{\tilde A})^\ell,
\end{eqnarray}
we can reproduce the second term in \eqref{pot4case2AS31odd} by assigning $R_{\tilde{Q}_{2,3}}=1$, while setting the $R$ charges of the other charged matter fields to zero. In this way, indeed, the 4d chirals that survive the twist in formula \eqref{eq:mapbenve} are $\Sigma_{1}^{(k)}$ and $\Sigma_3^{(k)}$. Such fields give rise to the 2d chirals $\varphi_{1,k}$ and $\varphi_{2,k}$ respectively. On the other hand, one component of the field $\Sigma_2^{(k+1)}$ has $R$ charge equal to two, which survives as the $\Psi_{k}$ Fermi.\\
In order to reproduce the first term in \eqref{pot4case2AS31odd}, we have to consider the electric superpotential \eqref{eq:sbenveno}. Observe that the structure for the flippers differs from the one in \cite{Bajeot:2022kwt}. Indeed, here we are flipping the operator $\tilde{\mathcal{A}}^n\widetilde{\mathcal{Q}}_a$ while in \cite{Bajeot:2022kwt} the authors flip the operator $\tilde{\mathcal{A}}^{n-1}\widetilde{\mathcal{Q}}^3$. In our case this gives rise to an extra term in the dual superpotential corresponding to 
\begin{equation}
    \Delta W = \tilde{B}_3B_1\Sigma_3^{(n-1)},
\end{equation}
where $\tilde{B}_3 = \tilde{\mathcal{A}}^{n-1}\widetilde{\mathcal{Q}}^3$ and $B_1=\mathcal{A}^{n} \mathcal{Q}$. The assignment of $R$ charges considered above imply that the 4d field $B_1$ becomes the 2d chiral $\varphi_3$, and the 4d field $\tilde{B}_3$ becomes the 2d Fermi $\hat{\Psi}$.

We proceed by checking the anomaly matching of the global symmetries. 
The charges of the fields in the electric and in the dual LG theory, including the flippers in (\ref{pot1case2AS31odd}),
are
\begin{equation}
\begin{array}{c|cccccc}
&\UU(1)_Q &\SU(3) &\UU(1)_{\tilde Q} &\UU(1)_A &\UU(1)_{\tilde A}&\UU(1)_R \\
\hline
Q &1&\square&0&0&0&0\\
\tilde Q&0&\cdot&1&0&0&0\\
A&0&\cdot&0&1&0&0\\
\tilde A&0&\cdot&0&0&1&0\\
\tilde \psi_\ell &0&\cdot&0&\ell&\ell&1\\
\psi_A&0&\cdot&-1&0&-n&1\\
 \psi_{\tilde A}&-3&\cdot&0&1-n&0&1\\
\hline
\varphi_{1,k}&1&\square&1&k&k&0\\
\varphi_{2,m}&2&\overline \square&0&m&m+1&0\\
\varphi_{3}&1&\square&0&n&0&0\\
\hat{\Psi} &-3&\cdot&0&1-2n&-n&1\\
\Psi^{(j)}&-3&\cdot&-1&2-2n+j&1-2n+j&1
\end{array}
\end{equation}
with $\ell=1,\dots,n$, $k=0,\dots,n-1$, $m=0,\dots,n-1$ and $j=0,\dots,n-1$.

The anomalies of the global symmetries are given by
\begin{equation}
\begin{array}{lll}
&
\kappa_{QQ}=6(n-1),& \quad
\kappa_{\tilde Q \tilde Q} = 2n, \\
&
\kappa_{AA} =\frac{1}{6} \left(-2 n^3+3 n^2+17 n-6\right),& \quad
\kappa_{AQ}=3 (1 - n),  \\
&
\kappa_{Q \tilde Q}=0 , & \quad
\kappa_{A \tilde Q}= 0  , \\
&\kappa_{\tilde A \tilde A}=\frac{1}{6} n \left(-2 n^2+3 n+5\right),& \quad
\kappa_{\tilde A  A}=-\frac{1}{6} n (n+1) (2 n+1),  \\
&\kappa_{\tilde A  Q}=0,& \quad
\kappa_{\tilde A \tilde Q}=-n , \\
&
\kappa_{\SU(3)^2} = n+\frac{1}{2}, & \quad
\end{array}
\end{equation}
and we checked that they match across the dual phases.

We conclude by showing that the identity between the elliptic genera of the gauge theory and of the LG dual descends from the basic identities for $\SU(n)$ and $\USp(2n)$ gauge theories with (anti-)fundamental matter.
The identity that we need to prove in this case is
\begin{eqnarray}
\label{ToProveodd2A3}
&&
 \theta(q/(t^{2n-2}x^3))
 \theta(q/(w^{2n}y))
\prod_{\ell=1}^n \theta(q/(t w )^{2\ell})
I_{\SU(2n+1)}^{(3;1;\cdot;1;1)} (x \vec u;y;\cdot;t^2;w^2)
 \nonumber \\
 =&&
 \frac{
 \theta(q/( t^{4n-2}x^{3}w^{2n}))
 \prod_{\ell=1}^{n}\theta(q/(w^2 x^3 y (tw)^{2(2n-1-\ell)})
 }
 {
 \prod_{\ell=0}^{n-1} \theta( u_a x y  (t w )^{2 \ell})
 \prod_{\ell=0}^{n-1} \theta(w^{2} u_a u_b x^2 (t w )^{2 \ell})
 \prod_{a=1}^3 \theta(t^{2n} u_a x)
 }\,,
\end{eqnarray}
with $u_1 u_2 u_3=1$ and where the $\theta$ functions in the LHS of (\ref{ToProveodd2A3})
refer to the Fermi flippers in the superpotential (\ref{pot1case2AS31odd}).

We proceed by deconfining the conjugated antisymmetric $\tilde A$ in the integrand on the LHS of   (\ref{ToProveodd2A3}) by using the substitution 
\begin{equation}
\frac{\theta(q/(yw^{2n}))}{ \prod_{i=1}^{2n+1} \theta(z_i^{-1} y)\prod_{1\leq i<j \leq 2n+1} \theta (z_i^{-1} z_j^{-1} w^2)} \rightarrow
I_{\USp(2n)}^{(1,2n+1;\cdot;\cdot)}(y/w, w/\vec{z};\cdot;\cdot)\,.
\end{equation}
Then we proceed by applying the identity (\ref{TOPROVE4}) to the integral associated to the $\SU(2n+1)$ gauge group.
We are left with the index of the $\USp(2n)$ gauge theory, that becomes
\begin{eqnarray}
\frac{\theta(q t^{2-4n}x^{-3}w^{-2n}) \prod_{\ell=1}^n \theta(q/(t w )^{2\ell})}{\prod_{a=1}^3 \theta(t^{2n} u_a x)}I_{\USp(2n)}^{(3,1;\cdot;1)}( w x\vec u,y/w;\cdot;w^2t^2)\,.
\end{eqnarray}
The last step consists of using the identity (\ref{USp(2n)A4}) and after applying the  formula  (\ref{inversion}) we arrive at the LHS of (\ref{ToProveodd2A3}).
%
%
%
%
%
%
\subsection{$\SU(2n)$ with $4$ fundamentals}
\label{Case6even}
%
%
%
%
%
%
\begin{figure}
\begin{center}
  \includegraphics[width=12cm]{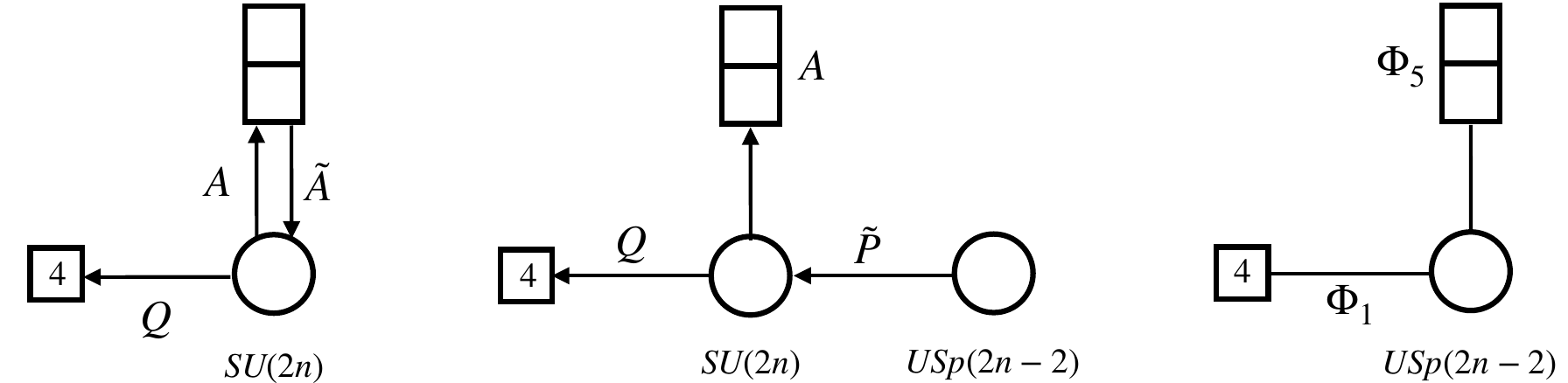}
  \end{center}
 \caption{The first quiver represents the original $\SU(2n)$ gauge theory with four fundamentals $Q$
  and one antisymmetric flavor $(A,\tilde A)$. We trade the conjugate antisymmetric $\tilde A$ with an $\USp(2n-2)$ gauge group with a bifundamental $\tilde P$.
  Then we dualize the $\SU(2n)$ group using the duality derived in subsection \ref{Case3even}, where the $\SU(2n-2)$ global symmetry is  gauged as $\USp(2n-2)$, obtaining the third $\USp(2n-2)$ quiver. We did not represent in this quiver gauge  the singlets, they can be found in the discussion in the main text.}
    \label{4AAteven}
\end{figure}
We conclude our analysis with a model corresponding to a $\SU(2n)$ with four fundamentals and an antisymmetric flavor. In this case the gauge invariant combinations that describe the LG theory are
\begin{equation}
    \begin{aligned}
        \varphi_{1,m} &=  \tilde A (A \tilde A)^m Q^2, \\
        \varphi_{4} &= A^{n-2}Q^4, 
    \end{aligned}
    \qquad
    \begin{aligned}
        \varphi_{2} &=\mathrm{Pf} A, \\
        \varphi_{5}  &=\mathrm{Pf} \tilde A,
    \end{aligned}
    \qquad
    \begin{aligned}
        \varphi_{3} &= A^{n-1}Q^2, \\
        \varphi_{6,\ell}&= (A \tilde A)^\ell,
    \end{aligned}
\end{equation}
with $m=0,\dots,n-2$ and $\ell=1,\dots,n-1$.
and a set of Fermi multiplets interacting with the chirals through
a superpotential.

Such superpotential in this case is a complicated function of the 
chiral fields, where the number of terms increases with the
rank of the gauge group.
However, we claim that by flipping some of the operators in the electric theory, through the superpotential 
\begin{equation}
\label{Wflipped401}
W = \psi_A \Pf A+\psi_{\tilde A} \Pf \tilde A+\sum_{\ell=1}^{n-1} \tilde \psi_\ell \Tr \big(A \tilde A \big)^\ell+\psi_0 A^{n-2} Q^4\,,
\end{equation}
the dual superpotential becomes cubic in the remaining chiral bosons $\varphi_{1,m}$ and $\varphi_{3}$
\begin{equation}
    \label{eq:polpot}
    W = \hat{\Psi}\,\varphi_{1,n-3}\,\varphi_{3} + \check{\Psi}\,\varphi_{3}^2 + \sum_{j_1,j_2,j_3=0}^{n-2}\Psi_{j_1}\,\varphi_{1,j_2}\,\varphi_{1,j_3}\,\delta_{j_1+j_2+j_3,2n-5}.
\end{equation}

The duality can be proven in presence of the flippers in (\ref{Wflipped401}) by trading the conjugate antisymmetric $\tilde A$ using an auxiliary $\USp(2n-2)$ gauge group as in the second quiver in Figure \ref{4AAteven}.
The superpotential of this gauge theory is 
\begin{equation}
\label{W402}
W = \psi_A \Pf A+\sum_{\ell=1}^{n-1} \tilde \psi_\ell \Tr \big(A \tilde P^2\big)^\ell+\psi_0 A^{n-2} Q^4\,.
\end{equation}

Then we observe that the $\SU(2n)$ gauge group has one antisymmetric, four fundamentals and $2n-2$ anti-fundamentals $\tilde P$, where we can use the results of subsection \ref{Case3even}.
In this case the $\SU(2n)$ singlets of the duality, defined in formula (\ref{fieldssu2n4}), become 
\begin{equation}
\Phi_1=Q \tilde P,\quad
\Phi_2=\Pf A,\quad
 \Phi_3=A^{n-1} Q^2,\quad
\Phi_4= A^{n-2} Q^4,\quad
\Phi_5= A \tilde P^2,\quad
\end{equation}
Using the duality of subsection \ref{Case3even}  we are left with the theory described by the third quiver in Figure \ref{4AAteven} where the superpotential, obtained after integrating out the massive fields, is
\begin{equation}
\label{W402}
W = \hat{\Psi} \Phi_5^{n-2}\Phi_1^2\Phi_3+\check{\Psi} \Phi_3^2+\sum_{\ell=2}^{n-1} \tilde \psi_\ell \Tr \Phi_5^\ell\,.
\end{equation}
 We are then left with an  $\USp(2n-2)$ gauge theory with a totally antisymmetric $\Phi_{5}$, and four fundamentals, 
$\Phi_1 $.
This theory is dual to a LG model, where the fields are the mesonic combinations
$\mathcal{M}_{a,b}^{(j)} \equiv \Phi_{1,a}  \Phi_{1,b} \Phi_{5}^{j-1}$ with $1\leq a <b\leq 4$ and $j=1,\dots,n-1$.
The superpotential of the dual LG model is
\begin{equation}
\label{W403}
W = \hat{\Psi} \mathcal{M}^{(n-2)} \Phi_3+\check{\Psi} \Phi_3^2+
\sum_{j_i}
\Psi_{\USp(2n)}^{(j_1)}\epsilon_{abcd} \mathcal{M}_{ab}^{(j_2)}  \mathcal{M}_{cd}^{(j_3)} \delta_{j_1+j_2+j_3,2n-2} \,,
\end{equation}
where $\Psi_{\USp(2n)}^{(j_1)}$ are Fermi fields.\\
The dictionary between the fields in \eqref{eq:polpot} and \eqref{W403} is
\begin{equation}
    \mathcal{M}^{(j)} \leftrightarrow \varphi_{1,j-1}, \qquad \Phi_3 \leftrightarrow \varphi_3, \qquad \Psi_{\mathrm{USp}(2n)}^{(j)} \leftrightarrow \Psi_{j-1}.
\end{equation}
We proceed by checking the anomaly matching of the global symmetries. The charges of the field of the electric and in the dual LG theory, including the flippers in (\ref{Wflipped401}), are
\begin{equation}
\begin{array}{c|cccccc}
&\UU(1)_Q &\SU(4) &\UU(1)_A &\UU(1)_{\tilde A}& \UU(1)_R\\
\hline
Q &1&\square&0&0&0\\
A&0&\cdot&1&0&0\\
\tilde A&0&\cdot&0&1&0\\
\psi_A &0&\cdot&-n&0&1\\
\psi_{\tilde A}&0&\cdot&0&-n&1\\
\tilde \psi_\ell&0&\cdot&-\ell&-\ell&1\\
\psi_0&-4&\cdot&2-n&0&1\\
\hline
\varphi_{1,k}&2&\begin{array}{c} \square\vspace{-3mm} \\ \square\end{array}&k&k+1&0\\
\varphi_3&2&\begin{array}{c} \square\vspace{-3mm} \\ \square\end{array}&n-1&0&0\\
\hat{\Psi} &-4&\cdot&2-2n&0&1\\
\check{\Psi}&-4&\cdot&3-2n&1-n&1\\
\Psi_{k} &-4&\cdot&4-2n+k\,\,&2-2n+k&1\\
\end{array}
\end{equation}
where $k=0,\dots,n-2$ and $\ell=1,\dots,n-1$.

The anomalies of the global symmetries are given by
\begin{equation}
\begin{array}{lll}
&
\kappa_{QQ}=8 (n-2) ,& \quad
\kappa_{\tilde A \tilde A} =\frac{n( 9n-2 n^2-7)}{6} ,   \\
&
\kappa_{ A  A}=\frac{3 n^2-2 n^3+17 n-24}{6} ,& \quad
\kappa_{ A Q}= 4(2-n), \\
&
\kappa_{\tilde A Q} =0 , & \quad
\kappa_{\tilde A  A}=  \frac{n( n-1) (1 - 2 n)}{6} ,  \\
&
\kappa_{\SU(4)^2} =n , &
\end{array}
\end{equation}
and we checked that they match across the dual phases.

We conclude by showing that the identity between the elliptic genera of the gauge theory and of the LG dual descends from the basic identities for $\SU(n)$ and $\USp(2n)$ gauge theories with (anti-)fundamental matter.
The identity that we need to prove in this case is
\begin{eqnarray}
\label{ToProveeven2A4}
&&
\theta(q/t^{2n})\theta(q/w^{2n})\theta(q/(t^{2(n-2)}x^4)\prod_{\ell=1}^{n-1} \theta(q/(t w)^{2 \ell})
I_{\SU(2n)}^{(4;\cdot;\cdot;1;1)} (x \vec u;\cdot;\cdot;t^2;w^2)
 \nonumber \\
=&&
\frac{\theta(q/(x^4 w^{2-2n}t^{6-4n}))
\theta(q/(x^4 t^{4-4n}))
\prod_{k=0}^{n-2}\theta(q/( x^4
w^{2(2n-2-k)}
t^{2(2n-4-k)}
))
}{\prod_{a<b}\theta(t^{2(n-1)} u_a u_b x^2) \cdot \prod_{\ell=0}^{n-2} \theta(u_a u_b x^2 t^{2 \ell}w^{2 \ell+2})}\,,
\end{eqnarray}
with $\prod_{a=1}^{4} u_i=1$ and where the $\theta$ functions in the LHS of (\ref{ToProveeven2A4})
refer to the Fermi flippers in the superpotential (\ref{Wflipped401}).

We proceed by deconfining the conjugated antisymmetric $\tilde A$ in the integrand on the LHS of   (\ref{ToProveeven2A4}) by using the substitution 
\begin{eqnarray}
\frac{\theta(q/w^{2n})}{\prod_{1\leq i<j \leq 2n} \theta (z_i^{-1} z_j^{-1} w^2)} \rightarrow
I_{\USp(2n-2)}^{(2n;\cdot;\cdot)}(w /\vec{z};\cdot;\cdot)\,.
\end{eqnarray}
Then we proceed by applying the identity (\ref{ToProve5})  to the integral associated to the $\SU(2n)$ gauge group.
We are left with the index of the $\USp(2n-2)$ gauge theory that becomes
\begin{eqnarray}
\frac{\theta(q/x^4 w^{2-2n}t^{6-4n})
\theta(q/x^4 t^{4-4n}) \prod_{\ell=1}^{n-1} \theta(q/(t w)^{2 \ell})}
{\prod_{a<b}\theta(t^{2(n-1)} u_a u_b x^2)}I_{\USp(2n-2)}^{(4;\cdot;1)}( x w \vec u;\cdot;w^2t^2)\,.
\end{eqnarray}
The last step consists of using the identity (\ref{USp(2n)A4}) and after applying the  formula  (\ref{inversion}) we arrive at the LHS of (\ref{ToProveeven2A4}).

\section{Beyond $\SU(n)$: $\USp(4)$ with two antisymmetrics and two $\square$}
\label{sec:usp4}

\begin{figure}
\begin{center}
\includegraphics[width=16cm]{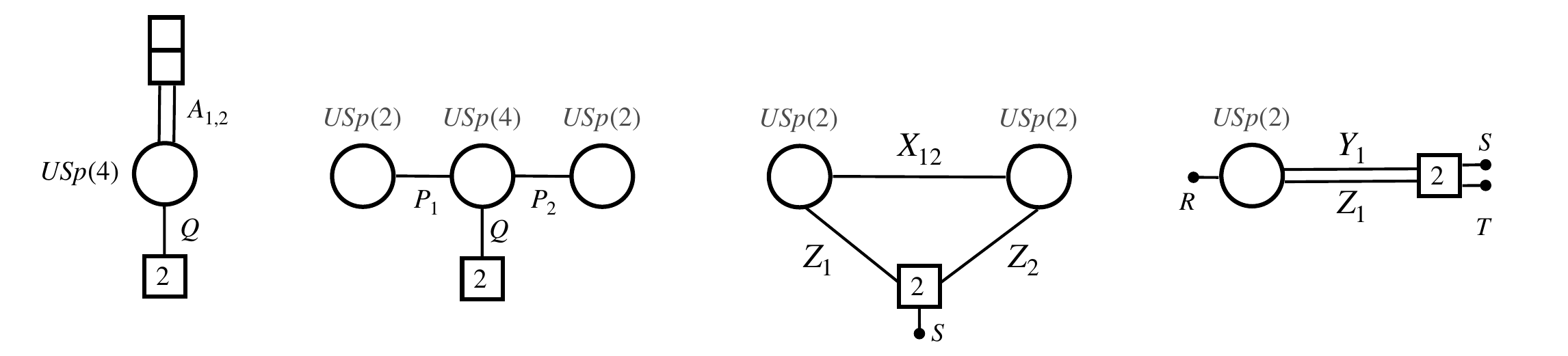}
  \end{center}
  \caption{The first quiver represents the $\USp(4)$ gauge theory with two fundamentals and two antisymmetrics. The second quiver is obtained by trading the two antisymmetrics with two $\USp(2)$ gauge groups. Then the third quiver is obtained by dualizing the original $\USp(4)$ gauge group and the last quiver is found after dualizing one of the two $\USp(2)$ groups.}
    \label{usp4case}
\end{figure}

We conclude our list of examples of new gauge/LG dualities in 2d $\mathcal{N}=(0,2)$ by studying another
case that cannot be derived from the 4d classification of \cite{Csaki:1996zb}.
The gauge theory corresponds to $\USp(4)$ with two fundamentals $Q_{1,2}$ and two antisymmetric tensors $A_{1,2}$.
In the following we will give evidences that this model is dual to a LG theory.

The charges of the fields under the flavor symmetries are 
\begin{equation}
\begin{array}{c|ccccc}
 &\SU(2)_{A} & \SU(2)_Q & \UU(1)_A & \UU(1)_Q & \UU(1)_R \\
 \hline
A & \square & \cdot & 1 & 0&0 \\
Q&\cdot & \square & 0& 1 & 0 \\
\end{array}
\end{equation}

We actually  consider on the gauge theory side the flipped  superpotential 
\begin{equation}
\label{WUSp4}
W = \sum_{a=1}^{2} (\psi_{A_a} \Pf A_a + \psi_\beta^{(a)} \Tr A_a)\,,
\end{equation}
where the two Fermi $\psi_{A_{1,2}}$ are charged under the $\SU(2)$ symmetry that rotates the two antisymmetrics.
We study the model by trading the two antisymmetric tensors $A_{1,2}$ with two $\USp(2)_{1,2}$ gauge groups.
The model corresponds to the second quiver in Figure \ref{usp4case} and it has superpotential
\begin{equation}
W = \psi_\beta^{(1)} P_1^2+ \psi_\beta^{(2)}  P_2^2\,.
\end{equation}

Next we dualize the $\USp(4)$ gauge theory with six fundamental into an LG model. We are left with the $\USp(2)_1\times \USp(2)_2$ theory depicted in the third quiver of Figure \ref{usp4case}  with superpotential
\begin{equation}
W = \Psi_{\USp(4)} \epsilon_{\alpha_1 \beta_1} \epsilon_{\alpha_2 \beta_2} \epsilon_{\alpha \beta}
\left(
X_{12}^{\alpha_1 \alpha_2}  Z_2^{\beta_2 \alpha} Z_1^{\beta \beta_1} +X_{12}^{\alpha_1 \alpha_2} X_{12}^{\beta_1 \beta_2 }S^{\alpha \beta}
\right).
\end{equation}

The following step consists of dualizing one of the two $\USp(2)$ gauge theories into a LG.
For example we can choose the $\USp(2)_2$ group, but the other choice is equivalent, due to the $\SU(2)_A$ symmetry rotating $A_{1}$ and $A_2$ in the original gauge theory.
Then we are left with the $\USp(2)$ theory in the last quiver in Figure \ref{usp4case}  with superpotential 
\begin{equation}
W =\epsilon_{\alpha_1 \beta_1}  \epsilon_{\alpha \beta}\left( \Psi_{\USp(4)} 
\left(
Y_1^{\alpha_1  \alpha} Z_1^{\beta \beta_1} +R^{\alpha_1 \beta_1  }S^{\alpha \beta}
\right)
+
\Psi_{\USp(2)_2} \left(Y_1^{\alpha_1  \alpha}  Y_1^{\beta_1  \beta} + R^{\alpha_1 \beta_1  }T^{\alpha \beta} \right)
\right),
\end{equation}
where $Y_1 = X_{12} Z_2 $, $R = X_{12}^2$ and $T=Z_2^2$.
The $\USp(2)_1$ gauge group has four fundamentals and it can be dualized into the final LG model.
The  singlets  that arise from this last duality are $\Phi_1 = Y_1^2$, $\Phi_2 = Z_1^2$ and $\Phi_3 = Z_1 Y_1$ and the superpotential is

\begin{equation}
W \!=\!    \epsilon_{\alpha \beta} \big(  \Psi_{\USp(4)} (\Phi_3^{\alpha \beta} \! +\! R S^{\alpha \beta} )
\!+\!
\Psi_{\USp(2)_2} (\Phi_1^{\alpha \beta} \!+ \!R T^{\alpha \beta} )
\!+\!
\Psi_{\USp(2)_1}  \epsilon_{\ell m} (\Phi_1^{\alpha \ell}  \Phi_2^{\beta m}\!+\!\Phi_3^{\alpha \ell} \Phi_3^{\beta m} )\!\big),
\end{equation}
where the component $\Phi_3^{12} -\Phi_3^{21} $ is massive. Defining $V_{\alpha \beta}$ as the massless component of $\Phi_3$ we are left with the superpotential 
\begin{equation}
W = 
\Psi_{\USp(2)_1}  (R T \Phi_2+R^2 S^2 + \det  V).
\end{equation}
Then we can read the final fields with respects of the original gauge invariant operators that are not set to zero by the flipped superpotential \eqref{WUSp4}.
We have $R = \Tr A_1 A_2$, $S = Q_1 Q_2$, $ T = Q_1 A_1 Q_2 $,
$ \Phi_2 = Q_1 A_2 Q_2 $ and $V_{\alpha \beta} = Q_\alpha A_1 A_2 Q_\beta$.
Once we have established the duality we want to test it by matching the 't Hooft anomalies and by studying the 
elliptic genus.

We start by assigning the global charges to the various fields of the model. The charges of the Fermi fields can be read from the superpotential, while the charges of the composite chirals in the dual LG theory are read from the duality map. The $\SU(2)_A$ symmetry is broken by the superpotential with the flippers in (\ref{WUSp4}) 
and only the two combinations $J_{\SU(2)_A} +  I_{\UU(1)_A}$ are leftover. These two combinations are rearranged in the two $\UU(1)_{1,2}$ symmetries in the table below.
\begin{equation}
\begin{array}{c|ccccc}
& \UU(1)_1 & \UU(1)_2 &\UU(1)_Q& \SU(2) & \UU(1)_{R_0} \\
\hline
Q & 0& 0 &1& \square & 0 \\
A_1& 1& 0 &0&\cdot & 0 \\
A_2& 0& 1 &0&\cdot &  0\\
\psi_{\beta}^{(1)}&-1 &  0&0&\cdot & 1 \\
\psi_{\beta}^{(2)}& 0&-1  &0&\cdot & 1 \\
\psi_{A_1}&-2 & 0 &0& \cdot& 1 \\
\psi_{A_2}&0 &-2 &0& \cdot& 1 \\
\hline
R&1 & 1 &0&\cdot & 0 \\
S& 0& 0 &2& \cdot&  0\\
T& 1&0  &2&\cdot &  0\\
\Phi_2&0 &1  &2& \cdot& 0 \\
V& 1& 1 &2& \square\!\square&  0\\
\Psi_{\USp(2)_1}& -2&-2  &-4& \cdot&  1
\end{array}
\end{equation}
We have computed the 't Hooft anomalies 
\begin{equation}
    \begin{aligned}
        &\kappa_{11}=\kappa_{22}=1,\\
        &\kappa_{12}=0,\\
        &\kappa_{1Q}=\kappa_{2Q}=0,\\
        &\kappa_{QQ}=8,
    \end{aligned}
    \qquad \qquad
    \begin{aligned}
        &\kappa_{1R_0} =\kappa_{2R_0}=-3,\\
        &\kappa_{QR_0} =-8,\\
        &\kappa_{R_0 R_0}=6,\\
        &\kappa_{\SU(2)^2}=2,
    \end{aligned}
\end{equation}
and showed that they match across the dual theories.
The other strong check of the duality consists of studying the elliptic genus.
In this case we start from the index of the theory with the flippers corresponding to
\begin{equation}
\label{prima}
\prod_{j=1}^{2} \theta(q/t_j^2,q/t_j^4)
I_{\USp(4)}^{(2;\cdot;2)} ( x u,x/u;\cdot; t_1^2,t_2^2).
\end{equation}
Then we substitute in the integrand the contribution of the antisymmetric tensors using two $\USp(2)$ gauge theories.
The substitution corresponds to 
\begin{equation}
\frac{\theta(q/t_j^4) }{\theta(z_1^{\pm 1} z_2^{\pm 1} t_j^2)} \rightarrow I_{\USp(2)_j}^{(4;\cdot;\cdot)} ( t_j \vec z;\cdot;\cdot)
\end{equation}
for $j=1,2$.
We iterate the application of the identity (\ref{USp(2n)A4}), first on the $\USp(4)$ integral and then on the $\USp(2)$ integrals.
Simplifying the various terms using the  formula (\ref{inversion})  we arrive to the final result, corresponding to 
\begin{equation}\label{seconda}
\frac{\theta (q/(t_1^4 t_2^4 x^4 ))}{\theta(t_1^2 t_2^2 x^2) \theta(x^2)  \theta(t_1^2 t_2^2 )
 \theta(t_1^2 t_2^2 x^2 u^{\pm 2})  \prod_{j=1}^{2} \theta(t_j^2 x^2) }\,.
\end{equation}

Observe that the final identity between (\ref{prima}) and  (\ref{seconda}) hides the $\SU(2)_A$ symmetry enhancement. Such enhancement can be explicitly shown by moving the $\theta$-functions associated to the Fermi flippers $\psi_{A_{1,2}}$ in (\ref{prima})  to the denominator on  (\ref{seconda}). Using the formula (\ref{inversion}) we can then rearrange the contributions 
$\theta(t_1^4)$, $\theta(t_2^4)$ and $\theta(t_1^2 t_2^2)$ into the adjoint of $\SU(2)_A$ reconstructing the global symmetry broken by the flippers. The other two flippers $\psi_{\beta}^{(1,2)}$ are similarly rearranged into a fundamental representation for $\SU(2)_A$.

We conclude this section by observing that, even if such duality is not derived from any known s-confining theory in 4d, there exists a similar duality  in 3d, originally worked out in \cite{Nii:2019ebv}.
The duality has been derived thereafter in \cite{Okazaki:2023hiv} by extending to
the 3d bulk a 2d boundary duality constructed from $\mathcal{N} = (0, 2)$ half-BPS boundary
conditions in 3d $\mathcal{N} = 2$.
Furthermore the duality has been shown to descend from other basic dualities  in 3d $\mathcal{N} = 2$  in  \cite{Amariti:2023wts}, using a strategy very similar to the one proposed here in 2d.

\section{Remarks on the deconfinement procedure}
\label{Andrea1}
Before moving on, we would like to comment on the relation between the deconfinement procedure at the level of field theory and the chains of identities at the level of the elliptic genus. 

The deconfinement of a tensor field in a gauge theory by means of a confining duality, thanks to which the tensor can be recast in terms of (anti-)fundamental fields of another gauge theory, translates into an identity at the level of elliptic genera and theta functions. In all the models studied in this paper, we applied a chain of identities which allowed us to start from the elliptic genus of the original model and end up with a Landau-Ginzburg model representing the evaluation of the starting elliptic genus. A crucial question arises whether the chain of identities at the level of the elliptic genus can be related to the actual sequence of intermediate gauge theories generated via deconfinement and dualities in the field-theoretic description\footnote{We are grateful to the referee for pointing out this subtlety in our procedure.}. Because the deconfinement acts locally on gauge nodes, while the elliptic genus of a gauge theory is defined in terms of a global JK cycle, the question is translated into asking whether we can provide a global extension of the JK prescription consistently with the local ones arising from the deconfinement procedure. We stress that this step is fundamental in order to guarantee that duality relations can be employed in the derivation of the Landau-Ginzburg model from the elliptic genus perspective. Being $\mathcal{G}_A$ the gauge group of the theory A and $\mathcal{G}_B$ the gauge group of the theory B, and considering flavor symmetries parameterized by the fugacities $\xi_a$ a duality is an identity of the form
\begin{equation}
    \oint\displaylimits_{\mathrm{JK}}\mathrm{d}u_i f(u_i,\xi_a)   = \oint\displaylimits_{\mathrm{JK}}\mathrm{d}\omega_j g(\omega_j,\xi_a).
    \label{eq:dualities}
\end{equation}

To make the discussion concrete let us consider a chiral\footnote{The discussion can be straightforwardly translated to the case of a Fermi multiplet.} multiplet $T$, transforming in some representation of the gauge group, which can be deconfined in terms of a fundamental confining duality. Let us denote the gauge group of the starting theory $\mathcal{G}_1$, with associated gauge holonomies $u_i$ and $\mathcal{G}_2$ the new gauge group generated by the deconfinement, with gauge holonomies $\omega_i$, so that the full gauge group is $\mathcal{G}_1 \times \mathcal{G}_2$. Let us denote the contribution to the elliptic genus due to the tensor field as $g(u_i,\xi_a)$. Then, upon deconfining $T$ we have
\begin{equation}
    \mathcal{I}(\mathbf{\xi}_a) = \oint\displaylimits_{\mathrm{JK}_u}\mathrm{d}u_i f(u_i,\xi_a) g(u_i,\xi_a) = \oint\displaylimits_{\mathrm{JK}_u}\mathrm{d}u_i f(u_i,\xi_a) \left(\,\oint\displaylimits_{\mathrm{JK}_\omega} \mathrm{d}\omega_j h(u_i,\omega_j,\xi_a)\right),
    \label{eq:deconf_generic}
\end{equation}
where $\xi_a$ parameterize the fugacities associated to the global flavor symmetries of the theory, while $h(u_i,\omega_j,\xi_a)$ denotes the contribution to the elliptic genus arising from the confining gauge theory description of the tensor multiplet $T$, for which $u_i$ are flavor-like fugacities. In this paper we will be interested to the case of \emph{fundamental} deconfinement for which the tensor multiplet is described in terms of a gauge theory where (anti-)fundamental fields only are present. 
While \eqref{eq:deconf_generic} is mathematically well-stated as it is simply a mathematical identity, physically it originates from the deconfinement of a field in the field theoretic description, which produces a new well-defined gauge theory. As such, one might be tempted to interpret \eqref{eq:deconf_generic} as the elliptic genus of the deconfined gauge theory with gauge group $\mathcal{G}_1\times \mathcal{G}_2$. However, such interpretation is valid only if the following relation holds
\begin{equation}
    \oint\displaylimits_{\mathrm{JK}_u}\mathrm{d}u_i f(u_i,\xi_a) \left(\,\oint\displaylimits_{\mathrm{JK}_\omega} \mathrm{d}\omega_j h(u_i,\omega_j,\xi_a)\right) = \oint\displaylimits_{\mathrm{JK}_{u,\omega}}\mathrm{d}u_i \mathrm{d}\omega_j f(u_i,\xi_a) h(u_i,\omega_j,\xi_a),
    \label{eq:JKextension}
\end{equation}
However, this interpretation is valid only if the following relation holds, ensuring that \eqref{eq:deconf_generic} indeed corresponds to the elliptic genus of a well-defined gauge theory and that dualities of the form \eqref{eq:dualities} can be consistently applied. Namely, we need to show that no obstruction arises in promoting the two local JK integrals to a single one in the whole space under the hypothesis of \emph{fundamental} deconfinement.
Consistency condition \eqref{eq:JKextension} holds if and only if one can extend the definition of the JK contour, so that it pick the same poles in both sides of \eqref{eq:JKextension}.

The JK prescription selects which poles contribute to the integral and it is defined in terms of a (co)vector $\eta_i \in \mathfrak{h_i}^*$ ($i=1,2$), where $\mathfrak{h_i}$ is the Cartan subalgebra of $\mathfrak{g}_i$ \cite{Benini:2013xpa}. The singularities of the integrand lie on hyperplanes of the form $Q(u) + c = 0$, where $Q \in \mathfrak{h}^*$ are charge covectors generating the hyperplanes. The set of charge covectors divide $\mathfrak{h}^*$ in chambers. Any pole for the $r=\mathrm{dim}\mathfrak{h}$ dimensional integral is obtained by intersections of at least $r$ hyperplanes at a common point and specifies a subset of charges $Q$, namely those involved in the intersection under consideration. Defined an $\eta \in \mathfrak{h}^*$ one can define a basis of $\mathfrak{h}^*$ for each pole, by picking exactly $r$ charge covector involved in the intersections of hyperplanes producing the specific pole considered\footnote{In the case of non-degenerate intersections (exactly $r$ hyperplanes meet at a point), there is no ambiguity in the choice of basis. When the intersection is degenerate, one needs to resolve the degeneracy to apply the JK prescription \cite{Benini:2013xpa}.}. The contributing poles are defined as the ones for which $\eta \in \mathrm{Cone}(Q_{i_1},...,Q_{i_r})=\left\{v \in \mathfrak{h}^* \big\vert v =  \sum_{j=1}^r \mathbb{R}_{\geq 0} Q_{i_j}  \right\}$. 

In order to promote two local JK prescriptions in $\mathbb{C}^{\mathrm{dim \mathcal{G}_i}}$ to a global one in the product space $\mathbb{C}^{\mathrm{dim \mathcal{G}_1}}\times \mathbb{C}^{\mathrm{dim \mathcal{G}_2}}$, one defines a global $\eta \coloneqq (\eta_1, \eta_2)$ out of the local $\eta_i$. 
Notice also that the set of charge covectors one can read from $h(u_i,\omega_j,\xi_a)$ for a fundamental deconfinement (and the hyperplanes) in the product space are given by $Q_{i,j}\coloneqq (Q_1)_i \otimes (Q_2)_j$, where $Q_{1,2}$ are (anti-)fundamental charge covectors for the first and second gauge group respectively. By virtue of the generic form of the deconfinement identity \eqref{eq:deconf_generic}, namely it produces a nested integral where the $u_i$ holonomies of the outer integral are fixed parameters for the inner one, the definition of $\eta$ almost immmediately ensures that the global JK prescription defines an integral of iterated residues in which one simply picks in the inner integral precisely the same residues of the left hand-side of \eqref{eq:JKextension}. The critical case regards the situation of degenerate intersections. However, in the case of deconfinement, no new mixed degenerate intersections between the outer and the inner integral can arise beyond the ones already present in the nested integrals. Degeneracies can at most occur within the inner integral or in $f(u_i,\xi_a)$ in the independent term of the outer integral. Those degenerate singularities are not problematic when extending JK to a global prescriptions they will be consistently picked in the same way both by $\eta_1$ and $\eta=(\eta_1,\eta_2)$ for the outer ones, or by $\eta_2$ and the second entry of $\eta=(\eta_1,\eta_2)$ for the inner integral. 

In the following we will not prove in full generality all the details of this discussion as it would go beyond the scopes of the present paper and limit our discussion to the simple but meaningful example of \ref{sec:usp4}, in which the duality of a gauge theory with a Landau-Ginzburg model does not receive a strong justification from the reduction of four dimensional dualities. We will further confirm our results by explicitly evaluating the elliptic genus and matching it with the dual Landau Ginzburg model.

We start from the following expression for the elliptic genus
\begin{equation}
       \mathcal{I}(m_a,t_j) = \frac{(q;q)_\infty^4}{8}\prod_{j=1}^2\theta \left(q/t_j^4\right)\oint\displaylimits_{\mathrm{JK}} \mathrm{d}{u_1}\mathrm{d}{u_2} \frac{\theta \left(z_1^{\pm 1}z_2^{\pm 1}\right)  \theta \left(z_1^{\pm 2}\right) \theta \left(z_2^{\pm 2}\right)}{\prod_{j=1}^{2}\theta\left(t_j^2\right)\theta\left( z_1^{\pm 1} z_2^{\pm 1} t_j^2\right)\prod\displaylimits_{a,i=1}^{2}\theta\left(z_i^{\pm 1} m_a\right)},
\end{equation}
with $\prod_{a=1}^{2}m_a = x^2$.
As discussed in Section \ref{sec:usp4}, after deconfining one antisymmetric we get
\begin{equation}
    \frac{(q;q)_\infty^6}{16}\theta \left(q/t_2^4\right)\oint\displaylimits_{\mathrm{JK}} \mathrm{d}{u_1}\mathrm{d}{u_2} 
    \frac{\theta \left(z_1^{\pm 1}z_2^{\pm 1}\right)  \theta \left(z_1^{\pm 2}\right) \theta \left(z_2^{\pm 2}\right)}
    {\theta\left(t_2^2\right)\theta\left( z_1^{\pm 1} z_2^{\pm 1} t_2^2\right)\prod\displaylimits_{a,i=1}^{2}\theta\left(z_i^{\pm 1} m_a\right)}
    \oint\displaylimits_{\mathrm{JK}} \mathrm{d}\omega \frac{\theta \left(z_3^{\pm 2}\right)}{\prod_{i=1}^{2}\theta\left( t_1 z_i^{\pm 1} z_3^{\pm 1}\right)},
    \label{eq:deconf_usp_ell_gen}
\end{equation}
with $z_3=e^{2\pi \mathrm{i} \omega}$.
In order to interpret this integral as the genuine elliptic genus of the deconfined gauge theory described in Section \ref{sec:usp4}, and so to promote this identity as a genuine deconfinement, we need to show that \eqref{eq:deconf_usp_ell_gen} can be written as
\begin{equation}
    \frac{(q;q)_\infty^6}{16}\theta \left(q/t_2^4\right)\oint\displaylimits_{\mathrm{JK}} \mathrm{d}{u_1}\mathrm{d}{u_2} \mathrm{d}\omega
    \frac{\theta \left(z_1^{\pm 1}z_2^{\pm 1}\right)  \theta \left(z_1^{\pm 2}\right) \theta \left(z_2^{\pm 2}\right)}
    {\theta\left(t_2^2\right)\theta\left( z_1^{\pm 1} z_2^{\pm 1} t_2^2\right)\prod\displaylimits_{a,i=1}^{2}\theta\left(z_i^{\pm 1} m_a\right)}
      \frac{\theta \left(z_3^{\pm 2}\right)}{\prod_{i=1}^{2}\theta\left( t_1 z_i^{\pm 1} z_3^{\pm 1}\right)}.
      \label{eq:uuusss}
\end{equation}
In the language of the previous paragraphs we have
\begin{align}
    f(u_i,\xi_a,t_2) =& \frac{\theta \left(z_1^{\pm 1}z_2^{\pm 1}\right)  \theta \left(z_1^{\pm 2}\right) \theta \left(z_2^{\pm 2}\right)}
    {\theta\left(t_2^2\right)\theta\left( z_1^{\pm 1} z_2^{\pm 1} t_2^2\right)\prod\displaylimits_{a,i=1}^{2}\theta\left(z_i^{\pm 1} m_a\right)}, \nonumber \\
    h(u_i,\omega,t_1) =& \frac{\theta \left(z_3^{\pm 2}\right)}{\prod_{i=1}^{2}\theta\left( t_1 z_i^{\pm 1} z_3^{\pm 1}\right)}.
\end{align}
One can immmediately notice that no degeneracies are produced by the deconfinement; the singular hyperplanes are divided into two groups: the ones defined by the singularities of $f(u_i,\xi_a,t_2)$ and the ones arising from $h(u_i,\omega,t_1)$.
If we choose $\eta_2=1$ for the inner integral in \eqref{eq:deconf_usp_ell_gen} the only contributing poles are those for which
\begin{equation}
    z_3=z_1^{\pm 1} t_1^{-1}.
\end{equation}
The poles of the outer integral will be divided into those arising from $f(u_i,\xi_a,t_1)$, and those involving the singularities of the inner integral. In any case, having chosen $\eta_2 = 1$, we can now choose $\eta_1$ freely and all the contributing poles to $\mathcal{I}$ will be selected by the JK prescription. Independently on our choice of $\eta_1$ the very same poles can be obtained by starting from \eqref{eq:uuusss} and defining $\eta=(\eta_1,1)$. In fact, such choice will pick the same poles for the integration variable $z_3$ and then will fix the poles in the remaining variables as the ones of \eqref{eq:deconf_usp_ell_gen}, thus proving that we can relate \eqref{eq:deconf_usp_ell_gen} with the corresponding gauge theory obtained by deconfinement. This discussion can be now iterated straightforwardly for each step of deconfinement. We will not discuss the details of each step and will provide an independent proof of the consistency of all our steps with the results obtained on the field theory side by explicitly evaluating the elliptic genus of the theory and matching it with the predicted Landau Ginzburg model.

\section{Evaluation of $\USp (4)$ with two antisymmetrics and two $\square $}
\label{Andrea2}
In this section, we present an independent evaluation of the elliptic genus and demonstrate its agreement with the expression for the Landau-Ginzburg model, providing explicit confirmation of the consistency of our sequence of duality and deconfinement steps.

The Landau-Ginzburg model is 
\begin{equation}
    \mathcal{I}(u,x,t_j) = \frac{\theta \left(q /\! \left(t_1 t_2x\right)^4\right)}{\theta(x^2)\theta(t_1^2 t_2^2) \theta(x^2 u^2 t_1^2 t_2^2) \theta(x^2 u^{-2} t_1^2 t_2^2) \prod_{j=1}^{2} \theta \left(t_j^2 x^2\right) }.
    \label{eq:land_gin}
\end{equation}
The elliptic genus is defined by 
\begin{equation}
    \mathcal{I}(m_a,t_j) = \frac{(q;q)_\infty^4}{8}\oint\displaylimits_{\mathrm{JK}} \mathrm{d}{u_1}\mathrm{d}{u_2} \mathcal{Z}_{\text{1-loop}}(u_i,m_a,t_j),
    \label{eq:ell_gen}
\end{equation}
where 
\begin{equation}
    \mathcal{Z}_{\text{1-loop}}(u_i,m_a,t_j)= \prod_{j=1}^2\theta \left(q/t_j^4\right)\frac{\theta \left(z_1^{\pm 1}z_2^{\pm 1}\right)  \theta \left(z_1^{\pm 2}\right) \theta \left(z_2^{\pm 2}\right)}{\prod_{j=1}^{2}\theta\left(t_j^2\right)\theta\left( z_1^{\pm 1} z_2^{\pm 1} t_j^2\right)\prod\displaylimits_{a,i=1}^{2}\theta\left(z_i^{\pm 1} m_a\right)}, 
\end{equation}
together with the dictionary $m_1\equiv x u$ and $m_2\equiv x / u$, solving the balancing condition 
\begin{equation}
\prod_{a=1}^{2}m_a = x^2.
\end{equation}
The singularities $u^*=\left(u_1^*,u_2^*\right)$ of $\mathcal{Z}_{\text{1-loop}}$ lie on the following hyperplanes of the form $Q(u)+ c = 0$, with $Q \in \mathfrak{h}^*$ charge covectors and $\mathfrak{h}$ Cartan subalgebra of $\mathfrak{usp}(4)$:
\begin{align}
    \mathrm{H}_{12,a}^{\pm,\pm} =\left\{ \pm u_1 \pm u_2 +2\rho_a = m + n \tau\right\}, && \mathrm{H}_{i,a}^{\pm}=\left\{ \pm u_i + \xi_a = m + n \tau \right\},
    \label{eq:planes}
\end{align}
where $e^{2\pi \mathrm{i} \xi_a} \coloneqq m_a$ and $e^{2\pi \mathrm{i} \rho_a} \coloneqq t_a $ and $m,n \in \mathbb{Z}$.

Two (or more) hyperplanes must intersect in order for $u^*$ to be a contributing pole to $\mathcal{I}$.
All the singularities are non-degenerate, meaning that exactly two hyperplanes intersect at any given singularity, assuming generic values for the chemical potentials $\xi_a$ and $\rho_a$. We will always assume this to be the case. Then, all the poles will also be simple poles.

The integral becomes
\begin{equation}
    \mathcal{I} = \frac{(q;q)_\infty^4}{8}\sum_{u^*}\text{JK-Res}\left(\mathrm{Q}(u^*),\eta\right) \mathcal{Z}_{\text{1-loop}}(u^*).
\end{equation}
The JK-Residue functional is evaluated upon specifying a covector $\eta \in \mathfrak{h}^*$ and defining the charge covectors $Q(u^*)=\left\{Q \big\vert u^* \in \mathrm{H}\right\}$, generators of the hyperplanes intersecting at each given singularity $u^*$. The charge covectors divide $\mathfrak{h}^*$ in chambers and $\text{JK-Res}\left(\mathrm{Q}(u^*),\eta\right)\neq 0$ when $\eta \in \mathrm{Cone}(Q^*_i,Q^*_j)$.

The charge covectors of this model and our choice for $\eta = (1,1/2)$ are summarized in the diagram above.

\begin{figure}[t]
    \centering
\begin{tikzpicture}[thick, scale=1.5, >=stealth]
  \draw[->, black!30] (-1.5,0) -- (1.6,0) node[right] {$u_1^*$};
  \draw[->, black!30] (0,-1.5) -- (0,1.6) node[above] {$u_2^*$};

  \draw[->] (0,0) -- (1,0) node[anchor=south west, font=\scriptsize] {$(1,0)$};
  \draw[->] (0,0) -- (0,1) node[anchor=south, font=\scriptsize] {$(0,1)$};
  \draw[->] (0,0) -- (1,1) node[anchor=west, font=\scriptsize] {$(1,1)$};
  \draw[->,red] (0,0) -- (1,1/2) node[anchor=west, font=\scriptsize] {$\eta$};
  \draw[->] (0,0) -- (1,-1) node[anchor=west, font=\scriptsize] {$(1,-1)$};
  \draw[->] (0,0) -- (-1,0) node[anchor=south east, font=\scriptsize] {$(-1,0)$};
  \draw[->] (0,0) -- (0,-1) node[anchor=north, font=\scriptsize] {$(0,-1)$};
  \draw[->] (0,0) -- (-1,-1) node[anchor=east, font=\scriptsize] {$(-1,-1)$};
  \draw[->] (0,0) -- (-1,1) node[anchor=east, font=\scriptsize] {$(-1,1)$};
\end{tikzpicture}
\caption{Charge covectors diagram for $\mathrm{USp}(4)$. Our choice of $\eta$ is highlighted in red.}
\label{}
\end{figure}

The non-zero contributions for this choice of $\eta$ arise from the following intersections of hyperplanes:
\begin{enumerate}
\item $\mathrm{H}_{2,a}^{-} \cap \mathrm{H}_{12,b}^{+,+}$: $(\mathbf{u}_1)_{a,b} =(-\xi_a \! - \! 2\rho_b,\, \xi_a) \quad a,b = 1,2$.

\item $\mathrm{H}_{12,a}^{+,-} \cap \mathrm{H}_{2,b}^{+}$: $(\mathbf{u}_2)_{a,b} =(-\xi_b \! - \! 2\rho_a,\, -\xi_b) \quad a,b = 1,2$.

\item $\mathrm{H}_{12,a}^{+,-} \cap \mathrm{H}_{12,b}^{+,+}$: $(\mathbf{u}_3)_{a,b}^{k,l} =\left(-(\rho_a \!+\! \rho_b) \!+\! \frac{k + l \tau}{2},\,(\rho_a\! - \! \rho_b) + \frac{k + l \tau}{2}\right) \; a,b = 1,2, \; k,l=0,1$. 

\item $\mathrm{H}_{1,a}^{+} \cap \mathrm{H}_{12,b}^{-,+}$: $(\mathbf{u}_4)_{a,b} =(-\xi_a,\, -\xi_a \! - \! 2\rho_b) \quad a,b = 1,2$.

\item $\mathrm{H}_{1,a}^{+} \cap \mathrm{H}_{2,b}^{+}$: $(\mathbf{u}_5)_{a,b} =(-\xi_a,\, -\xi_b) \quad a,b = 1,2$.

\item $\mathrm{H}_{1,a}^{+} \cap \mathrm{H}_{12,b}^{+,+}$: $(\mathbf{u}_6)_{a,b} =(-\xi_a,\, \xi_a \! - \! 2\rho_b) \quad a,b = 1,2$.
\end{enumerate}

Additionally, we notice the further facts.
\begin{itemize}
 \item For each $a,b$ fixed the poles in $1.$, $2.$ and $4.$ are Weyl equivalent between themselves, thus they contribute equally to \eqref{eq:ell_gen}.
 \item Of all the poles in 3. and 5., the ones with $a=b$ provide vanishing contribution to \eqref{eq:ell_gen}, since such poles get canceled by a corresponding zero of the same order in the vector multiplet.
 \item An extra $1/2$ factor arises in the evaluation of the poles in 3. from the JK formula, due to the fact that the charge matrix obtained by $Q_1=(1,1),\,Q_2=(1,-1)$ generators of the intersecting hyperplanes has determinant 2. 
\end{itemize}
Furthermore, to evaluate the elliptic genus we notice that when picking the residue we have, from the quasi-double periodicity properties of the $q$-theta function,
\begin{equation}
\theta(z) \underset{z\to1}{\sim} (-1)(q;q)_\infty^2 \frac{z-1}{z}, \quad \theta(z) \underset{z\to q}{\sim} (q;q)_\infty^2 \frac{z-q}{z}.
\label{eq:theta_res}
\end{equation}
Then,
\begin{align}
    \mathcal{I} = \!\kappa \!\left( \mathcal{Z}_{\text{1-loop}}(\mathbf{u}_{3}) + \mathcal{Z}_{\text{1-loop}}(\mathbf{u}_5) + \sum_{a,b=1}^{2}\left(3\mathcal{Z}_{\text{1-loop}}(\mathbf{u}_1)_{a,b} + \mathcal{Z}_{\text{1-loop}}(\mathbf{u}_6)_{a,b} \right)\!\right),
    \label{eq:usp4_gauge_eval}
\end{align}
where $\displaystyle \mathcal{Z}_{\text{1-loop}}(\mathbf{u}_{3})\!=\!\!\sum_{a=1}^2 \!\sum_{k,l=0}^{1} \! \mathcal{Z}_{\text{1-loop}}(\mathbf{u}_{3})_{a,a^c}^{k,l}$,  $\displaystyle \mathcal{Z}_{\text{1-loop}}(\mathbf{u}_{5})\!=\!\!\sum_{a=1}^2 \mathcal{Z}_{\text{1-loop}}(\mathbf{u}_{5})_{a,a^c}$ and $\kappa \!= \!\frac{1}{8 \theta \left(x^2\right) \theta \left(t_1^2 t_2^2\right)}$.
We also define $a^c \in \{1,2\}$ as the unique element in $\{1,2\} \setminus \{a\}$. Explicitly we have,
\begin{align}
&  \mathcal{Z}_{\text{1-loop}}(\mathbf{u}_{3}) = \theta \left(x^2\right) \theta \left(t_1^{-2}\right) \theta \left(t_2^{-2}\right) \theta \left(t_1^{-2} t_2^{-2}\right)\!\!\!\!\!\!
\sum _{\epsilon \in \{\pm 1,\pm \sqrt{q}\}} \frac{q x^{-4} \delta_{|\epsilon|,\sqrt{q}}}{\prod _{a=1}^2 \theta \left(\epsilon\, m_a \, t_1^{\pm 1} t_2^{\pm 1}\right)}, \nonumber \\
& \mathcal{Z}_{\text{1-loop}}(\mathbf{u}_{5}) = \frac{ 2\, \theta\left( t_1^2t_2^2 \right)\theta \left(m_1^{-2}\right) \theta \left(m_2^{-2}\right) \theta \left(x^{-2}\right) \theta \left(t_1^4\right) \theta \left(t_2^4\right)}
{\prod_{a=1}^{2} \theta \left(t_a^2\right) \theta \left(t_a^2 \, m_1^{\pm 1} m_2^{\pm 1}\right) }, \nonumber\\
& \mathcal{Z}_{\text{1-loop}}(\mathbf{u}_{1})_{a,b} = 
\frac {3\,\theta\left(t_b^{4}\right)\theta\left(m_a^{-2}t_{b^c}^{-4}\right)\theta\left(m_a^{-2}t_{b^c}^{-2}\right)}
{\theta\left(\frac{m_{a^c}}{m_a}\right) \theta \left(\frac{m_{a^c}}{m_{a} t_{b^c}^2}\right)\theta\left(m_a^2t_1^2t_2^2\right) \theta \left(t_b^2\right) \theta \left(\frac{t_b^2}{t_{b^c}^2}\right)  \theta \left(\frac{t_b^2}{m_a^2 t_{b^c}^2}\right) \theta \left(x^2 t_{b^c}^2\right)}, \nonumber \\
&  \mathcal{Z}_{\text{1-loop}}(\mathbf{u}_{6})_{a,b} = \frac{\theta \left(m_a^{-2}\right) \theta \left(t_b^{-2}\right) \theta \left(t_{b^c}^4\right) \theta \left(m_a^2 t_b^{-4}\right) \theta \left(t_b^2 m_a^{-2}\right)}
{\theta \left(t_1^2\right) \theta \left(t_2^2\right) \theta \left(m_a^2\right) \theta \left(\frac{m_{a^c}}{m_a}\right)\theta \left(\frac{t_1^2 t_2^2}{m_a^2}\right) \theta \left(\frac{t_{b^c}^2}{t_b^2}\right) \theta \left(\frac{x^2}{t_b^2}\right) \theta \left(\frac{t_b^2 m_{a^c}}{m_a}\right) \theta \left(\frac{m_a^2 t_{b^c}^2}{t_b^2}\right)}. \nonumber \\
\end{align}
One can check that \eqref{eq:usp4_gauge_eval} precisely matches the singular behaviors and symmetries of \eqref{eq:land_gin}. Additionally, we verified that their $q$-series expansion in coincide up to order $q^6$, thus providing a strong evidence that \eqref{eq:land_gin} and \eqref{eq:ell_gen} are equal.

\section{Comments on c-extremization}
\label{commentsR} 

We conclude this section by commenting on c-extremization in the various cases studied above.
The choice of $R$ charges that we have made here corresponds to $R=0$ for the chirals and $R=1$ for the Fermi.
This implies that the central charges corresponding to such choice, i.e. $c_R = 3 \kappa_{R_0 R_0}$, are always positive. However, in general, if we allow the mixing $R = R_0 +\alpha_i F_i$, where $F_i$ are the $\UU(1)$ abelian generators for the various flavor symmetries, the exact $R$ symmetry has to be determined by extremizing the function $\kappa_{RR}$ with respect to the $\alpha_i$ coefficients. Nevertheless, in most of the cases studied above, such mixing gives rise to a negative central charge (either $c_L$ or both $c_R$ and $c_L$) at least for some values of the rank $N$ of the $\SU(N)$ gauge group. This situation is similar to the one discussed in \cite{Sacchi:2020pet}, where the interpretation of this fact is related to the presence of non-compact direction in the target space, that indeed cannot be included in the extremization problem. This implies in general that the exact $R$ charge is the one with $R=0$ for the chirals and $R=1$ for the Fermi, and that it is then determined from the  $\kappa_{R_o R_0}$ anomalies.

Actually we studied the c-extremization by allowing general mixing in each case. Sporadically we found cases where the central charges are  positive for some ranges of the gauge symmetry rank, generically by turning off only some of the mixing factors for the $\UU(1)$ symmetries.
However, we have found that such symmetries are associated to non-compact directions in the target space by looking at the equatios of motion for the chirals in the LG superpotential. It implies that the relative mixing factors have to be turned off in the extremization problem. 

This discussion has a counterpart in the analysis of the elliptic genus. Indeed, by turning off the non-abelian fugacities, we have observed divergencies, due to such non-compact directions, induced by the leftover abelian fugacities associated to non-compact directions of the target space discussed above.
A similar discussion has appeared in \cite{Sacchi:2020pet}.

\section{Remarks}
\label{sec:remarks}
We would like to emphasize some conceptual aspects of this work.
In this paper we discussed new examples of gauge/LG dualities for theories with $\mathcal{N}=(0,2)$ supersymmetry. Such dualities are obtained from the dimensional reduction of 4d confining gauge theories compactified on two sphere with a topological twist using the prescription of \cite{Gadde:2015wta}. This prescription guarantees the validity of the identity at the level of  the elliptic genus, provided the fact that the 4d $S^2\times T^2$ indices match. In this sense providing a derivation of the conjectural identity in 2d is a check of the 4d identity as well (at least for the zero flux sector).

Throughout the paper we have showed such 2d identities by using the trick denoted as tensor deconfinement in 4d (and then in 3d), and in addition we have been forced to apply the trick in some flipped version of the duality, i.e. where some gauge invariant combinations of the 2d $\mathcal{N}=(0,2)$ chirals is set to zero by some J-term.
We have then further checked the validity of the 't Hooft anomaly matching for the proposed dualities. Such check has been performed in the case where the flippers are turned on, even if we have performed the same check in the model obtained directly from the 4d reduction (where the $S^2 \times T^2$ identity guarantees the validity of the matching).

One key question regards the IR dynamics of such theories. Even if the 4d dynamics is confining we claim that the 2d model becomes conformal, similarly to the results obtained by \cite{Dedushenko:2017osi,Sacchi:2020pet}, for other cases of gauge/LG dualities. A first concern regards a negative value for the central charge arising from a naive application of the c-extremization procedure. However, we still expect that no chiral symmetry breaking occurs and the dynamics is indeed conformal. The main motivations behind such expectation is related to the application of the c-extremization procedure in presence of non-compact target space. Often in these situations one ends up with negative value for the central charge from a naive evaluation of such quantity. The presence of non-compact directions in the target space represents a caveat to the procedure discussed in \cite{Benini:2012cz}, which then requires a modification of the c-extremization procedure as discussed in \cite{Sacchi:2020pet}. Then, it may be possible that the theory is still conformal with a well defined central charge, however one must identify the exact central charge of the theory removing the mixing with currents associated to the non-compact directions in the target space. Compatibly with the divergent structure of the elliptic genus we have observed that the trial R-charges that we have used in each case correspond to the exact charges as well. In this way we have computed the left and the right central charges in each case, obtaining a consistent picture with respect to the c-theorem. Another argument supporting our hypothesis is that we have verified that the unitarity bounds are never violated in any case, thus avoiding potential issues arising from the presence of accidental IR symmetries.

Another crucial point regards the physics of the flippers. Such fields have been often used in 4d and in 3d in order to study models with accidental symmetries, and more recently they have been used to simplify the analysis of dualities where the tensor has been necessary in order to derive a duality in terms of other, somehow more fundamental, dualities.
Here the flippers have been necessary and not only useful for our analysis. Furthermore, due to the chiral nature of supersymmetry in the 2d models under investigations, such flippers corresponds to Fermi fields. We have restricted our analysis to the case were E-terms are set to zero, even if the condition may be relaxed by converting the E-terms into J-terms. Nevertheless, the conditions $E\cdot J = 0$ does not always allow us to have both $E$ and $J$ non vanishing. 
It would be interesting to enlarge our analysis to such cases as well, and we are currently investigating situations of this type, but in the paper at hand such situation never arises.
An important aspect of the physics of such flippers is that they originate from the reduction of the s-confining limit of the IP duality, where an $\USp(2n)$ gauge theory with $2n-2$ fundamental chirals is dual to an antisymmetric chiral meson, with a non trivial J-term, setting its Pfaffian to zero. The Fermi that emerges in the dual LG enforcing the J-term is (up to our knowledge) not clearly understood in the literature, in the sense that the duality map has  not been stated for such Fermi field in terms of gauge invariant operators of the electric phase. Nevertheless, its charges can be read from the J-term and from the duality map relating the chiral singlet in the dual phase and the antisymmetric chiral meson of the electric phase.
Another evidence of such duality is given by the fact that it can be worked out as a boundary duality starting for the 3d confining limit of $\USp(2n)$ Aharony duality, and in such case the 2d ``superpotential" is identical to the 3d one, provided the fact that the monopole operator acting as a singlet in the Aharony dual phase is traded with the 2d Fermi, forcing the same equation of motion for the antisymmetric meson. Then, such Fermi corresponds to the 2d boundary monopole in the language of \cite{Dimofte:2017tpi}.

The invariance under duality of the deconfining procedure in presence of non-compact target space is in general a delicate issue that requires a complete understanding of the target space at quantum level. Indeed even if at classical level the various steps (deconfining tensors and dualizing) preserve the target space, at quantum level this is not guaranteed in presence of non-compact target space. For this reason the dualities and the procedure used to derive them adopted in the paper have to be considered in the mass deformed case, i.e. where only isolated vacua are leftover. Observe that this does not imply that the dualities survive in the massless case, where indeed the identities between the elliptic genera relate divergent quantities, and they cannot be used.
This is consistent with similar observations appearing in the literature \cite{Sacchi:2020pet, Jiang:2024ifv} for similar 2d $\mathcal{N}=(0,2)$ dualities.

Here, we can provide further arguments supporting the robustness of our results.
The crucial point is that we can fully rely on the validity of the 4d duality and the 4d deconfining procedure on which we have much better control. Indeed, all our cases can be mapped in each step of deconfinement and duality to the 4d ``parent" ones performed in \cite{Bajeot:2022kwt}. This means that by assigning to each step of \cite{Bajeot:2022kwt} the integer non-negative R-charges as we have fixed here, consistently with the 4d duality map, we can reduce to the corresponding 2d intermediate steps discussed here. This guarantees the robustness of the 2d deconfining procedure in all scenarios we explored.

\section{Conclusions}
\label{sec:conclusions}

In this paper we have studied 2d $\mathcal{N}=(0,2)$ gauge theories with a LG dual description in terms of chiral and Fermi multiplets.
A generic feature of the gauge theories studied here is that they only have charged matter associated to chiral multiplets, and the possible Fermi fields on the gauge theory side are introduced only to flip some gauge invariant combinations of the charged matter fields themselves.
The dual LG models have instead chiral multiplets associated to the gauge invariant combinations that are not set to zero on the gauge theory side by the Fermi flippers.
There are also Fermi fields in the dual LG models that allow for the presence of J-terms. The global charges of such Fermi are then read from the superpotentials themselves even if their origin in the duality map is unclear at this level.

This last feature is common to other similar models discussed in the literature \cite{Gadde:2015wta,Putrov:2015jpa,Dedushenko:2017osi,Sacchi:2020pet} that can be derived by twisted compactification of 4d $\mathcal{N}=1$ confining gauge theories.
Similarly to the results of \cite{Gadde:2015wta,Putrov:2015jpa,Dedushenko:2017osi,Sacchi:2020pet}, most of the models studied here can be derived from 4d by considering two s-confining dualities studied in \cite{Csaki:1996zb} involving $\SU(N)$ SQCD with an antisymmetric or an antisymmetric flavor.

In this sense most of the dualities found here can be ``derived" from 4d as indeed we showed in the body of the paper.
However, the 2d proofs  of our dualities allowed us to go beyond the relation with the 4d models.
Indeed we have proposed that another model, corresponding to $\SU(2n)$ with an antisymmetric flavor and four fundamentals is dual to a LG model.
The interesting fact of the proof is that this model can be studied using only dualities that have a 4d origin (i.e. from twisted compactification on $S^2$), despite the fact that the model itself is not originating from the compactification of any s-confining theory.
We have found a further duality without an immediate 4d origin, involving $\USp(4)$ with two antisymmetrics and two fundamentals.

In addition, all the models found here have a 3d counterpart, extensively studied in \cite{Nii:2019ebv}, when the 3d dual picture has two types of gauge invariant fields appearing in the confining superpotential.
These last are mesonic and baryonic combinations of the charged matter fields (singlets) and possibly dressed monopoles that describe the Coulomb branch.
One can observe that the 2d LG found here are almost identical to the 3d duals of \cite{Nii:2019ebv}, provided the relation of the 3d singlets with the 2d chiral multiplets\footnote{This is the reason why we decided to present the 2d superpotentials instead of the J-terms, setting as explained above the E-terms to zero. Indeed, we wanted to stress the similarity between such 2d superpotentials and other 3d superpotentials appearing for models with the same field content on the gauge side of the duality (provided the fact that the 3d  $\mathcal{N}=2$ chirals of the 3d models are "identified" with the charged $\mathcal{N}=(0,2)$ chirals of the 2d models studied here).} and of the 3d monopoles with the 2d Fermi fields (with the correct normalization of the $R$ symmetry of the superpotential, i.e. $R[W_{3\dd}] =2$ and $R[W_{2\dd}] =1$).

We hope that such observation  can 
be helpful in the understanding  of the 
reason why the models discussed here can be derived from 4d through the topological twist procedure, that is indeed not guaranteed a priori. Indeed in general one might expect that the 4d duality is preserved in 2d by the presence of finite size effects, in analogy with the 4d/3d reduction where such effects are captured by the KK monopoles.
Here such roles is expected \cite{Gadde:2015wta} to be  played by surface defect of Gukov-Witten \cite{Gukov:2006jk,Gukov:2008sn} type. The fact that removing the 3d KK monopoles through real mass flow lead to the dressed monopoles and the similarity of these last with the Fermi fields that we obtained in the LG description  may be relevant in order to understand the role of the finite size effects from the 2d perspective. 
It would be interesting also to connect the 3d and the 2d dynamics along the lines of the dual boundary conditions studied in \cite{Dimofte:2017tpi}. For the dualities studied here a relevant discussion appeared in \cite{Okazaki:2023hiv}, as discussed above in Section \ref{sec:usp4}.

There are many other possible developments that we are planning to investigate.
For example the similarity between the models found here and the higher dimensional confining gauge theories suggests the existence of other
2d gauge theories with a LG dual that have not been conjectured so far in the literature.
In a recent paper \cite{Jiang:2024ifv} some of such models have been proposed by twisted compactification of 4d $\mathcal{N}=2$ gauge theories.
The structure of the identities for the elliptic genera of such models remind similar structures found in 3d for the matching of the three sphere partition functions.
For many of these cases it should be possible to give a pure 2d derivation of these dualities along the lines of the analysis performed here.
 
Another class of 2d $\mathcal{N}=(0,2)$ dualities was obtained by compactifying 4d dualities on a magnetized torus \cite{Kutasov:2013ffl,Kutasov:2014hha, Tatar:2015sga,Tatar:2017pnm, Gao:2024syg}. It would be interesting to see if the ADE type dualities of \cite{Kutasov:2014hha} can follow from the basic ones in absence of tensor(s), in the same spirit of the recent analysis of \cite{Benvenuti:2024glr,Hwang:2024hhy} in higher dimension.

A last comment regards the existence of star-triangle type relations for the dualities obtained here. In the case of $\USp(2N)$ dualities (either with $2N+2$ fundamentals or with four fundamentals and one antisymmetric) such
relations have been extensively discussed in \cite{de-la-Cruz-Moreno:2020xop}. It would be interesting to investigate similar relations associated to the dualities discussed here.

\section*{Acknowledgments}
The work of the authors has been supported in part by the Italian Ministero dell'Istruzione, Universit\'a e Ricerca (MIUR), in part by Istituto Nazionale di Fisica Nucleare (INFN) through the “Gauge Theories, Strings, Supergravity” (GSS) research project.

\appendix

\section{Basic dualities}
\label{sec:basic}
Here we review the basic dualities that we have used in order to prove  the new dualities in the body of the paper.
Such dualities have been discussed in the literature, and they have been derived by the $S^2$ reduction of 4d dualities using the prescription of  \cite{Gadde:2015wta}.

\subsection{$\SU(N)$ with $N$ fundamental and $N$ anti-fundamental chirals}
\label{sec:SUNNN}

This duality originates from the limiting case of 4d $\SU(N)$ Seiberg duality with $N+1$ flavors.
The model has been discussed in \cite{Gadde:2015wta,Putrov:2015jpa}. It can be derived from 4d by twisting the superfields by assigning one $R$ charge equal to one to a fundamental and an anti-fundamental and 
a vanishing $R$ charge to the other fundamentals.

In the dual description such assignment of $R$ charges allows the existence of a chiral meson $\Phi_M$ of the leftover non-abelian flavor symmetries, two other chirals corresponding to the baryon $\Phi_B$ and the antibaryon $\Phi_{\tilde B}$ and a Fermi  field, corresponding to the $M_{N+1,N+1}$ component of the 4d meson, that has indeed $R$ charge 2.
The 4d superpotentials $W  = B M \tilde B+\det M$ becomes 
\begin{equation}
W = \Psi (\Phi_{\phantom{\tilde{B}}\!\!\!\!\! B} \Phi_{\tilde B} + \det \Phi_M),
\end{equation}
and one can verify that the global anomalies among the gauge theory and the dual LG model match.
Furthermore the duality translates into a matching between the elliptic genera, as discussed in \cite{Putrov:2015jpa}
\begin{eqnarray}
\label{ellipticdualSU}
I_{\SU(N)}^{(N;N;\cdot;\cdot;\cdot)}(\vec u;\vec v;\cdot;\cdot;\cdot)
=
\frac{\theta\left(q/\prod_{a=1}^N u_a v_a\right)}{
\theta\left(\prod_{a=1}^N u_a\right)
\theta\left(\prod_{a=1}^N v_a\right)
\prod_{a,b=1}^N \theta (u_a v_b)}.
\nonumber \\
\end{eqnarray}

\subsection{$\SU(N)$ with $N+1$ fundamental and $N-1$ anti-fundamental chirals}
\label{confGWR}
Here we provide evidences of another duality involving a 2d $\mathcal{N}=(0,2)$ $\SU(N)$ gauge theory with $N+1$ fundamental chirals $Q$ and $N-1$ anti-fundamental chirals  $\tilde Q$.
We claim that the dual LG involves chiral meson $\Phi_M = Q \tilde Q$,
a chiral baryon $\Phi_B = Q^N$ and a Fermi $\Psi$, with superpotential 
\begin{equation}
\label{claimappendix}
W = \Psi \Phi_M \Phi_B.
\end{equation}

The global charges of the fields are
\begin{equation}
\begin{array}{c|cccc}
&\UU(1)_Q&\UU(1)_{\tilde Q}& \SU(N+1)&\SU(N-1) \\
Q &1&0&\square&\cdot\\
\tilde Q&0&1&\cdot&\square\\
\hline 
\Phi_M & 1&1&\square&\square \\
\Phi_B& N&0 &\overline \square&\cdot\\ 
\Psi&-N-1&-1&\cdot&\overline \square 
\end{array}
\end{equation}
A first check of this duality consists of matching the global anomalies. 
They are indeed 
\begin{equation}
    \kappa_{QQ}=N(N+1), \quad
    \kappa_{Q \tilde Q}=0, \quad
    \kappa_{\tilde Q \tilde Q}=N(N-1), \quad
    \kappa_{\SU(N+1)^2}=\kappa_{\SU(N-1)^2}=\frac{N}{2},
\end{equation}
in both the electric and magnetic phase.

We can also provide a derivation of the duality from 4d by topologically twisting the theory on a two-sphere.
Starting from 4d $\SU(N)$ with $N+1$ fundamental flavors, the twist is done along the non-anomalous $R$ symmetry that assigns $R$ charge 0 to the all the fundamentals  and to $N-1$ anti-fundamentals and  $R$ charge 1 to the remaining  two anti-fundamentals.

On the dual side we have three gauge singlets, the meson $M=Q \tilde Q$, the baryon $B = Q^{2N}$ and the anti-baryon $\tilde B = \tilde Q^{2N}$.
We can see that $N^2-1$ components out of the $(N+1)^2$ components of the mesons have $R$ charge zero while the remaining components have $R$ charge 1. The $N+1$ dimensional baryon has $R$ charge zero as well, while $(N-1)$ components of the anti-baryon have $R$ charge 2 and the remaining two components have $R$ charge 1.

At the level of the 2d field content this tells us that the electric theory has an $\SU(N+1) \times \SU(N-1)$ non-abelian flavor symmetry with $N+1$ 
fundamental chirals and $N-1$ anti-fundamental chirals. On the other hand the dual LG model has a meson $\Phi_M = Q \tilde Q$, a baryon $\Phi_B=Q^{2N}$ and a Fermi $\Psi$. We can also construct the 2d superpotential starting from the   4d one, $W = \det M + B M \tilde B $. The first term disappears while the second term becomes the 2d superpotential  (\ref{claimappendix}) as claimed above.

A further check of the duality consists of studying the case $N=2$, where the $\SU(2)$ gauge theory can be regarded as $\USp(2)$. In this case the 
four fundamentals $Q_{1,2,3}$ and $\tilde Q$ on the gauge theory side reconstruct an $\SU(4)$ fundamentals, that we can denote as $P_{1,2,3,4}$. This can be seen also on the dual side, where the superpotential can be written in terms of the contractions of the charged fields as
\begin{equation}
W = \Psi \epsilon_{ijk} (\tilde Q Q_i) ( Q_j  Q_k)
\propto \Psi  \epsilon_{\ell ijk} (P_\ell  P_i) (P_j P_k)
= \Psi \Pf A ,
\end{equation}
where $A=P^2$ is the antisymmetric meson of the $\USp(2)$ gauge theory.

At the level of the elliptic genus the duality translates in the conjectural identity 
\begin{equation}
\label{chirallSU}
I_{\SU(N)}^{(N+1;N-1;\cdot;\cdot;\cdot)}(\vec u;\vec v;\cdot;\cdot;\cdot)
=
\frac{\prod_{b=1}^{N-1} \theta\left(q/(v_b\prod_{a=1}^{N+1} u_a )\right)}{
\prod_{c=1}^{N+1} \theta\left(\prod_{a=1}^{N+1} u_a/u_c\right)
\prod_{a=1}^{N+1} \prod_{b=1}^{N-1}  \theta (u_a v_b)}\,.
\end{equation}

In this case we further checked the identity for higher rank by expanding the index at finite $N$.
We have computed the index  by using the JK-residue prescription and then by expanding the result either at order  $q^0$ by turning on the non-abelian fugacities or at higher order in $q$ but setting to one the other fugacities. 
For example, for the first non-trivial case\footnote{The case $N=2$ is actually the case of $\USp(2)$ with four fundamentals already discussed in the literature. We will provide a full derivation of this last case in Appendix \ref{sec:USP2Np2}.}, corresponding to $N=3$, we have evaluated the index 
by combining  the poles in the form $(z_1,z_2)$, where $z_1$ and $z_2$ are taken from the sets 
below 
\begin{eqnarray}
\label{list1}
(z_1 \in \{  u_i^{-1}, v_j {z_2^*}^{-1} \}, z_2 \in \{ u_i^{-1},  v_1^{-1} v_2^{-1}  \}),
\end{eqnarray}
with $i=1,\dots,4$ and $j=1,2$ and where the $z_2^*$  are the ones taken the from the second set.

For example at order $q^0$ we found that the index (\ref{chirallSU}) becomes 
\begin{equation}
I_{\SU(3)}^{(4;2;\cdot;\cdot;\cdot)}(\vec u;\vec v;\cdot;\cdot;\cdot)
\xrightarrow{q \rightarrow 0}
\frac{\displaystyle \prod _{i=1,2}^2 \big(1-v_i \prod _{j=1}^4 u_j\big)}
{\displaystyle \prod _{i=1}^2 \prod _{j=1}^4 \left(1-v_i u_j\right) \cdot \!\!\! \!\!\! 
\prod _{1\leq i<j<k\leq 4} \!\!\!\! \!\! \left(1-u_i u_j u_k\right)
}
\end{equation}
in both the gauge theory and in the LG model.
At higher orders in $q$ we kept only the abelian fugacities by defining $u_i = x m_i$ and $v_i = y  n_i$ 
with $m_1 m_2 m_3 m_4= n_1 n_2= 1$.  
In this case we found 
\begin{eqnarray}
I_{\SU(3)}^{(4;2;\cdot;\cdot;\cdot)}(x \vec m;y \vec n;\cdot;\cdot;\cdot)
&&\xrightarrow{ \vec m,\vec n \rightarrow 1}
\phantom{+} \frac{\left(x^4 y-1\right)^2}{\left(x^3-1\right)^4 (x y-1)^8} \\
- &&
q \,\frac{2\left(x^4 y-1\right)^2 \left(x^8 y^2-2 x^7 y-4 x^5 y^2-4 x^3-2 x y+1\right)}{x^4 \left(x^3-1\right)^4 y (x y-1)^8}+ \order{q^2},\nonumber
\end{eqnarray}
where we omit the higher orders because they are not very illuminating, but we checked explicitly the matching in the dual phases up to $q^4$.

\subsection{$\SU(N)$ with $N+2$ fundamental and $N-2$ anti-fundamental chirals}
\label{appdualGRW}

This duality is a subcase of a more general duality studied in \cite{Gadde:2015wta}  for $\SU(N)$  with fundamentals and  anti-fundamental chiral and fundamental Fermi multiplets.
Here we discuss the explicit derivation of the duality in order to obtain the relation between the charges and the matching of the elliptic genera.

We start by considering 4d $\SU(N)$ SQCD with $N+2$ fundamental flavors and we parametrize the $R$ symmetries of the fundamentals and the anti-fundamentals in terms of the global symmetries
\begin{equation}
R_{Q_a} = R_0 + b + t_a,\quad
R_{\tilde Q_a} = R_0 - b + w_a,
\end{equation}
where $R_0$ is a trial $R$ symmetry, $b$ represents the baryonic symmetry 
and $t_a$ and $w_a$ refer to  the abelian generators of the $\SU(N+2)^2$ flavor symmetry, imposing the constraints $\sum_{a=1}^{N+2} t_a = \sum_{a=1}^{N+2} w_a = 0$.
There is a further constraint from the requirement that the $R$ symmetry is anomaly free, corresponding to $\sum_{a=1}^{N+2} (R_{Q_a}+R_{\tilde Q_a})=4$.

The charge assignation where all the fundamentals and $N-2$ anti-fundamentals have $R$ charge $0$ and the remaining anti-fundamentals have $R$ charge $1$ is then anomaly free and gives rise to $N+2$ chiral bosons in the fundamentals and $N-2$ chiral bosons in the anti-fundamental of the $\SU(N)$ gauge group in the reduced 2d $\mathcal{N}=(0,2)$ model.

On the other hand the $R$ charges of the flavors of the dual $\SU(2)$ gauge theory are
\begin{equation}
R_{\tilde q_a} = \frac{1}{2} \sum_{c=1}^{N+2}  R_{Q_c}-R_{Q_a}, \quad
R_{q_a} =  \frac{1}{2} \sum_{c=1}^{N+2}  R_{\tilde Q_c}-R_{\tilde Q_a}.
\end{equation}
In this dual theory the $N+2$ anti-fundamentals survive as chiral bosons $\tilde q$ while only $N-2$ anti-fundamentals survive, but this time as Fermi multiplets $\Psi_q$.
There is also a  meson $M$ in the bifundamental of the  $\SU(N+2) \times \SU(N-2)$ flavor symmetry that survives and the 2d sueperpotential read from the 4d one is
\begin{equation}
W = M \tilde q \Psi_q\,.
\end{equation}
The $R$ charge assignment discussed above allows us also to read the global charges of the fields in the dual phases. They are summarized in the following table
\begin{equation}
\label{tablecharges}
\begin{array}{c|ccccc}
& \SU(N+2) & \SU(N-2)&\UU(1)_B&\UU(1)_A&\UU(1)_R \\
Q&\square&\cdot&1&1&0\\
\tilde Q&\cdot&\square&-1&1&0 \\
\hline
\tilde q&\overline \square&\cdot&\frac{N}{2}&\frac{N}{2}&0\\
\Psi_q&\cdot&\overline \square&-\frac{N}{2}&-\frac{N+4}{2}&1\\
M&\square&\square&0&2&0
\end{array}
\end{equation}

One can check that the abelian global anomalies match between the two phases. 
They are 
\begin{equation}
    \kappa_{AA}=\kappa_{BB}=2N^2, \qquad
    \kappa_{AB}=4N.
\end{equation}
The anomalies of the non-abelian symmetries are 
$\kappa_{\SU(N+2)^2}=\kappa_{\SU(N-2)^2}=\frac{N}{2}$
and they match as well.

The identity among the elliptic genera in this case becomes
\begin{equation}
\label{idbrutta}
I_{\SU(N)}^{(N+2;N-2;\cdot;\cdot;\cdot)}(\vec t;\vec w;\cdot;\cdot;\cdot)
=
\prod_{\ell=1}^{N+2}\prod_{j=1}^{N-2} \frac{1}{\theta_0(t_\ell w_j)}
I_{\SU(2)}^{(\cdot;N+2;N-2\cdot;\cdot)}\left(\cdot;\vec{\tilde t};\vec {\tilde w};\cdot;\cdot\right),
\end{equation}
where
$\tilde t_j = \sqrt{\prod_{\ell=1}^{N+2} t_\ell}/ t_j$ and $\tilde w_j =1/\left(\sqrt{\prod_{\ell=1}^{N+2} t_\ell}  w_j\right)$.
Observe that in this formula the fugacities $u_a$ and $v_a$ can be represented also in terms of the fugacities of the global symmetries in formula  (\ref{tablecharges}).
Denoting the fugacity of the baryonic symmetry by $b$, the fugacity of the axial symmetry by $a$, the fugacities of the $\SU(N+2)$ as $u_{\ell=1,\dots,N+2}$ and the  fugacities of the $\SU(N-2)$ as $v_{\ell=1,\dots,N-2}$  with $\prod_{\ell=1}^{N+2} u_\ell= \prod_{\ell=1}^{N-2} v_\ell=1$, we can use the new fugacities mapped to the ones in (\ref{idbrutta}) through the relation 
$u_\ell a b = t_\ell$ and $v_\ell = a/b w_{\ell}$, such that the identity becomes
\begin{equation}
\label{idbrutta2}
I_{\SU(N)}^{(N+2;N-2;\cdot;\cdot;\cdot)}(a b \vec u;a/b \vec v;\cdot;\cdot;\cdot)
=\!\!
\prod_{\ell=1}^{N+2}\prod_{j=1}^{N-2} \frac{1}{\theta_0(a^2 u_\ell v_j)}
I_{\SU(2)}^{(\cdot;N+2;N-2\cdot;\cdot)}\bigg(\!\cdot;(ab)^{\frac{N}{2}} \vec{ \tilde u};\frac{\vec {\tilde v}}{a^{\frac{N+4}{2}} b^{\frac{n}{2}}};\cdot;\cdot\!\bigg),
\end{equation}
with $\tilde u_\ell = u_\ell^{-1}$ and $\tilde v_\ell = v_\ell^{-1}$

We can provide some explicit checks for this identity by evaluating the index at finite $N$.
It is more convenient to  parameterize the fugacities $a$ and $b$ as $x \equiv a b $ and $y \equiv a/b$,
such that $Q$ has charge $1$ under $U(1)_x$ and zero under $U(1)_y$ while  $\tilde Q$ has charge $1$ under $U(1)_y$ and zero under $U(1)_x$.

Then we compute the index at order $q^0$ with all the global fugacities turned on. We consider the case $N=3$, because the case $N=2$ corresponds to a trivial self-duality. In this case we have only one anti-fundamental field. Then, the left-hand side of the identity is evaluated using the JK-res prescription on the poles of the form $(z_1,z_2)$ with
\begin{eqnarray}
    \label{list2}
    (z_1 \in \{  u_i^{-1}x^{-1}, y^{-1} {z^*_2}^{-1} \}, z_2 \in \{ u_i^{-1}x^{-1}  \}),
\end{eqnarray}
where $i=1,\dots,5$ and again $z^*_2$ is evaluated on each element of the second set of \eqref{list2}.

The result is not particularly illuminating, but we report it here for completeness 
\begin{eqnarray}
    &&
    I_{\SU(3)}^{(5;1;\cdot;\cdot;\cdot)}(x \vec u; y;\cdot;\cdot;\cdot)\xrightarrow{q \rightarrow 0} u_1^3 u_2^3 u_3^3u_4^3\Big(u_3 x^4 (y-x^{10}y^2-x^5)\nonumber  \\ && + x^6 (x y (u_4 (u_3 x^6-1)-u_3)+x^5 y (y-x^5)+1)+u_2 x^4 \big(x^3 y (u_3 x^6-1) \nonumber \\ 
    &&  
    + u_4 (x^9 y-u_3 (x^{10} y^2+x^5-y))\big)+u_1^2 u_2 u_3 u_4 x^4 \big(x^2 \big(x y (u_4 (u_3 x^6-1)-u_3) \nonumber \\
    && 
    + x^5 y (y-x^5)+1\big)
    u_2 \big(x^3 y (u_3 x^6-1)
    u_4 (x^9 y-u_3 (x^{10} y^2+x^5-y))\big)\big) \nonumber \\
    && 
    + u_1 \big(x^7 y (u_3 x^6-1)+
    u_2^2 u_3 u_4 x^6 (x y (u_4 (u_3 x^6-1)-u_3)+x^5 y (y-x^5)+1) \nonumber \\
    && 
    + u_4 x^4 (x^9 y-u_3 (x^{10} y^2+x^5-y))+
    u_2 \big(u_3 u_4^2 x^6 (-u_3 x y+x^5 y (y-x^5)+1) \nonumber \\
    &&+
    u_4 (u_3 x^6-1) \big(u_3 (x^5 y (y-x^5)+1)+ x^{14} y^2+x^9-x^4 y \big) + x^{13} y\big)\big) \Big)\Big{/} \nonumber \\
    &&
    \big((x y-u_1 u_2 u_3 u_4)
    \prod _{1\leq a<b<c\leq 4}  (x^3 u_a u_b u_c-1)
    \prod _{1\leq a < b \leq 4} (x^3-u_a u_b)
    \prod _{a=1}^4(x y u_a - 1)\big)
    \nonumber \\
    &&
    \phantom{ I_{\SU(3)}^{(5;1;\cdot;\cdot;\cdot)}(x \vec u; y;\cdot;\cdot;\cdot)}\xleftarrow{q \rightarrow 0}\prod_{\ell=1}^{5} \frac{1}{\theta_0(x y u_\ell)}I_{\SU(2)}^{(\cdot;4;1;\cdot;\cdot)}(\cdot; x \vec u;y;\cdot;\cdot).
\end{eqnarray}

We also checked the identity for higher orders in $q$, by turning off the non-abelian fugacities, \textit{i.e.} by setting $u_a,\, v_b$ to $1$.
We obtained a matching between the left-hand side of the identity and the right-hand side up to $q^4$ but the result is not very illuminating. We write explicitly only the first order in $q$:
\begin{eqnarray}
    I_{\SU(3)}^{(5;1;\cdot;\cdot;\cdot)}(x \vec u; y;\cdot;\cdot;\cdot)= 
    &&
    \frac{x^{11} y^2+3 x^8 y^2-5 x^7 y+x^6+x^5 y^2-5 x^4 y+3 x^3+1}{\left(x^3-1\right)^7 (x y-1)^5} \nonumber \\
    &&
    + q\Big(x^{21} \left(-y^2\right)+7 x^{18} y^2+26 x^{16} y^3-21 x^{15} y^2+10 x^{14} y^4 \nonumber \\ 
    && 
    + 13 x^{13} y^3-40 x^{12} y^2+x^{11}
    \left(5 y^4+y\right)+x^{10} \left(y^3+5\right)-40 x^9 y^2\nonumber \\
    &&
    + 13 x^8 y+10 x^7-21 x^6 y^2+26 x^5 y+7 x^3 y^2-y^2\Big)/\nonumber \\
    &&
    \Big(x^5 \left(x^3-1\right)^7 y (x y-1)^5\Big) +
    \mathcal{O}(q^2).
\end{eqnarray}
In addition, we verified the identity also for $N=4$ at order $q^0$ turning off the non-abelian fugacities. The result is
\begin{align}
    &I_{\SU(4)}^{(6;2;\cdot;\cdot;\cdot)}(x \vec u; y;\cdot;\cdot;\cdot)= 
    \Big(x^{24} y^4+6 x^{20} y^4\!-\! 12 x^{19} y^3+3 x^{18} y^2+6 x^{16} y^4\!-\!32 x^{15} y^3+39 x^{14} y^2 \nonumber \\
    &\!-\!12 x^{13} y+x^{12} \left(y^4+1\right)
    \!-\!12 x^{11} y^3 + 
    39 x^{10} y^2-32 x^9 y+6 x^8+3 x^6 y^2-12 x^5 y+6 x^4+1\Big)
    \nonumber \\
    &
    /\left(\left(x^4-1\right)^9 (x y-1)^{12}\right),
\end{align}
matching precisely the dual phase.

\subsection{$\USp(2N)$ with $2N+2$ fundamental chirals }
\label{sec:USP2Np2}

This duality has been proposed in \cite{Gadde:2015wta,Dedushenko:2017osi} by reducing the 4d confining $\USp(2N)$ SQCD with $2N+4$ fundamentals. The twist requires two fundamentals to have $R=1$ and all the others $R=0$.
The 2d duality obtained in this way  relates an $\USp(2N)$ gauge theory with $2N+2$ fundamental chiral bosons $Q$ to a LG theory with an antisymmetric chiral boson $\Phi = Q^2$ and a Fermi $\Psi$ with superpotential $W = \Psi \Pf \Phi$. This duality translates on the elliptic genus into the conjectured identity 
\begin{equation}
\label{ellipticdualUSP}
I_{\USp(2N)}^{(2N+2,\cdot)}(x\vec u;\cdot;\cdot)
=
\frac{\theta\left(q/x^{2N+2} \right)}{\prod_{1\leq a<b\leq 2N+2} \theta(u_a u_b x^2)}\, ,
\end{equation}
where $\prod_{a=1}^{2N+2}u_a=1$.
We computed explicitly the identity \eqref{ellipticdualUSP} for the case $N=1$. Here, the gauge theory side is given by
\begin{equation}
    I_{\USp(2)}^{(4,\cdot)}(\vec{u},x) = \frac{(q;q)_\infty^{2}}{2}\jkint\frac{\dd{z}}{2\pi \mi z}\frac{\theta\qty(z^{\pm 2})}{ \prod_{a=1}^4\theta\qty(z^{\pm1}u_a x)}.
    \label{eq:USP2_4}
\end{equation}
This integral has been already evaluated in \cite{Putrov:2015jpa}. In the following we provide an alternative derivation of the result.
The poles contributing to the integral are of the form $z=u_a^{-1}x^{-1}$, for $a=1,\ldots 4$. The singular behavior of the $\theta$ function near the poles can be extracted from the first order expansion 
\begin{align}
    \theta(z u_a x) &= \prod_{n=1}^\infty (1-zu_a x q^k)(1-(zu_a x)^{-1}q^{k+1})\nonumber\\
    &\sim(1-zu_a x)\prod_{k=1}^\infty (1-q^k)^2=(1-zu_a x)(q;q)_\infty^2\\
    \label{eq:sing_theta}
\end{align}
The contribution $(q;q)_\infty^2/z$ in \eqref{eq:USP2_4} cancels with the corresponding terms in \eqref{eq:sing_theta} when evaluated at the poles. Summing over the residues, the integral becomes
\begin{equation}
    I_{\USp(2)}^{(4,\cdot)}(\vec{u},x) = \frac
    { \sum_{a=1}^4\theta(1/(x u_a)^2) \prod_{b\neq a}\theta\qty(u_a/u_b)  \prod_{\substack{c<b,\,c\neq a}}\theta\qty(u_b/u_c)\theta\qty(u_c/u_b)\theta(x^2 u_b u_c)}
    { \prod_{a<b}\theta\qty(u_a/u_b)\theta\qty(u_b/u_a)\theta(u_au_b x^2)},
\end{equation} 
thus for the identity \eqref{ellipticdualUSP} to hold, we need to prove that
\begin{equation}
    \theta(x^4)=\frac{
         \sum_{a=1}^4\theta(1/(x u_a)^2) \prod_{b\neq a}\theta\qty(u_a/u_b) \prod_{\substack{c<b,\,c\neq a}}\theta\qty(u_b/u_c)\theta\qty(u_c/u_b)\theta(x^2 u_b u_c)
    }{\prod_{a<b}\theta(u_a u_b^{-1})\theta(u_b u_a^{-1})}\,,
    \label{eq:equality}
\end{equation}
where we used \eqref{inversion} on the LHS.

We start by giving the following definitions. The Jacobi $\theta$ functions can be defined in terms of infinite products as \cite{Kharchev_2015}
\begin{equation}
    \begin{aligned}
        &\theta_1(\xi|\tau) \coloneqq 
        2 q^{\frac{1}{8}} \sin \pi \xi \, 
        \prod_{n=1}^{\infty} 
        \left( 1-q^n \right)
        \left( 1-q^n s\right)
        \left( 1-q^n/s\right)\,, \\
        &\theta_2(\xi|\tau) \coloneqq 
        2 q^{\frac{1}{8}} \cos \pi \xi \, 
        \prod_{n=1}^{\infty} 
        \left( 1-q^n \right)
        \left( 1+q^n s\right)
        \left( 1+q^n /s\right)\,, \\
        &\theta_3(\xi|\tau) \coloneqq 
        \prod_{n=1}^{\infty} 
        \left( 1-q^n \right)
        \left( 1+q^{n-\frac{1}{2}} s\right)
        \left( 1+q^{n-\frac{1}{2}} /s\right)\,, \\
        &\theta_3(\xi|\tau) \coloneqq 
        \prod_{n=1}^{\infty} 
        \left( 1-q^n \right)
        \left( 1-q^{n-\frac{1}{2}} s\right)
        \left( 1-q^{n-\frac{1}{2}} /s\right)\,, \\
    \end{aligned}
\end{equation}
where
\begin{equation}
    q= \ee^{2\pi \mi\tau}\,, \qquad s= \ee^{2\pi \mi \xi}\,.
\end{equation}
The two functions $\theta(\xi|\tau)$ and $\theta_1(\xi|\tau)$ are related by
\begin{equation}
    \theta_1(\xi|\tau)=  \mi \, (q;q)_{\infty} \,  \ee^{\pi  \mi \left(\frac{\tau}{4}-\xi\right)} \, \theta(\xi|\tau)\,.
    \label{eq:theta0theta1}
\end{equation}
In what follows we suppress the dependence on the modular parameter $\tau$. Using \eqref{eq:theta0theta1}, equation \eqref{eq:equality} is then written as
\begin{equation}
    \label{toprov}
    \begin{aligned}
        2\,\theta_1(4x)\,=\,&\frac{
        \theta_1(2x+\xi_1+\xi_2)\,\theta_1(2x+\xi_1+\xi_3)\,
        \theta_1(2x+\xi_2+\xi_3)\,\theta_1(-2x-2\xi_4)
        }{
        \theta_1(\xi_1-\xi_4)\,\theta_1(\xi_2-\xi_4)\,\theta_1(\xi_3-\xi_4)
        }\,+\\ 
        +\,&\frac{
        \theta_1(2x+\xi_1+\xi_2)\,\theta_1(2x+\xi_1+\xi_4)\,
        \theta_1(2x+\xi_2+\xi_4)\,\theta_1(-2x-2\xi_3)
        }{
        \theta_1(\xi_1-\xi_3)\,\theta_1(\xi_2-\xi_3)\,\theta_1(\xi_4-\xi_3)
        }\,+\\ 
        +\,&\frac{
        \theta_1(2x+\xi_1+\xi_3)\,\theta_1(2x+\xi_1+\xi_4)\,
        \theta_1(2x+\xi_3+\xi_4)\,\theta_1(-2x-2\xi_2)
        }{
        \theta_1(\xi_1-\xi_2)\,\theta_1(\xi_3-\xi_2)\,\theta_1(\xi_4-\xi_2)
        }\,+\\ 
        +\,&\frac{
        \theta_1(2x+\xi_2+\xi_3)\,\theta_1(2x+\xi_2+\xi_4)\,
        \theta_1(2x+\xi_2+\xi_4)\,\theta_1(-2x-2\xi_1)
        }{
        \theta_1(\xi_2-\xi_1)\,\theta_1(\xi_3-\xi_1)\,\theta_1(\xi_4-\xi_1)
        }\,.
    \end{aligned}
\end{equation}
Borrowing the notation of \cite{Kharchev_2015}
\begin{equation}
    [r] \equiv \theta_r(X)\,\theta_r(Y)\,\theta_r(Z)\,\theta_r(W)\,, \qquad
    r=1,2,3,4\,,
\end{equation}
we can express the terms in \eqref{toprov} in a more convenient form. For example
\begin{equation}
    \theta_1(2x+\xi_1+\xi_2)\,\theta_1(2x+\xi_1+\xi_3)\,
        \theta_1(2x+\xi_2+\xi_3)\,\theta_1(-2x-2\xi_4)\,,
\end{equation}
can be written as $[1]$, upon defining 
\begin{equation}
    X=2x+\xi_1+\xi_2\,, \quad Y=2x+\xi_1+\xi_3\,, \quad Z=2x+\xi_2+\xi_3\,, \quad W=-2x-\xi_4\,.
\end{equation}
Through a five-term Riemann identity \cite{Kharchev_2015}, we rewrite $[1]$ as
\begin{equation}
    \label{5tr}
    2\,[1]=[1]'+[2]'-[3]'+[4]',
\end{equation}
where 
\begin{equation}
    [r]'\equiv \theta_r(X')\,\theta_r(Y')\,\theta_r(Z')\,\theta_r(W')\,,
    \qquad
    r=1,2,3,4\,,
\end{equation}
which depend on the following ``dual variables" \cite{Kharchev_2015}
\begin{equation}
    \begin{aligned}
        X'&=\frac{1}{2}(-X+Y+Z+W)=\xi_3-\xi_4 \,,\\
        Y'&=\frac{1}{2}(X-Y+Z+W)=\xi_2-\xi_4 \,,\\
        Z'&=\frac{1}{2}(X+Y-Z+W)=\xi_1-\xi_4 \,,\\
        W'&=\frac{1}{2}(X+Y+Z-W)=4x+\xi_1+\xi_2+\xi_3+\xi_4=4x\,.
    \end{aligned}
\end{equation}
In the last equation, we used the $\SU(N)$ constraint $\sum_{a=1}^{4}\xi_a=0$. By using \eqref{5tr}, the first term of \eqref{toprov} becomes
\begin{equation}
    \frac{1}{2}\,\theta_1(4x) + \sum_{r=2}^{4} \frac{
        \theta_r(4x)\,\theta_r(\xi_1-\xi_4)\,
        \theta_r(\xi_2-\xi_4)\,\theta_r(\xi_3-\xi_4)
    }{
        2 \, \theta_1(\xi_1-\xi_4)\,
        \theta_1(\xi_2-\xi_4)\,\theta_1(\xi_3-\xi_4)
    }.
\end{equation}
If we repeat this procedure also for the other three terms of \eqref{toprov}, and we use the parity of the $\theta$ functions
\begin{equation}
    \theta_r(-\xi)=(-1)^{\delta_{1r}}\,\theta_r(\xi)\,,
\end{equation}
we see that the proof of \eqref{toprov} is equivalent to prove that
\begin{equation}
    \label{toprov2}
    \begin{aligned}
        \sum_{r=2}^{4}(-1)^{\delta_{3r}}\,\Biggl(\,
        &\frac{
        \theta_r(4x)\,\theta_r(\xi_1-\xi_4)\,
        \theta_r(\xi_2-\xi_4)\,\theta_r(\xi_3-\xi_4)
        }{
        \theta_1(\xi_1-\xi_4)\,
        \theta_1(\xi_2-\xi_4)\,\theta_1(\xi_3-\xi_4)
        }\,+\\
        +&\,\frac{
        \theta_r(4x)\,\theta_r(\xi_1-\xi_2)\,
        \theta_r(\xi_2-\xi_3)\,\theta_r(\xi_2-\xi_4)
        }{
        \theta_1(\xi_1-\xi_2)\,
        \theta_1(\xi_2-\xi_3)\,\theta_1(\xi_2-\xi_4)
        }\,+\\
        -&\,\frac{
        \theta_r(4x)\,\theta_r(\xi_1-\xi_2)\,
        \theta_r(\xi_1-\xi_3)\,\theta_r(\xi_1-\xi_4)
        }{
        \theta_1(\xi_1-\xi_2)\,
        \theta_1(\xi_1-\xi_3)\,\theta_1(\xi_1-\xi_4)
        }\,+\\
        -&\,\frac{
        \theta_r(4x)\,\theta_r(\xi_1-\xi_3)\,
        \theta_r(\xi_2-\xi_3)\,\theta_r(\xi_3-\xi_4)
        }{
        \theta_1(\xi_1-\xi_3)\,
        \theta_1(\xi_2-\xi_3)\,\theta_1(\xi_3-\xi_4)
        }
    \,\Biggr)\,=\,0\,.
    \end{aligned}
\end{equation}
Now, if we focus on the first two terms of \eqref{toprov2}, we have that their sum amounts to 
\begin{equation}
    \label{pgiuda1}
    \frac{
        \theta_r(4x)\,\theta_r(\xi_2-\xi_4)\,R(\xi)
        }{
        \theta_1(\xi_1-\xi_2)\,\theta_1(\xi_1-\xi_4)\,
        \theta_1(\xi_2-\xi_4)\,\theta_1(\xi_2-\xi_3)\,
        \theta_1(\xi_3-\xi_4)
        }\,,
\end{equation}
with
\begin{equation}
    R(\xi)=[11rr]+[rr11]\,,
\end{equation}
where we used the notation
\begin{equation}
    \begin{aligned}
        &[11rr] \equiv \theta_1(X)\,\theta_1(Y)\,\theta_r(Z)\,\theta_r(W)\,, \\
        &[rr11] \equiv \theta_r(X)\,\theta_r(Y)\,\theta_1(Z)\,\theta_1(W)\,,
    \end{aligned}
\end{equation}
with 
\begin{equation}
    X=\xi_1-\xi_2\,,\quad
    Y=\xi_2-\xi_3\,,\quad
    Z=\xi_1-\xi_4\,,\quad
    W=\xi_3-\xi_4\,.
\end{equation}
In this case we use a four-term Jacobi identity \cite{Kharchev_2015},
\begin{equation}
    [11rr]+[rr11]=[11rr]'+[rr11]'\,,
    \qquad r=2,3,4\,,
\end{equation}
where the dual variables then become
\begin{equation}
    X'=0\,,\quad
    Y'=\xi_1-\xi_3\,,\quad
    Z'=\xi_2-\xi_4\,,\quad
    W'=\xi_1-\xi_2+\xi_3-\xi_4\,.\quad
\end{equation}
Hence, the term \eqref{pgiuda1} becomes, for $r=2,3,4$,
\begin{equation}
    \label{eq:bello}
    \frac{
        \theta_r(0)\,\theta_r(4x)\,\theta_r(\xi_1-\xi_3)\,\theta_r(\xi_2-\xi_4)\,
        \theta_1(\xi_1-\xi_2+\xi_3-\xi_4)
    }{
        \theta_1(\xi_1-\xi_2)\,\theta_1(\xi_1-\xi_4)\,
        \theta_1(\xi_2-\xi_3)\,\theta_1(\xi_3-\xi_4) 
    }\,.
\end{equation}
Lastly, by applying the same procedure to the sum of the last two terms of \eqref{toprov2}, they simplify to
\begin{equation}
    -\frac{
        \theta_r(0)\,\theta_r(4x)\,\theta_r(\xi_1-\xi_3)\,\theta_r(\xi_2-\xi_4)\,
        \theta_1(\xi_1-\xi_2+\xi_3-\xi_4)
    }{
        \theta_1(\xi_1-\xi_2)\,\theta_1(\xi_1-\xi_4)\,
        \theta_1(\xi_2-\xi_3)\,\theta_1(\xi_3-\xi_4) 
    }\,,
\end{equation}
for $r=2,3,4$ which cancels precisely with \eqref{eq:bello}. We conclude that the identity \eqref{toprov} is proven.

For the higher rank cases, we checked the results perturbatively in the modular parameter $q$. One could try to compute the identity exactly, but the computation is highly dependent on $N$.

\subsection{$\USp(2N)$ with  one antisymmetric and four fundamental chirals}
\label{appSacchi}

This duality has been derived in \cite{Sacchi:2020pet} through the same approach that we have largely used in this paper.
The model can be derived   by topologically twisting the s-confining  model with $\USp(2N)$ gauge group,   
six fundamentals $Q$ and one totally (traceless) antisymmetric two index tensor $A$, originally studied in \cite{Cho:1996bi,Csaki:1996eu}. 
If two fundamentals have $R$ charge $R=1$, while the other $R$ charges for the remaining fields are set to $R=0$, 
the 2d model has a LG dual in terms of dressed mesons and Fermi multiplet.

The duality has been proven through an iterative procedure by trading the antisymmetric matter with another $\USp(2N-2)$ 
gauge group and by dualizing the original $\USp(2N)$ node. By iterating this process one arrives to the expected LG theory.
The superpotential for the LG is a simple function of towers of mesons and Fermi multiplets if the traces $\Tr A^{j}$ in the electric theory are  flipped by a tower of Fermi fields $\hat \Psi_i$
through the superpotential 
\begin{equation}
W = \sum_{j=2}^{N} \hat \Psi_j \Tr A^j.
\end{equation}

In this case the LG dual is described by the mesons $\Phi_{ab} ^{(j)} = Q_a A^{j-1} Q_b$ with $1\leq a<b\leq 4$ and $j=1,\dots,N$.
The dual superpotential is then
\begin{equation}
\label{sacchiUSp4ASspot}
W = \sum_{j_1,j_2,j_3=1}^{N} \epsilon_{abcd}  \Psi_{j_1} \Phi_{ab}^{(j_2)}\Phi_{cd}^{(j_3)} \delta_{j_1+j_2+j_3,2N+1}.
\end{equation}

The identity relating the elliptic genera of the dual phase can be derived following the same iterative process spelled out above,
i.e. by using only the relation (\ref{ellipticdualUSP}).
The identity has been derived in  \cite{Sacchi:2020pet}  and it is
\begin{eqnarray}
\label{USp(2n)A4}
I_{\USp(2N)}^{(4;\cdot;1)}(\vec u;\cdot; t)
=
\frac{\prod_{\ell=1}^{N} \theta(q/(t^{2N-1-\ell}\prod_{a=1}^4 u_a))}{\prod_{\ell=0}^{N-1}\prod_{a<b} \theta(u_a u_b t^\ell)}.
\end{eqnarray}

\bibliographystyle{JHEP}
\bibliography{ref.bib}

\end{document}